\documentclass[%
reprint,
amsmath,
amssymb,
aps,
pra,
]{revtex4-2}

\usepackage[english]{babel}
\usepackage{silence}
\usepackage{graphicx}
\usepackage{braket}
\usepackage{dcolumn}
\usepackage{bm}
\usepackage{esvect}
\usepackage{mathrsfs}
\usepackage{xcolor}
\usepackage[unicode=true, bookmarks=true,bookmarksnumbered=false,bookmarksopen=false,
breaklinks=false,pdfborder={0 0 1},backref=false,colorlinks=true, linkcolor=magenta, urlcolor=blue, citecolor=blue, pdfstartview={FitH}, hyperfootnotes=true, linktocpage]{hyperref}
\usepackage[capitalise]{cleveref}
\usepackage{appendix}
\usepackage{tikz}
\usetikzlibrary{matrix}
\usepackage{makecell}
\usepackage{array}
\usepackage{colortbl}

\makeatletter

\DeclareMathAlphabet\mathbfcal{OMS}{cmsy}{b}{n}

\newcommand{\tr}{\text{Tr}}
\newcommand{\sket}[1]{\ket{#1}\rangle}
\newcommand{\sbra}[1]{\langle\bra{#1}}
\newcommand{\sbraket}[2]{\langle\langle #1 | #2 \rangle \rangle}
\newcommand\scalemath[2]{\scalebox{#1}{\mbox{\ensuremath{\displaystyle #2}}}}

\newcommand{\bD}{{\mathbf D}}
\newcommand{\bP}{{\mathbf P}}
\newcommand{\bx}{{\mathbf x}}
\newcommand{\bG}{{\mathbf G}}
\newcommand{\bmm}{{\mathbf m}}

\usepackage{soul}
\usepackage[normalem]{ulem}

\colorlet{lightred}{red!40!}
\colorlet{lightblue}{cyan!30!}
\definecolor{anti-flashwhite}{rgb}{0.95, 0.95, 0.96}
\definecolor{beige}{rgb}{0.96, 0.96, 0.86}
\definecolor{khaki}{rgb}{0.76, 0.69, 0.57}
\definecolor{lavenderblush}{rgb}{1.0, 0.94, 0.96}
\definecolor{manatee}{rgb}{0.59, 0.6, 0.67}
\definecolor{beaublue}{rgb}{0.74, 0.83, 0.9}

\makeatletter
    \def\CT@@do@color{%
      \global\let\CT@do@color\relax
            \@tempdima\wd\z@
            \advance\@tempdima\@tempdimb
            \advance\@tempdima\@tempdimc
    \advance\@tempdimb\tabcolsep
    \advance\@tempdimc\tabcolsep
    \advance\@tempdima2\tabcolsep
            \kern-\@tempdimb
            \leaders\vrule
                    \hskip\@tempdima\@plus  1fill
            \kern-\@tempdimc
            \hskip-\wd\z@ \@plus -1fill }
    \makeatother

\begin{document}

\preprint{APS/123-QED}

\title{Driven-Dissipative Ising Model: An exact field-theoretical analysis}

\author{Daniel A. Paz}
\email{pazdanie@msu.edu}
\author{Mohammad F. Maghrebi}
\affiliation{%
Department of Physics \& Astronomy, Michigan State University,  East Lansing Michigan 48824
}%

\date{\today}

\begin{abstract}
Driven-dissipative many-body systems are difficult to analyze analytically due to their non-equilibrium dynamics, dissipation and many-body interactions.
In this paper, we consider a driven-dissipative infinite-range Ising model with local spontaneous emission, which naturally emerges from the open Dicke model in the large-detuning limit. Utilizing an adaptation of the Suzuki-Trotter quantum-to-classical mapping, we develop an exact field-theoretical analysis and a diagrammatic representation of the spin model
that can be understood from a simple scattering picture. With this representation, we are able to analyze critical behavior, finite-size scaling and the effective temperature near the respective phase transition. 
Our formalism further allows a detailed study of the ordered phase where we find a ``heating'' region within which the effective temperature becomes negative, thereby exhibiting a truly non-equilibrium behavior. 
At the phase transition, we find two distinct critical behaviors with overdamped and underdamped critical dynamics at generic and weakly-dissipative critical points, respectively. 
We further show that the  underdamped critical behavior is robust against short-range perturbations and is not an artifact of the mean-field nature of the model. 
To treat such perturbations, we extend our diagrammatic representation to include the coupling to spin waves due to the short-range interactions. 
The field-theoretical approach and the diagrammatics developed in this work should prove useful in applications to generic short-range driven-dissipative spin systems.
\end{abstract}

\maketitle

\tableofcontents

\section{Introduction}
Open quantum systems that are coherently driven, widely known as \textit{driven-dissipative systems}, have received a great deal of attention in recent years. The interplay between drive, dissipation, and many-body interactions gives rise to rich physics of both fundamental interest and practical importance. Various driven-dissipative systems have been experimentally realized in numerous platforms such as cavity QED and cold atoms \cite{baumann_dicke_2010,baden_realization_2014, muniz_exploring_2020}, circuit QED \cite{fitzpatrick_observation_2017}, and trapped ions \cite{genway_generalized_2014, barreiro_open-system_2011}, thanks to the rapid advancement of experimental techniques in the past twenty years. In addition, they are prominent platforms for quantum simulation and quantum computation \cite{verstraete_quantum_2009}, especially relevant to quantum computing systems in the NISQ era \cite{preskill_quantum_2018}. However, driven-dissipative systems remain difficult to treat numerically due to their exponentially large Hilbert space, and analytically due to the presence of many-body interactions and non-equilibrium dynamics. Any physically relevant model that is amenable to theoretical treatment is therefore of vital importance to better understand the physics behind driven-dissipative systems.\par 
We present here a thorough investigation of a minimal many-body driven-dissipative system, the driven-dissipative Ising model (DDIM) with infinite-range interactions \cite{paz_critical_2019}, a many-body model that admits analytical solutions and is numerically tractable. This model describes a system of coherently driven atoms interacting via an infinite-ranged Ising-type interaction and in the presence of a transverse magnetic field. Each atom is also weakly coupled to a zero-temperature Markovian bath giving rise to individual atomic spontaneous emission. The DDIM can be directly realized in experiments \cite{kim_entanglement_2009} and is closely related to the open Dicke model \cite{dimer_proposed_2007,baumann_dicke_2010, torre_keldysh_2013, muniz_exploring_2020} that itself is one of the most well-studied driven-dissipative systems.
In fact, in the limit of large laser-cavity detuning, the infinite-range DDIM can be derived from the open Dicke model through adiabatic elimination of the cavity mode \cite{damanet_atom-only_2019, muniz_exploring_2020}. In this paper, we present a comprehensive study of the DDIM in various regimes: deep in the ordered phase as well as the phase transition; at weak or strong dissipation; and, with or without short-range perturbations that spoil the mean-field character of the model. \par 
The model being infinite-ranged means a mean-field analysis 
is exact in the thermodynamic limit, providing access to the exact phase diagram. To investigate fluctuations, however, we must go beyond mean field. To this end, we introduce a non-equilibrium quantum-to-classical mapping formalism which allows for an exact mapping of the driven-dissipative model to a Keldysh field theory \cite{sieberer_keldysh_2016, torre_keldysh_2013}, a non-trivial task due to the local nature of the dissipation. Representing the model as a field theory allows for the study of fluctuations as well as their finite-size scaling near the phase transition, from which the static and dynamical critical exponents characterizing the phase transition are extracted. The phase transition in the limit of weak dissipation yields a different set of exponents from anywhere else along the phase boundary, representing the fundamentally distinct underdamped critical dynamics in contrast with the relaxational dynamics at a generic critical point.\par

The distinct critical behavior at the weakly dissipative point might be viewed as an artifact of the infinite-range interactions, hence the mean-field nature of the model.
To analytically test the robustness of the weakly-dissipative critical behavior, we set out on a detailed study of the  DDIM model perturbed by short-range interactions. We develop a powerful diagrammatic representation that captures the interactions with spin waves due to the short-range perturbation. With this technology, we show that the underdamped critical behavior persists even in the presence of short-range interactions. We believe that our field-theoretical techniques will prove useful in the analytical study of more generic driven-dissipative spin systems with local dissipation. \par 
This paper is structured as follows: In Sec. \ref{model}, we briefly review the DDIM, derive it from the open Dicke model and provide mean-field solutions. Next, we provide an introduction to a general non-equilibrium quantum-to-classical mapping and derive the non-equilibrium Keldysh field theory in Sec. \ref{mapping}. Section \ref{analysis} covers the field-theoretical analysis of the DDIM including its critical properties, exponents, finite-size scaling, effective temperature, and the emergence of distinct stochastic Langevin equations at generic and weakly-dissipative critical points. In Sec. \ref{integrability}, we introduce perturbative short-range interactions and study its effect on the critical behavior using the diagrammatic techniques. Finally, we summarize our main results and discuss  future extensions and applications in Sec. \ref{conclusion}. 

\section{Model}\label{model}
The infinite-range DDIM is recovered from a natural limit of  the open Dicke model. The latter model is one of the quintessential many-body non-equilibrium models, having been experimentally realized in multiple contexts \cite{baumann_dicke_2010, baumann_exploring_2011, baden_realization_2014, klinder_dynamical_2015, zhiqiang_nonequilibrium_2017, safavi-naini_verification_2018, muniz_exploring_2020} and extensively studied theoretically \cite{torre_keldysh_2013, dalla_torre_dicke_2016, keeling_collective_2010, damanet_atom-only_2019, hwang_dissipative_2018, konya_finite-size_2012, nagy_critical_2011, liu_light-shift-induced_2011, gegg_superradiant_2018}. The Hamiltonian describes a collection of atoms interacting with a single cavity mode. In this driven-dissipative variant,
the interaction is mediated by an external drive to boost the achievable interaction strength compared to the single-atom detuning \cite{dimer_proposed_2007}. Besides the coherent dynamics due to the Hamiltonian, the cavity itself is lossy, leading to dissipative dynamics. Furthermore, atomic spontaneous emission may not be neglected as it can change the onset of the phase transition  \cite{baden_realization_2014}. The ensuing dynamics is governed by a quantum master equation that incorporates the effects of loss alongside the coherent dynamics. For the open Dicke model with cavity loss and atomic spontaneous emission, the Hamiltonian, in the rotating frame of the drive \cite{dimer_proposed_2007, torre_keldysh_2013},  takes the form
\begin{equation}\label{ODM}
    \dot \rho=\mathcal{L}[\rho] = -i[H_{\text{Dicke}}, \rho] + \kappa\mathcal{D}_a [\rho] + \Gamma \sum_i \mathcal{D}_{\sigma^-_i} [\rho]\,.
\end{equation}
Here, the operator $\mathcal{L}$ denotes the Liouvillian and the (curly) brackets denote the (anti-)commutator. The Dicke model's Hamiltonian $H_{\rm Dicke}$ is given by
\begin{equation}
    H_{\rm Dicke} = \omega_0 a^\dagger a + \Delta S_z + \frac{2g}{\sqrt{N}}(a + a^\dagger)S_x\,.
\end{equation}
Here, $a$ denotes the cavity mode operator and the atoms are represented by Pauli operators. Given the collective nature of the model, we have introduced the total spin $S_{\alpha} = \sum_i \sigma^\alpha_i$ with $\alpha=x,y,z$. Note that $\omega_0$ is the cavity detuning, $\Delta$ is the transverse field, and $g$ is the drive-mediated spin-cavity interaction. Finally, the dissipative dynamics is described by
\begin{equation}\label{dissipator}
    \mathcal{D}_L[\rho] = L \rho L^\dagger - \frac{1}{2}\{L^\dagger L, \rho\}\,,
\end{equation}
with the Lindblad operator $L$. Dissipation is comprised of photon loss of the leaky cavity at the rate $\kappa$, corresponding to the Lindblad operator $L= a$, and spontaneous emission of individual atoms via $L=\sigma_i^-$ at the rate $\Gamma$. 

Since \cref{ODM} is quadratic in the photon field, one can naively ``integrate out'' the photon field, which in turn generates an Ising-type interaction. More rigorously, the cavity mode can be adiabatically eliminated in the large-detuning limit and with large $\kappa$ ($\omega_0, \kappa  \gg g$), giving rise to a driven-dissipative infinite-ranged Ising model with spontaneous emission as well as dephasing  \cite{damanet_atom-only_2019}:
\begin{equation}
\dot \rho_S =\mathcal{L}_S[\rho_S] =  -i[H_S, \rho_S] + \Gamma\sum_i \mathcal{D}_{\sigma^-_i} [\rho_S] + \frac{\Gamma_x}{N}\mathcal{D}_{S_x} [\rho_S]\,,
\end{equation}
with the system Hamiltonian
\begin{equation}\label{Hamiltonian}
    H_S = -\frac{J}{N}S_x^2 + \Delta S_z\,.
\end{equation}
The subscript $S$ denotes the system of atoms upon integrating out photons, and the parameters in the above equations are related to those in the Dicke model as  $J = 16g^2 \omega_0/(\kappa^2 + 4\omega_0^2)$, and $\Gamma_x = J \kappa/\omega_0$. Note that the factor $1/N$ in front of the Ising term makes the Hamiltonian extensive in the system size. This Hamiltonian is also known as the Lipkin-Meshkov-Gorkov model, and has been studied extensively in a variety of contexts \cite{pan_analytical_1999, morita_exact_2006, das_infinite-range_2006, botet_size_1982, muniz_exploring_2020, louw_thermalization_2019}. 
An important consequence of integrating out the cavity mode is the introduction of $x$-dephasing via $L = S_x$ due to the microscopic photon loss. If one takes the additional limit where $\omega_0 \gg \kappa \gg g$, we have $\Gamma_x \ll J$ and the contribution from the dephasing becomes negligible. The transverse field $\Delta$ and the decay rate $\Gamma$ are microscopic parameters and are unaffected by this limit, and can be chosen to be of the same order as $J$. The resulting  model is the driven-dissipative  infinite-ranged Ising model with spontaneous emission:
\begin{equation}\label{DDIM}
    \mathcal{L_S}[\rho] = -i[H_S, \rho_S] + \Gamma \sum_i (\sigma^-_i \rho_S \sigma^+_i - \frac{1}{2}\{\sigma^+_i \sigma^-_i, \rho_S\})\,.
\end{equation}
For a rigorous derivation, see Appendix \ref{ODM to DDIM}.
For notational convenience, we shall drop the subscript $S$ in the rest of the paper. 
\par

Equation \eqref{DDIM} defines a minimal model of driven-dissipative spin systems, also directly relevant to experiments realizing the Dicke model. The infinite-range interaction makes a myriad of analytical and numerical techniques available, and makes this model an ideal setting to explore questions which are otherwise difficult or rather intractable in more complex models. This model has a $\mathbb{Z}_2$ symmetry upon changing $\sigma_{x,y} \to -\sigma_{x,y}$ and exhibits a phase transition from a normal phase ($\langle S_x \rangle = 0 )$ to an ordered phase ($\langle S_x \rangle \neq 0)$ where the symmetry is broken. The presence of drive and dissipation means that the long-time state is not a thermal state, but is instead a non-equilibrium steady state. \par

Before introducing an exact treatment of the DDIM, we begin with a simple mean-field analysis of Eq. \eqref{DDIM}. The mean-field equations of motion are obtained by calculating the expectation values $\langle \sigma^\alpha_i \rangle$ and assuming that the density matrix is factorized in space and is uniform:
\begin{equation}
    \rho = \bigotimes_i \rho_i = \rho_{\text{MF}}^{\otimes N}\,,
\end{equation}
where $\rho_{\text{MF}}$ is the mean-field density matrix, uniform across all sites. 
Using this approximation, we find the mean-field Heisenberg equations of motion (in the $N \to \infty $ limit)
\begin{subequations}
\begin{align}
    \partial_t \langle \sigma^x \rangle &= -2\Delta \langle \sigma^y \rangle - \frac{\Gamma}{2}\langle \sigma^x\rangle\,, \\
    \partial_t \langle \sigma^y \rangle& = (4J \langle \sigma^z \rangle + 2\Delta )\langle \sigma^x \rangle - \frac{\Gamma}{2} \langle \sigma^y \rangle\,, \\
    \partial_t \langle \sigma^z \rangle &= -4J \langle \sigma^y \rangle \langle \sigma^x \rangle - \Gamma(1 + \langle \sigma^z \rangle)\,,
\end{align}
\end{subequations}
where we have dropped the spatial index due to the uniform ansatz. By setting the LHS to zero, we can solve for the non-equilibrium steady-state values of the three observables. In the normal phase, the only solution is the trivial one: $\langle \sigma^x \rangle_{\text{ss}} = 0, \langle \sigma^y \rangle_{\text{ss}} = 0, \langle \sigma^z \rangle_{\text{ss}} = -1$ with the subscript indicating the steady state. In the ordered phase, we identify two stable solutions as
\begin{subequations}
\begin{align}
    \langle \sigma^x \rangle_{\text{ss}} &= \pm \frac{\sqrt{32 J \Delta - 16\Delta^2 - \Gamma^2}}{4 \sqrt{2}J}\,, \\
    \langle \sigma^y \rangle_{\text{ss}} &= \mp \frac{\Gamma \sqrt{32 J \Delta - 16\Delta^2 - \Gamma^2}}{16 \sqrt{2}J \Delta}\,,  \\
    \langle \sigma^z \rangle_{\text{ss}} &= -\frac{\Gamma^2 + 16\Delta^2}{32J\Delta}\,,
\end{align}
\end{subequations}
from which the phase boundary follows as 
\begin{equation}
    \Gamma^2 + 16\Delta^2 - 32J\Delta = 0\,.
\end{equation}
The phase diagram of this model is given in Fig.~\ref{phase space}(a) and is contrasted against that of equilibrium.
The mean-field solution is exact in the thermodynamic limit due to the collective interactions; however, to characterize fluctuations and to identify the critical behavior of the model, we need to go beyond mean field. Using a quantum-to-classical mapping, we shall provide an exact field-theoretical description, allowing us to make a systematic study of fluctuations beyond mean field.

\section{Mapping to Keldysh Field Theory}\label{mapping}
A natural framework to describe the critical behavior is through a field-theoretical analysis. 
An immediate challenge, however, is to describe the driven-dissipative spin model in a terms of a field theory. Previous works \cite{das_infinite-range_2006, torre_keldysh_2013, lerose_chaotic_2018, persico_coherence_1975} utilized the Holstein-Primakoff transformation to bosons to much success; however, local spontaneous emission in \cref{DDIM} breaks the total-spin conservation, in which case this transformation is no longer applicable. 
More recently, a mapping of spins to composite fermions \cite{dalla_torre_dicke_2016} was used to tackle local spontaneous emission, however, the fermionic model becomes rather complex. Here, we seek a alternative route to tackle the dynamics in the presence of local spontaneous emission.
To this end, we take inspiration from the equilibrium quantum-to-classical mapping, also known as the Suzuki-Trotter mapping \cite{suzuki_relationship_1976}. The Suzuki-Trotter decomposition involves mapping the partition function of the quantum system in $d$ dimensions to that of a classical model in one higher dimension. To set up a similar mapping, our starting point is the non-equilibrium partition function
\begin{equation}\label{partition function}
    Z = \tr(\rho(t)) = 1\,.
\end{equation}
It is important to note that $Z = 1$ at all times, representing the conservation of probability from the Liouvillian dynamics governed by Eq.~\eqref{DDIM}. The first step  will be a non-equilibrium extension of the Suzuki-Trotter decomposition. In equilibrium, the decomposition is performed by decomposing the thermal state into many ``imaginary-time'' slices and inserting a resolution of the identity at each time slice. In our non-equilibrium setting, the evolution operator $\exp(t\mathcal{L})$ is a \textit{superoperator}, as can be seen from the fact that the Liouvillian takes the form $\mathcal{L}[\bullet] = \sum_i A_i \bullet B_i$ for some matrices $A_i, B_i$. To adapt the Suzuki-Trotter decomposition to Liouvillian dynamics, we must first ``vectorize'' the density matrix $\rho \to \sket{\rho}$, such that the Liouvillian superoperator $\mathcal{L}$ is transformed into a non-Hermitian matrix $\mathbb{L}$. More explicitly, we vectorize the density matrix by performing the transformation
\begin{equation}
    A |i \rangle \langle j| B \to A \ket{i} \otimes B^T \ket{j},
\end{equation}
where the element $|i \rangle \langle j|$ of the density matrix is mapped to the \textit{superket} $\ket{i}\otimes\ket{j} = \ket{i}\ket{j} \equiv \sket{i, j}$. Inner products in this vectorized form are equivalent to the Hilbert-Schmidt norm in the original operator space,
\begin{equation}
    \sbraket{A}{B} = \tr(A^\dagger B)\,.
\end{equation}
The non-equilibrium partition function upon vectorization take the form
\begin{equation}\label{vectorized partition}
    Z = \tr(e^{t \mathcal{L}} \rho_0) \to \sbra{I} e^{t \mathbb{L}} \sket{\rho_0}\,,
\end{equation}
where the matrix $\mathbb{L}$ is given by
\begin{align}\label{vectorized liouvillian}
    \mathbb{L} = & -i\left(H \otimes I - I \otimes H \right)\\
    & + \Gamma \sum_i \left[ \sigma^-_i \otimes \sigma^-_i - \frac{1}{2} \left(\sigma^+_i \sigma^-_i \otimes I + I \otimes \sigma^+_i \sigma^-_i\right)\right]\,. \nonumber
\end{align}
\par 
The process of vectorization can be interpreted as the purification of the density matrix, which is achieved by doubling the system of spins. In this picture, the Liouvillian is a non-Hermitian Hamiltonian governing the dynamics of the doubled spin system. As an example, consider a chain of quantum spins evolving under Liouvillian dynamics. In the process of vectorization, this chain is mapped to a one-dimensional ladder of quantum spins governed by the non-Hermitian Hamiltonian $\mathbb{L}$ where dissipation couples the two legs of the ladder.\par 
The vectorized partition function is now in a form amenable to the Suzuki-Trotter decomposition. We first apply the decomposition to the evolution operator, choosing to split the exponential into two parts: 
\begin{equation}
    e^{t \mathbb L} = \lim_{M \to \infty} \left(e^{\delta t \mathbb{L}_0} e^{\delta t \mathbb{L}_1}\right)^M.
\end{equation}
Here the time step is $\delta t = t/M$, the matrix $\mathbb{L}_0$ contains the Ising interaction, and the matrix $\mathbb{L}_1$ contains all the other terms in Eq. \eqref{vectorized liouvillian}. This will be a convenient choice for the next step, where we insert a resolution of the identity in the basis that diagonalizes the Ising interaction, 
\begin{equation}\label{basis}
    \mathbb{I}_n = \sum_{\sigma} \bigotimes_i |\sigma^{(u)}_{i,n}, \sigma^{(l)}_{i,n} \rangle \langle \sigma^{(u)}_{i,n}, \sigma^{(l)}_{i,n}|\,,
\end{equation}
at each time step $n$, with $\sigma^x |\sigma \rangle = \sigma |\sigma \rangle, \sigma \in \{1, -1\}$. We have introduced the upper/lower notation to denote spins on the upper and lower legs of the ladder, i.e. $\sigma^{x(u)} = \sigma^x \otimes I$.
For a single time step, the corresponding matrix element is
\begin{equation}
\bra{\{\sigma^{(u)}_n\}}\bra{\{\sigma^{(l)}_n\}}e^{\delta t \mathbb{L}_0} e^{\delta t \mathbb{L}_1} \ket{\{\sigma^{(u)}_{n-1}\}}\ket{\{\sigma^{(l)}_{n-1}\}},
\end{equation}
where the collection $\{\sigma^{(u/l)}_n\}$ denotes the set of all ``classical spins'' at time step $n$. Now, $\mathbb{L}_0$ is diagonal in our basis resulting in an action $ \mathcal{S}_0$ in terms of the classical spins:
\begin{equation}
    \mathcal{S}_0 = \frac{\delta t J}{N}\sum_n\left((S^{(u)}_n)^2 -(S^{(l)}_n)^2\right) =\sum _n \mathcal{S}_{0,n}\,,
\end{equation}
with $S_n^{(u/l)} = \sum_i \sigma_{i,n}^{(u/l)}$ the collective classical spin. On the other hand, $\mathbb{L}_1$ is not diagonal and acts nontrivially in this basis; however, it will not be necessary to calculate the matrix elements of $\exp(\delta t \mathbb{L}_1)$, as we shall see shortly.

\subsection{The DDIM Action}
With the Ising interaction becoming a  $c$-number, we can utilize the standard techniques to obtain a field-theoretical description of the DDIM. First, we perform a Hubbard-Stratonovich transformation to decouple the Ising interaction and introduce real scalar fields $m^{(u/l)}_n$, one for each leg of the ladder. This is given by (up to a normalization constant)
\begin{align}\label{hs transformation}
    e^{i\mathcal{S}_{0,n}} \sim \int \mathcal{D}^{(M)}&[m] \exp\bigg[ - i J \delta t N \left((m^{(u)}_n)^2 - (m^{(l)}_n)^2 \right) \nonumber \\
    &+ i 2J\delta t\left( m^{(u)}_n S^{(u)}_n -m^{(l)}_n S^{(l)}_n\right)\bigg]\,.
\end{align}
For compactness, we have defined the measure (up to a normalization constant) $\mathcal{D}^{(M)} \sim \prod_n^M dm^{(u)}_n dm^{(l)}_n$. 
Since the spins are decoupled, they can be traced out. This procedure gives the partition function in terms of the scalar fields $m^{(u/l)}_n$,
\begin{equation}\label{discrete action}
\begin{split}
    Z = \lim_{M \to \infty}\int \mathcal{D}^{(M)}&[m]\, e^{-i 2 J \delta t N \left((m^{(u)}_n)^2 - (m^{(l)}_n)^2 \right)}\\
    &\times\left(\sbra{I}\prod_{n'=0}^M e^{\delta t \mathbb{T}_{n'}}\sket{\rho_0}\right)^N\,.
\end{split}
\end{equation}
The matrix $\mathbb{T}(m^{(u/l)}_n) \equiv \mathbb{T}_n$ results from tracing out a single spin, and will be defined shortly. 
For convenience, we have assumed that all spins are in the same initial state; this will not affect the properties of the unique non-equilibrium steady state.
Notice that the contribution from all sites gives rise to the power of $N$ in the first exponential. Finally, the matrix $\mathbb{T}_n$ captures the effects of dissipation, transverse field, as well as the order parameter via $m$, and it is obtained by taking advantage of the $M \to \infty$ limit to combine all single-site operators. In terms of the (single-site) spin operators, this matrix takes the form
\begin{align}
    &\mathbb{T}_n = i2J\left( m^{(u)}_n \sigma^{x(u)}- m^{(l)}_n \sigma^{x(l)} \right) -i \Delta(\sigma^{z(u)}-\sigma^{z(l)})\nonumber \\
    &\, + \Gamma \sigma^{-(u)}\sigma^{-(l)} - \frac{\Gamma}{2}\left(\sigma^{+(u)}\sigma^{-(u)} + \sigma^{+(l)}\sigma^{-(l)}\right).
\end{align}
The object $\exp(\delta t \mathbb{T}_n)$ is akin to a transfer matrix for a single rung of the spin ladder. More explicitly, the matrix $\mathbb{T}(m^{u/l}(t))\equiv \mathbb{T}(t)$ in the $\ket{\sigma^{(u)}}\ket{\sigma^{(l)}}$ basis is given by
\[
    \mathbb{T} = 
    \renewcommand*{\arraystretch}{1.5}
    \scalemath{.69}{
    \begin{pmatrix}
    -\frac{\Gamma}{4}+i2\sqrt{2}J m_q & i\Delta & -i\Delta & \frac{\Gamma}{4} \\
    i\Delta - \frac{\Gamma}{2} & -\frac{3\Gamma}{4} + i 2\sqrt{2} J m_c  & -\frac{\Gamma}{4} & -i\Delta -\frac{\Gamma}{2} \\
    -i\Delta - \frac{\Gamma}{2} &-\frac{\Gamma}{4} &- \frac{3\Gamma}{4} -i2\sqrt{2} J m_c & i\Delta - \frac{\Gamma}{2} \\
    \frac{\Gamma}{4} & -i\Delta & i\Delta & -\frac{\Gamma}{4} -i 2\sqrt{2} J m_q
    \end{pmatrix}
    }\,.
\]

Finally, we take the continuum limit ($\delta t \to 0$) with the initial state given at $t \to -\infty$. This leads to a path-integral formulation of the non-equilibrium partition function:
\begin{equation}
Z = \int\mathcal{D}[m_c(t), m_q(t)] e^{i\mathcal{S}[m_{c/q}(t)]}\,, 
\end{equation}
with the Keldysh action
\begin{equation}\label{Keldysh Action}
     \mathcal{S} =  -2JN\int_t m_{c}(t)m_{q}(t) -iN\ln\tr\Big[\mathcal{T}e^{\int_t \mathbb{T}(m_{c/q}(t))}\Big].
\end{equation}
In obtaining Eq. \eqref{Keldysh Action}, we have performed the Keldysh rotation $m_{c/q} = (m^{(u)} \pm m^{(l)})/\sqrt{2}$ to bring the action into the conventional Keldysh form, and have absorbed all prefactors into the measure $\mathcal{D}[m]$. The trace in the last term is chosen for convenience as we are only interested in the non-equilibrium steady state at late times with no memory of the initial state. The time-ordering operator $\mathcal{T}$ also makes its appearance in the continuum limit to properly treat the time-dependence of the fields. 

\subsection{Field-Spin Relationship}\label{field spin relationship}
The quantum-to-classical mapping utilizes the Hubbard-Stratonovich transformation to introduce a real field $m$ in place of the classical total spin $S$. Therefore, expectation values of $m$ should be naturally related to those of the original spin operator $S_x$. To derive this relationship, we introduce time-dependent source fields $\alpha^{(u/l)}(t)$ coupled to $S_x$ on both the upper and lower legs, $i(\alpha^{(u)}S_x^{(u)} - \alpha^{(l)}S_x^{(l)})$, where we can have $\alpha^{(u)} \neq \alpha^{(l)}$ so that the non-equilibrium partition function $Z \neq 1$  \cite{sieberer_keldysh_2016}. The source fields do not alter the quantum-to-classical mapping derivation; they simply introduce new elements to the matrix $\mathbb{T}$ as
\begin{equation}
   \mathbb{T}^{'}(t) = \mathbb{T}(m_{c/q}(t)) + i \sqrt{2}\begin{pmatrix} \alpha_q & 0 & 0 & 0\\ 0 & \alpha_c & 0 & 0 \\ 0 & 0 & -\alpha_c & 0 \\ 0 & 0 & 0 & -\alpha_q \end{pmatrix}\,,
\end{equation}
where we have performed the Keldysh rotation $\alpha_{c/q} = (\alpha^{(u)} \pm \alpha^{(l)})/\sqrt{2}$ on the source fields.
The fields $m_{c/q}$ are dummy variables under the path integral. Making the change of variables  $m_{c/q}(t) \to m_{c/q}(t) + \alpha_{c/q}(t)/2J$, we can move the source terms out into the quadratic portion of the action to find (using the same field variables)
\begin{align}
 \mathcal{S} = &N \int_t \bigg(  m_c(t) \alpha_q (t)  + m_q(t)\alpha_c(t) - \frac{\alpha_q(t)\alpha_c(t)}{2J}\bigg) \nonumber \\ 
 &
 + \mathcal{S}_0 [m_{c/q}],
\end{align}
where $\mathcal{S}_0$ is the original action without the source fields in Eq. \eqref{Keldysh Action}. Taking derivatives of the generating functional $Z[\alpha(t)]$ with respect to the source fields generates correlation functions \cite{sieberer_keldysh_2016}. Specifically, taking a derivative with respect to $\alpha_q$ yields 
\begin{equation}\label{spin field mag}
   \frac{\sqrt{2}}{N}\langle S_x (t) \rangle = -i \frac{\partial Z}{\partial \alpha_q (t) }\bigg\rvert_{\alpha_{c/q} = 0} = \langle m_c(t) \rangle\,,
\end{equation}
which provides a clear translation between the two descriptions (the factor of $\sqrt{2}$ arises due to the Keldysh rotation). Next, we consider the two-point correlation function and response function, respectively:
\begin{equation}
\begin{split}
   \frac{1}{N^2}\langle \{S_{x}(t), S_{x}(t')\} \rangle &= -\frac{\delta Z}{\delta \alpha_q(t) \delta \alpha_q(t')}\bigg\rvert_{\alpha_{c/q} = 0} \\
   &= \langle m_c(t) m_c(t') \rangle,
\end{split}
\end{equation}
and
\begin{equation}\label{DDIMspinResponse}
\begin{split}
    \frac{1}{ N^2}\langle &[S_{x}(t), S_{x}(t')] \rangle \\
    &= -\left(\frac{\delta Z}{\delta \alpha_q(t) \delta \alpha_c(t')} - \frac{\delta Z}{\delta \alpha_c(t) \delta \alpha_q(t')}\right)\bigg\rvert_{\alpha_{c/q} = 0}\\
    &=  \langle m_c(t) m_q(t') \rangle - \langle m_q(t) m_c(t') \rangle \,.
\end{split}
\end{equation}
This establishes the  relationships between the spin operator and the fields. It is straightforward to find the analogs of these relations at higher orders by taking  appropriate derivatives with respect to the source fields.\par

\section{Field-Theoretical Analysis}\label{analysis}
Having mapped a driven-dissipative spin model to a non-equilibrium Keldysh action, we can now take advantage of the field-theoretical toolbox available to us. Equation \eqref{Keldysh Action} appears formidable due to the log-trace term; however, the overall factor of $N$ in the action (residual of the collective nature of the DDIM) means that the saddle-point approximation becomes exact in the thermodynamic limit. Using this approximation, we find the steady-state expectation value of the order parameter and expand the action in powers of fluctuations around the order parameter, both in the ordered and normal phases.
\subsection{Saddle-Point Solution}
We seek the solutions to the saddle-point equations
\begin{equation}\label{saddlepoint equations}
    \frac{\delta \mathcal{S}}{\delta m_c (t)} = 0, \quad \frac{\delta \mathcal{S}}{\delta m_q (t)} = 0\,,
\end{equation}
with $m_c (t) = m \equiv \text{const}$ and $m_q (t)= 0$. Equations \eqref{saddlepoint equations} are essentially a semi-classical approximation to the problem, neglecting statistical and quantum fluctuations, but they constitute a first step before considering critical properties. Carrying out the explicit calculations for the order parameter $m$, we find
\begin{align}\label{saddlepoint}
     \frac{\delta \mathcal{S}}{\delta m_{q}(t)}\bigg|_{\substack{m_c = m\\m_q = 0}} &= -2JN m - iN \frac{\tr\left(e^{(t_f-t)\mathbb{T}_0}\mathbb{T}_q e^{(t-t_i) \mathbb{T}_0} \right)}{\tr{\left(e^{(t_f-t_i) \mathbb{T}_0}\right)}} \nonumber \\
     &=-2JN m - iN \sbra{I}\mathbb{T}_q\sket{\rho_{\text{ss}}} = 0\,,
\end{align}
where $t_{i/f}$ denote  the initial and final times, respectively. Here, we introduce the notation
\begin{subequations}
\begin{align}\label{T matrix no m}
    \mathbb{T}_0 &= \mathbb{T}(m_c = m, m_q = 0)\\ 
    \mathbb{T}_{c} &= \frac{\partial \mathbb{T}}{\partial m_{c}(t)} {\Big |}_{\substack{m_c = m\\ m_q = 0}} = i2\sqrt{2}J\,{\rm diag}\{0,1,-1,0\}\,,\\
    \mathbb{T}_{q} &= \frac{\partial \mathbb{T}}{\partial m_{q}(t)} {\Big |}_{\substack{m_c = m\\ m_q = 0}} = i2\sqrt{2}J\,{\rm diag}\{1,0,0,-1\}\,,
\end{align}
\end{subequations}
with all the matrices evaluated at the saddle-point stationary values. To obtain the second line of Eq.~\eqref{saddlepoint}, we have conveniently taken $t_i \to - \infty$, and used the fact that the only non-negative eigenvalue of $\mathbb{T}_0$ is zero (corresponding to the non-equilibrium steady state), leaving us with an inner product of the corresponding left and right eigenvectors $\sbra{I}$ (representing the identity) and $\sket{\rho_{\text{ss}}}$ (denoting the steady state). The identity vector is simply $\sbra{I} = (1,0,0,1)$, while the steady-state vector is given by
\begin{widetext}
\begin{equation}\label{steadystate vector}
    \sket{\rho_{\text{ss}}} = \bigg( \scalemath{.85}{ \frac{8 \sqrt{2} \Delta  J m}{\Gamma ^2+16 \Delta ^2+16 J^2m^2}+\frac{1}{2},
    -\frac{\Gamma ^2+16 \Delta ^2+4 i \sqrt{2} \Gamma J m}{2 \Gamma ^2+32 \Delta ^2+32 J^2 m^2},
    -\frac{\Gamma ^2+16\Delta ^2-4 i \sqrt{2} \Gamma  J m}{2 \left(\Gamma ^2+16 \Delta^2+16 J^2 m^2\right)},
   \frac{1}{2}-\frac{8 \sqrt{2} \Delta  Jm}{\Gamma ^2+16 \Delta ^2+16 J^2 m^2} } \bigg)^T\,.
\end{equation}
\end{widetext}
These two vectors are normalized such that $\sbraket{I}{\rho_{\text{ss}}} = 1$. Evaluating the second line of Eq. \eqref{saddlepoint}, we find that $m = 0$ in the normal phase and 
\begin{equation}\label{magnetization}
m = \pm \sqrt{-\Gamma^2-16\Delta^2+32\Delta J}/4J
\end{equation}
in the ordered phase. The phase boundary is located where the latter solutions are trivial (i.e., zero), and coincides  with that of  mean field. In a sense, our analysis here is the mean-field treatment at the level of the field theory.
As we show explicitly later, these solutions and the phase boundary are exact in the thermodynamic limit, as expected due to the infinite-ranged nature of the model.

\subsection{Quadratic Expansion}
Equipped with the saddle-point solutions, we can now investigate Gaussian fluctuations. Expanding Eq. \eqref{Keldysh Action} to second order around the saddle-point solutions in the normal phase ($m = 0$), we have
\begin{equation}
\label{action expansion}
      \mathcal{S}^{(2)} = \frac{1}{2}\int_{t, t'} \begin{pmatrix}m_c,\,
      m_q\end{pmatrix}_{t}
      \begin{pmatrix}0 & P^A\\ P^R & P^K\end{pmatrix}_{t-t'}
      \begin{pmatrix}m_c\\m_q\end{pmatrix}_{t'},
\end{equation}
where a factor of $\sqrt{N}$ has been absorbed into the fields for convenience. Note that the kernel is a function of the time difference only, reflecting the fact that time translation symmetry is restored in the non-equilibrium steady state. The kernel also exhibits the Keldysh structure \cite{kamenev_field_2011, torre_keldysh_2013, sieberer_keldysh_2016}, therefore the elements $P^{R/A}$ can be interpreted as the retarded/advanced inverse Green's functions and $P^K$ as the Keldysh component. These terms are given by
\begin{subequations}
\begin{align}\label{inverse response}
    P^{R}(t) &= P^A(-t) = \frac{\delta \mathcal{S}}{\delta m_q(t) \delta m_c(0)}\bigg\rvert_{\substack{m_c=0\\m_q=0}} \nonumber \\
    &= -2J\delta(t) -i\Theta(t)\sbra{I}\mathbb{T}_{q} e^{t\, \mathbb{T}_0} \mathbb{T}_{c}\sket{\rho_{\text{ss}}}\nonumber \\
    &=-2J\delta(t) + \Theta(t)8J^2 e^{-\frac{\Gamma}{2} t}\sin{(2\Delta t)}\,,
\end{align}
and 
\begin{align}\label{inverse keldysh}
    P^K(t) &= \frac{\delta \mathcal{S}}{\delta m_q(t) \delta m_q(0)}\bigg\rvert_{\substack{m_c=0\\m_q=0}}= -i\sbra{I}\mathbb{T}_{q} e^{|t|\, \mathbb{T}_0} \mathbb{T}_{q}\sket{\rho_{\text{ss}}} \nonumber \\
    &= i8J^2e^{-\frac{\Gamma}{2} |t|}\cos{(2\Delta t)}\,.
\end{align}
\end{subequations}
The above elements take a relatively simple form, with the dissipation leading to the exponential decay and the transverse field to oscillations. In addition, a delta function emerges in \cref{inverse response} as a remnant of the Hubbard-Stratonovich transformation and ensures the proper normalization of the partition function. The step function in the latter equation stems from $\sbra{I}\mathbb{T}_c \exp(t \mathbb{T}_0) \mathbb{T}_q \sket{\rho_{ss}} = 0$, and enforces the proper time ordering of the matrices. Because we absorbed a factor of $\sqrt{N}$ into the fields, higher-order terms in the expansion are at least of the order $\mathcal{O}(1/N)$, rendering Eq. \eqref{action expansion} exact in the thermodynamic limit. 

It will be convenient to recast these expressions in frequency space. With the Fourier transform $m_{c/q}(t) =\int \frac{d\omega}{2\pi}m_{c/q} (\omega)e^{-i\omega t}$, the kernel elements are
 \begin{subequations}
 \begin{align}\label{inverse response frequency} 
 \begin{split}
    P^R(\omega) = -2J - 4i J^2\bigg( &\frac{1}{-\Gamma/2 -i(2\Delta - \omega)}\\
    &-\frac{1}{-\Gamma/2 + i (2\Delta + \omega)}\bigg),
\end{split}
\end{align}
and
\begin{align}\label{inverse correlation frequency}
\begin{split}
    P^K(\omega) = 4 i  J^2 \Gamma \bigg( &\frac{1}{\Gamma^2/4 + (\omega - 2\Delta)^2}\\
    &+ \frac{1}{\Gamma^2/4 + (\omega + 2\Delta)^2}\bigg).
\end{split}
\end{align}
\end{subequations}
These analytic expressions follow from the simple form of the steady-state vector in the normal phase, $\sket{\rho_{ss}} = (1, -1, -1, 1)^T/2$. In the ordered phase, evaluating expressions like the second line in \cref{inverse response,inverse keldysh} become difficult as they require the nontrivial form of $\sket{\rho_{ss}}$ shown in Eq. \eqref{steadystate vector}, as well as the fact that the $\mathbb{T}$ matrix is now evaluated at finite $m$. We nevertheless derive formal expressions for the above functions as follows. We first decompose the exponential matrix $e^{t \mathbb{T}_m}$, where $\mathbb{T}_m = \mathbb{T}\left(m_c(t) = m, m_q(t) = 0\right)$, into its spectral form
\begin{equation}
\begin{split}
    e^{t \mathbb{T}_m} &= \sum_{i=0}^3 e^{\lvert \delta t \rvert \lambda_i}\sket{\lambda_i^R}\sbra{\lambda_i^L}\\
    &= \sket{\rho_{ss}}\sbra{I} + \sum_{i=1}^3 e^{\lvert \delta t \rvert \lambda_i}\sket{\lambda_i^R}\sbra{\lambda_i^L}\,.
\end{split}
\end{equation}
The vectors $\sbra{\lambda_i^L}$ and $\sket{\lambda_i^R}$ denote the $i$'th left and right eigenvectors of $\mathbb{T}_m$ corresponding to the eigenvalue $\lambda_i$, , respectively, and are normalized as $\sbraket{\lambda_i^L}{\lambda^R_j} = \delta_{ij}$; the biorthogonal structure is due to $\mathbb{T}_m$ being non-Hermitian. The expressions for the inverse response and Keldysh components in the frequency domain are then
\begin{subequations}
\begin{align}
    P^R(\omega) &= -2J -i \sum^3_{i=1} C_i \int_{\delta t}e^{i\omega \delta t} \Theta(\delta t)e^{\lvert \delta t\rvert \lambda_i} \nonumber \\
    &= -2J + i\sum^3_{i=1} C_i \frac{1}{\lambda_i + i \omega}\,, \label{op inv response}
\end{align}
and 
\begin{align}
    P^K(\omega) = -i\sum^3_{i=1} \tilde{C}_i \int_{\delta t}e^{i\omega \delta t} e^{\lvert \delta t \rvert \lambda_i} 
    &= 2i\sum_{i=1}^3 \tilde{C}_i \frac{\lambda_i}{\lambda_i^2 + \omega^2}\,. \label{op inv keldysh}
\end{align}
\end{subequations}
The accompanying coefficients are given by 
\begin{subequations}
\begin{align}
    C_i &= \sbra{I}\mathbb{T}_q\sket{\lambda_i^R} \sbra{\lambda_i^L}\mathbb{T}_c \sket{\rho_{ss}},  \\
\tilde{C}_i &= \sbra{I}\mathbb{T}_q\sket{\lambda_i^R} \sbra{\lambda_i^L}\mathbb{T}_q \sket{\rho_{ss}}.
\end{align}
\end{subequations}
$P^R(\omega)$ and $P^K(\omega)$ can be obtained by numerically solving for the eigenvalues and the eigenvectors of $\mathbb{T}_m$.

\subsection{Correlation and Response Functions}
The physical observables of interest in terms of the kernel elements describing the field $m$ are the symmetrized correlation function
\begin{subequations}
\begin{align}
    C(\omega) &= \frac{1}{N}{\cal F}_\omega \langle \{S_x(t), S_x(0)\}\rangle = \langle m_c(\omega) m_c(-\omega)\rangle \nonumber \\
    &=\frac{-iP^K(\omega)}{P^R(\omega)P^A(\omega)}\,, \label{SP Correlation}
\end{align}
and the response function
\begin{align}
    \chi(\omega) &= \frac{1}{i N}\mathcal{F}_\omega\langle [S_x(t), S_x(0)]\rangle \nonumber \\
    &= \frac{1}{i}\langle m_q(\omega)m_c(-\omega) - m_c(\omega) m_q(-\omega) \rangle \nonumber \\
    &=\frac{1}{P^R(\omega)} - \frac{1}{P^A(\omega)}\,. \label{SP Response}
\end{align}
\end{subequations}
Their explicit forms can be obtained from \cref{inverse response frequency,inverse correlation frequency} as
\begin{subequations}
\begin{equation}\label{correlation definition}
  C(\omega) = \frac{\Gamma(\Gamma^2 + 4(4\Delta^2 + \omega^2))}{2 (\omega - \omega_1)(\omega-\omega_2)(\omega-\omega_1^*)(\omega-\omega_2^*)}\,,
\end{equation}
and
\begin{equation}\label{response definition}
    \chi(\omega) = 4\Delta\left(\frac{(\omega-\omega_1^*)(\omega-\omega_2^*) - (\omega - \omega_1)(\omega-\omega_2)}{(\omega - \omega_1)(\omega-\omega_2)(\omega-\omega_1^*)(\omega-\omega_2^*)}\right).
\end{equation}
\end{subequations}
The poles in these equations are given by
\begin{equation}\label{poles}
\omega_1 = -\frac{i}{2}(\Gamma - \Gamma_c)\,,\qquad \omega_2 = -\frac{i}{2}(\Gamma + \Gamma_c)\,,
\end{equation}
where $\Gamma_c = 4\sqrt{(2J-\Delta)\Delta}$. We thus observe that $\omega_1$ is the ``soft mode'' which vanishes at the phase transition, while we can identify $\omega_2$ as the ``fast mode'' that remains finite. The soft mode is responsible for critical dynamics and signifies the critical slowdown as we approach the phase transition. The limit $\Gamma_c \to 0$, where both poles become soft, gives rise to qualitatively different behavior as we shall discuss later.\par 

In the time domain, the correlation and response function are given by
\begin{subequations}
\begin{equation}\label{correlation function}
\begin{split}
     C(t) = \frac{ e^{-\Gamma |t|/2}}{\Gamma_c}\bigg(&\frac{\Gamma \Gamma_c + 16(J-\Delta)\Delta }{\Gamma + \Gamma_c}e^{-\Gamma_c |t|/2} \\
     &+ \frac{\Gamma \Gamma_c -16(J-\Delta)\Delta}{\Gamma - \Gamma_c}e^{\Gamma_c |t|/2}\bigg)\,,
\end{split}
\end{equation}
and
\begin{equation}\label{response function}
\chi(t)=\text{sgn}(t) \frac{4 \Delta}{\Gamma_c}e^{-\Gamma |t|/2}\left(e^{-\Gamma_c |t|/2} - e^{\Gamma_c |t|/2}\right).
\end{equation}
\end{subequations}
We can identify two distinct regimes in the disordered phase. For $\Delta < 2J$, we see that $\Gamma_c$ is real, and that both $C(t)$ and $\chi(t)$ are purely relaxational. On the other hand, $\Gamma_c$ becomes imaginary for $\Delta > 2J$, hence complex-valued poles, and the dynamics becomes underdamped. In this regime, the overall decay rate is controlled by $\Gamma$, and the oscillation time scale is set by $\Gamma_c$. This behavior arises due to the competition between the interaction $J$ and the transverse field $\Delta$. For sufficiently large $\Delta$, the transverse field is dominant and causes the large spin to precess about the $z$-axis; while on average the longitudinal spin components are zero, their temporal correlations expose the oscillations.\par
\begin{figure}[tp]
    \centering
    \includegraphics[width=\linewidth]{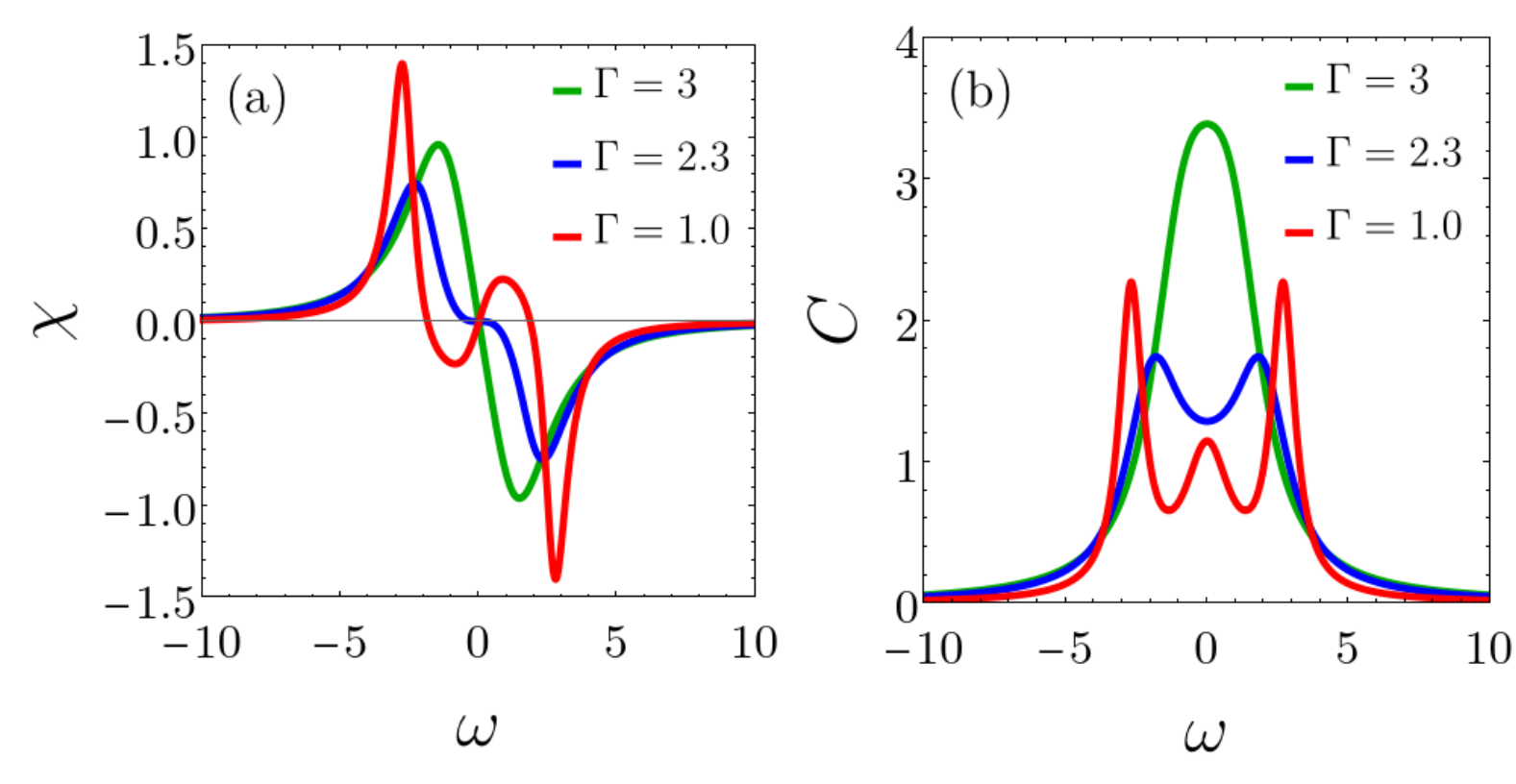}
    \caption{The response function (a) and correlation function (b) in the ordered phase ($J = 1, \Delta = 1$) for different values of $\Gamma$. (a) As we move away from the phase boundary, $\chi(\omega)$ at low frequencies plateaus before changing sign, indicating a gainy rather than lossy behavior. (b) The peak at $\omega = 0$, signifying the dominant soft mode near the phase boundary, splits into two as $\Gamma$ is decreased. For sufficiently small $\Gamma$ ($\lesssim 2.3$), another peak appears at $\omega = 0$.}
    \label{ordered resp fig}
\end{figure}
In the ordered phase, the correlation and response function can be evaluated numerically starting with the inverse response and Keldysh elements in \cref{op inv response,op inv keldysh}. In Fig.~\ref{ordered resp fig}, we plot $\chi(\omega)$ and $C(\omega)$ within the ordered phase and for different values of $\Gamma$. 
As $\Gamma$ is decreased, the low-frequency region of $\chi(\omega)$ changes sign, indicating that the system is no longer lossy and is rather ``gainy'' at low frequencies. This behavior is of course related to the driven nature of the system. 
Similarly, the correlation function shows a single peak at $\omega = 0$ for larger $\Gamma$ close to the phase boundary (within the ordered phase); this behavior can be attributed to the soft mode. As we move away from the phase boundary, this peak splits into two and eventually gives rise to a smaller peak at $\omega=0$. Indeed, this appears at the same point where the low-frequency behavior of $\chi$ changes qualitatively. In Sec. \ref{thermalization}, we show that this behavior can be interpreted as the emergence of a negative effective temperature.

\begin{figure}[tp]
    \centering
    \includegraphics[scale=.4]{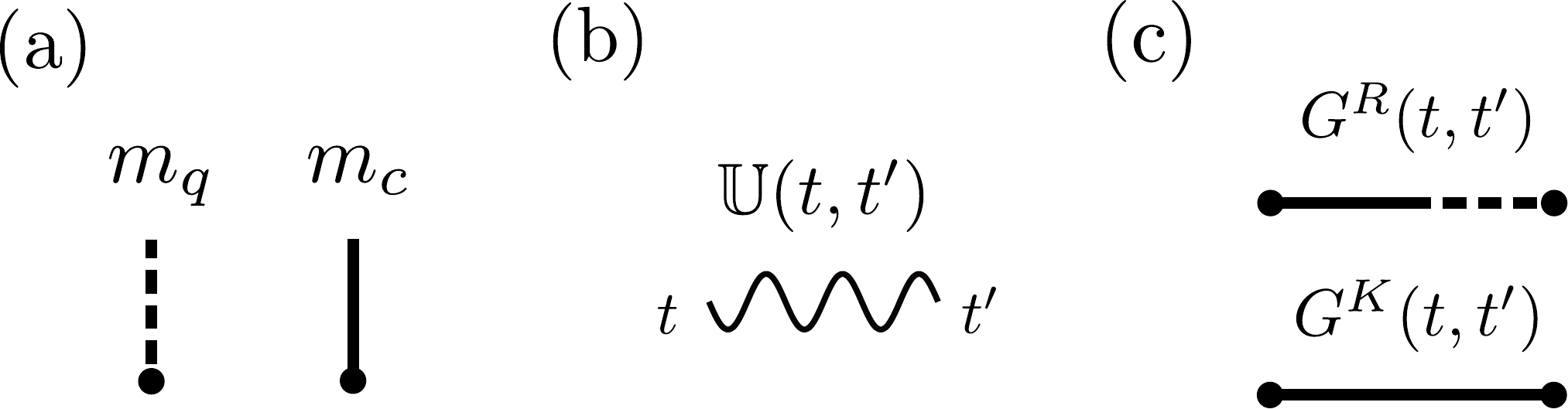}
    \caption{Diagrammatic representation. (a) The solid and dashed legs in (a) represent classical ($m_c$) and quantum ($m_q$) fields, respectively. (b) The wavy line represents the time evolution where time ordering is understood from right to left. (c) Connected legs correspond to Green's functions with $G^R$  the response function and $G^K$ the Keldysh correlation function.}
    \label{feynman rules}
\end{figure}
\subsection{Diagrammatics}\label{Diagrammatics}
To go beyond the quadratic action, we introduce here a diagrammatic representation of interaction terms in the expansion of the action. These terms can be found by first expanding the argument of the logarithm in Eq. \eqref{Action} in powers of the fields as
\begin{equation}\label{log expansion}
    \mathcal{S} = -2J\int_t m_c(t) m_q(t) - i N\ln \left(1 + \sum_{i, \bm{\alpha}} {\cal D}_{i, \bm{\alpha}}\right)\,,
\end{equation}
where, as stated before, a factor of $\sqrt{N}$ has been absorbed into $m_{c/q}$, and ${\cal D}_{i,\bm{\alpha}}$ is the $i$'th-order connected diagram:
\begin{equation}\label{log diagrams}
    {\cal D}_{i, \bm{\alpha}} = \frac{1}{N^{\frac{i}{2}}}\int_{\bm{t}} \overline{u}_{\bm{\alpha}}(\bm{t})m_{\alpha_1}(t_1) \cdots m_{\alpha_{i-1}}(t_{i-1})  m_{\alpha_i}(t_i).
\end{equation}
Here, we have used $\bm{t}$ as a shorthand for $\{t_1, \cdots, t_{i-1},  t_i\}$ and similarly for $\bm{\alpha}$. The latter indices take the values $c/q$ representing classical/quantum fields, respectively, and the sum over $\bm{\alpha}$ in Eq. \eqref{log expansion} is only over distinct orderings of $c/q$ to avoid overcounting. 
The rules for constructing these diagrams can be found Fig. \ref{feynman rules}.
\begin{figure}[tp]
    \centering
    \includegraphics[scale=.3]{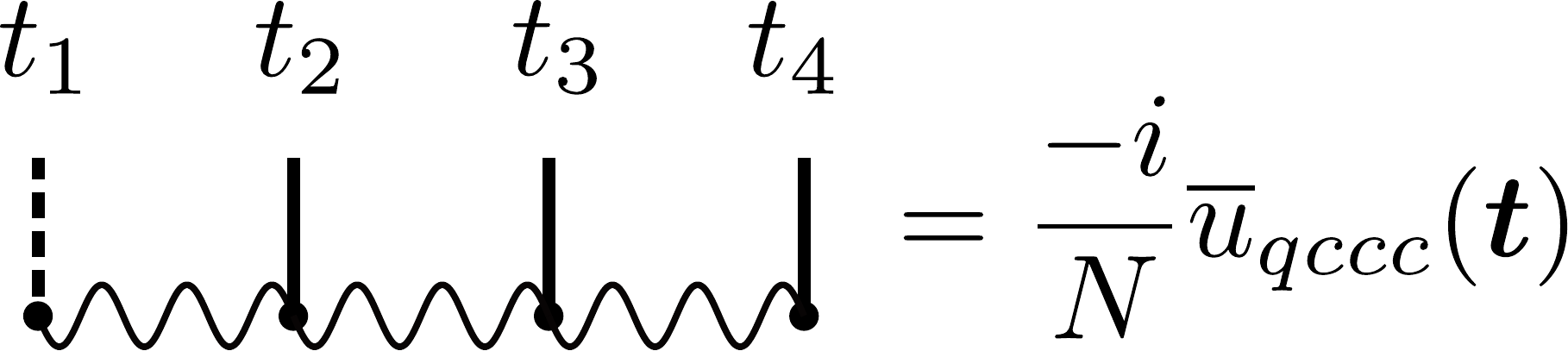}
    \caption{A representative (classical) vertex. The interaction coefficient $\overline{u}_{qccc}(\bm{t})$ is time ordered such that $t_1\geq t_2\geq t_3\geq t_4$, and is given explicitly by \cref{interaction coefficients}.} 
    \label{classical vertex figure}
\end{figure}
Connected diagrams are time ordered from right to left, therefore the interaction coefficient $\overline{u}_{\bm{\alpha}_j}$ is time ordered too with $t_1\geq  \cdots \geq t_{i-1} \geq t_i$. These coefficients are given by 
\begin{equation}\label{interaction coefficients}
    \overline{u}_{\bm{\alpha}} = \overline{\tr}\left( \mathbb{T}_{\alpha_{1}} \mathbb{U}(t_1, t_2) \mathbb{T}_{\alpha_{2}} \cdots \mathbb{U}(t_{i-1}, t_{i}) \mathbb{T}_{\alpha_{i}}\right)\,,
\end{equation}
where we have defined the ``trace'' operation $\overline{\tr}(\bullet) = \sbra{I} \bullet \sket{\rho_{ss}} $, have utilized the matrices $\mathbb{T}_{c/q}$ defined earlier, and have introduced the \textit{propagators} $\mathbb{U}(t-t') = \exp((t-t')\mathbb{T}_0)$ with $\mathbb{T}_0 = \mathbb{T}(m = 0)$. The latter propagators are depicted as wavy lines in our diagrammatic notation; see Fig. \ref{feynman rules}. This time-ordered representation of the interaction coefficients (and diagrams) is best understood in a scattering picture. The superket $\sket{\rho_{ss}}$ describes the non-equilibrium steady state of a pair of spins on the upper and lower leg of the spin ladder, and is taken as the ``in'' state. This state propagates freely (via $\mathbb{U}$) while  scattering off the mean-field ($m_{c/q}$) intermittently. In other words, the interaction coefficients follow from the time-dependent perturbation theory in the expansion of  the evolution operator $\mathcal{T} e^{\int_t \mathbb{T}(t)}$, with  $\mathbb{T}(t) = \mathbb{T}_0 + m_c(t) \mathbb{T}_c + m_q(t) \mathbb{T}_q$, in powers of the scattering potentials $m_{c/q}\mathbb{T}_{c/q}$.\par
The scattering interpretation becomes manifest in Fourier space. Let's first consider the free propagator in Fourier space: 
\begin{equation}
    \mathbb{U}(\omega) = \int_{t>0} e^{-i \omega t} e^{ t \mathbb{T}_0} = - \frac{1}{\mathbb{T}_0 -i \omega}.
\end{equation}
Here, we have used the fact that the matrix $\mathbb{T}_0$ is diagonalizable, and that the real part of its eigenvalues $\lambda$ is non-positive. For an eigenvalue with a zero real part, we substitute $\omega \to \omega-i\epsilon$ due to causality with the understanding that the limit $\epsilon \to 0$ is taken at the end of the calculation.
The above expression is reminiscent of the Lippmann-Schwinger equation with $\mathbb{T}_0$ taking the role of the Hamiltonian, though we must recall that $\mathbb{T}_0$ is non-Hermitian and acts on two copies of a spin. It is often convenient to compute the interaction coefficient in the Fourier space. Some algebra yields
\begin{equation}\label{u ft}
   \overline{u}(\bm{\omega}) = \overline{\tr}\left( \mathbb{T}_{\alpha_{1}} \mathbb{U}(\tilde \omega_1) \mathbb{T}_{\alpha_{2}} \cdots \mathbb{U}(\tilde \omega_i) \mathbb{T}_{\alpha_{i}}\right)\,,
\end{equation}
where $\widetilde{\omega}_j = \omega_1 +...+\omega_j - \omega_{j+1}-...-\omega_i$.
\par 

So far, we have considered the connected diagrams that arise inside the logarithm in \cref{log expansion}. However, the full diagrammatic expansion of the action requires an expansion of the logarithm too. 
Expanding Eq. \eqref{log expansion} in powers of the connected diagrams, we obtain all interaction vertices comprising connected as well as disconnected diagrams. Formally, a multi-legged diagram with $M = \sum_{i=1}^p l_i$ disconnected parts is given by
\begin{equation}\label{interaction vertex}
\begin{split}
    i\frac{(M-1)! (-1)^M}{N^{-1} \prod_{j}^p l_j !}& 
    ({\cal D}_{i_1, \bm{\alpha}_1})^{l_{1}}({\cal D}_{i_2, \bm{\alpha}_2})^{l_{2}}\cdots ({\cal D}_{i_p, \bm{\alpha}_p})^{l_{p}}
\end{split}
\end{equation}
where each ${\cal D}_{i, \bm{\alpha}}$, integrated over the corresponding time coordinates, represents one of the $p$ unique connected diagrams with multiplicity $l_j$.
The combinatorial factor $\frac{1}{M}\frac{M!}{\prod^p_j  l_j!} = \frac{(M-1)}{\prod^p_j  l_j!}$ is included, where the factor of $\frac{1}{M}$ is due to the expansion of the logarithm, and $\frac{M!}{\prod^p_j  l_j!}$ accounts for each set of identical disconnected diagrams with multiplicity $l_j$. As an example, Fig.~\ref{classical vertex figure} depicts the diagrammatic representation of the ``classical vertex'' $\frac{-i}{N}\int_{\bm{t}} \overline{u}_{qccc}(\bm{t}) m_q(t_1) m_c(t_2)m_c(t_3)m_c(t_4)$ with the time integral constrained as $t_1\geq t_2\geq t_3\geq t_4$. We remark that the disconnected diagrams discussed here emerge at the level of the action, before expanding the exponential factor in the partition function. Expanding the latter exponential factor will further generate disconnected diagrams whose coefficients should be properly determined from the combinatorial factors reported above. In this sense, we must keep the origin of various disconnected diagrams (whether they appear in the action itself or result from the expansion of the exponential factor). This pattern is in contrast with the standard diagrammatic representation and is a unique feature of our nonequilibrium setting.

The diagrams discussed here have certain causal properties. First, each diagram must come with at least one quantum leg (dashed line), reflecting the property of the Keldysh action that $S(m_c,m_q = 0) = 0$. Furthermore, the last leg of all connected diagrams is \textit{always} a quantum field which enforces causality and ensures that the partition function retains its normalization ($Z=1$). 
Curiously, only certain orderings of classical and quantum legs are allowed. The diagrammatic notation developed here will prove very useful when calculating quantities such as self-energy corrections as well as expanding the action in the ordered phase. The former can be determined systematically by contracting quantum and classical fields in these diagrams.

\subsection{Effective Thermalization}\label{thermalization}
Driven-dissipative systems are inherently non-equilibrium, and therefore there is no intrinsic notion of temperature. However, an \textit{effective} temperature can be defined by imposing a fluctuation-dissipation relation (FDR)\cite{torre_keldysh_2013, sieberer_keldysh_2016, gelhausen_many-body_2017, chiocchetta_short-time_2015, smith_prethermalization_2013, titum_non-equilibrium_2019},
\begin{equation}\label{FDR}
    P^K(\omega) = F(\omega)\left(P^R(\omega) - P^A(\omega)\right),
\end{equation}
where $F(\omega)$ is a distribution function defined by this equation. In equilibrium and at finite temperature, the distribution function only depends on temperature and takes the form $F(\omega) = \coth(\omega/2T)$. Specifically, the low-frequency limit of the distribution function yields the classical FDR with $F(\omega) = 2T/\omega$. While there is no intrinsic temperature in our driven-dissipative system, we can still impose the classical form of the FDR (to be justified later) to identify the effective temperature; in the normal phase, we find
\begin{figure}[tp]
    \centering
    \includegraphics[scale=.4]{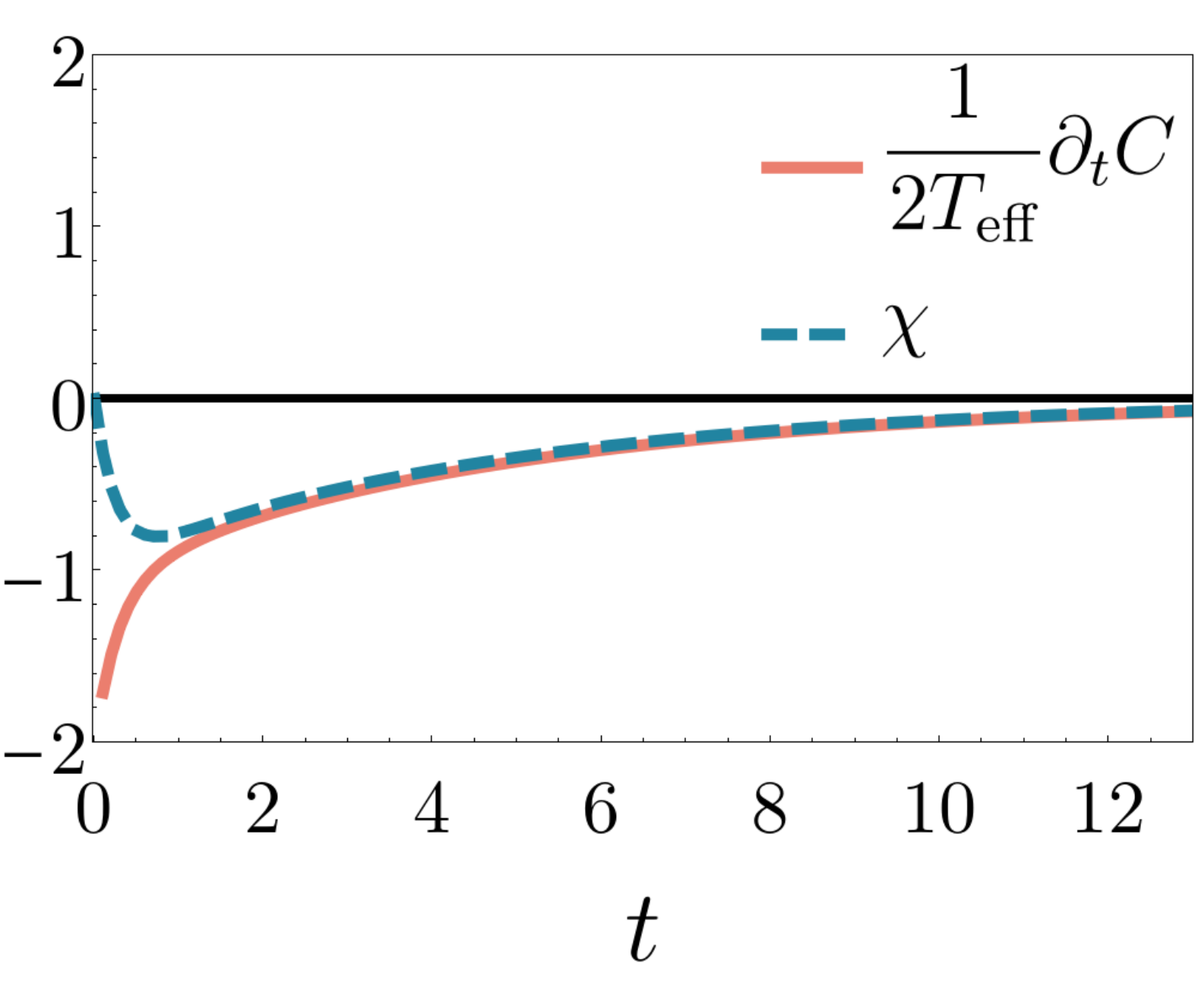}
    \caption{Numerical plot of the correlation and response functions with a system size of $N = 100$ near a generic critical point ($J=1, \Delta = 1, \Gamma = 4$). The classical FDR $\chi(t) = \partial_t C(t)/2 T_{\text{eff}}$ holds at long times ($t \gtrsim \Gamma^{-1}$) with $T_{\text{eff}} = J$.}
    \label{FDR GP plot}
\end{figure}
\begin{figure}[tp]
    \centering
    \includegraphics[scale=.4]{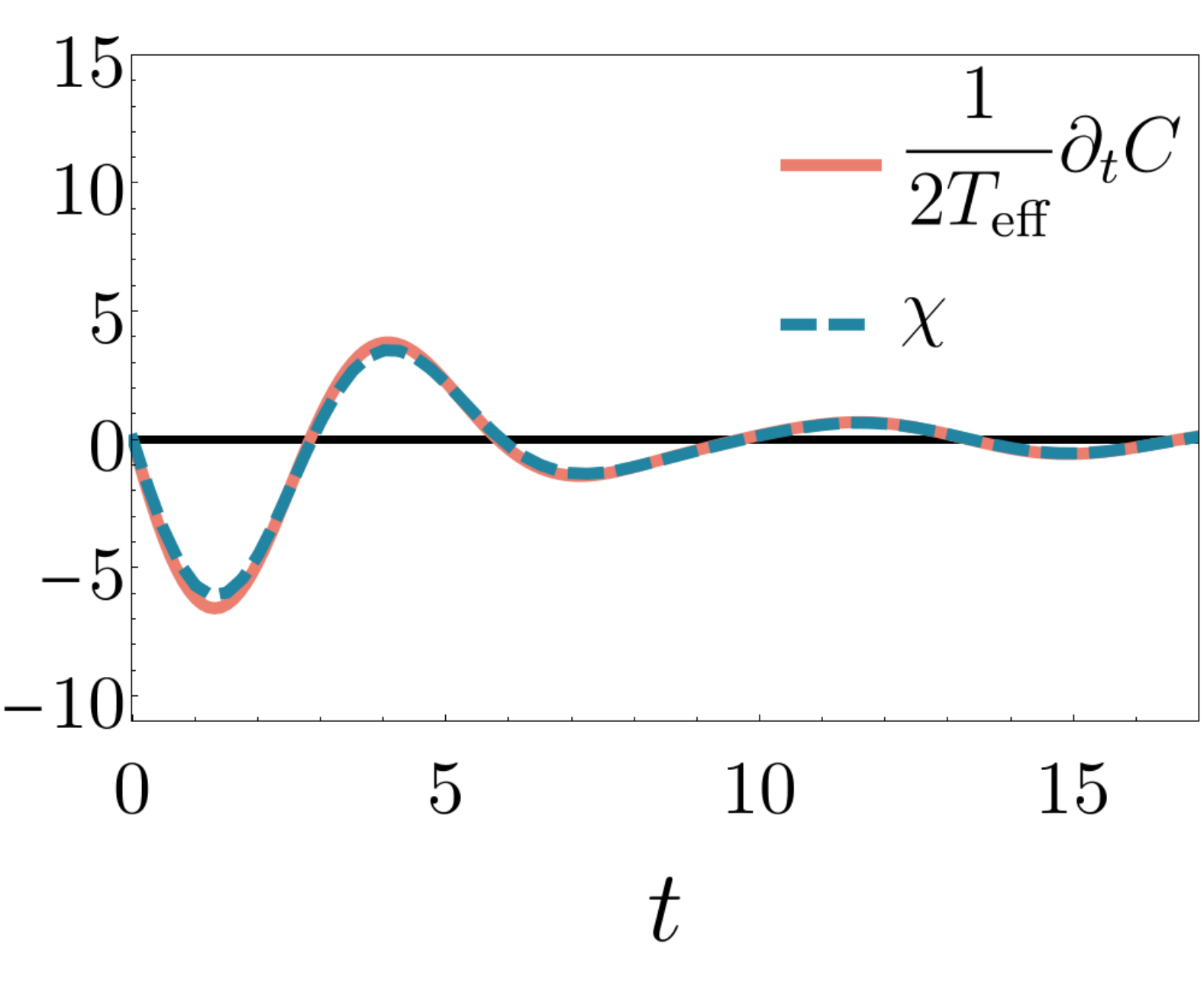}
    \caption{Numerical plot of the correlation and response functions with a system size of $N = 100$ near the weakly-dissipative critical point ($J=1, \Delta = 2, \Gamma = 0.1$). The classical FDR $\chi(t) = \partial_t C(t)/2 T_{\text{eff}}$ holds for exact numerics with $T_{\text{eff}} = J$ holds almost perfectly at all times.}
    \label{FDR SP plot}
\end{figure}
\begin{equation}\label{effective temperature}
    T_{\text{eff}} = \lim_{\omega \to 0} \frac{\omega}{2} \frac{P^K}{P^R - P^A} 
    =\frac{\Gamma^2 + 16\Delta^2}{32 \Delta}\,.
\end{equation}
The effective temperature diverges as $\Delta \to 0$ in harmony with the observation in Ref.~\cite{foss-feig_solvable_2017} that, in the absence of a transverse field, the population (in the $S_x$ basis) is that of a fully mixed state, hence infinite temperature. 
We must note however that an effective temperature defined at low frequencies is only sensible near a critical point where a slow mode dominates the dynamics. In contrast, various modes contribute to the effective temperature away from criticality, i.e., away from the phase boundary, which further complicates the interpretation of the low-frequency effective temperature. Exactly at the phase transition [see \cref{magnetization}], we find that the effective temperature is simply given by $T_\text{eff} = J$ everywhere along the phase boundary. Equation \eqref{effective temperature} can also be expressed in the time domain, $\chi(t) = \partial_t C(t)/2 T_{\text{eff}}$, which provides another form of the classical FDR \cite{tauber_critical_2014}. This relationship holds analytically for the correlation and response functions in \cref{correlation function,response function} with $T_{\text{eff}} = J$.  The latter analytical functions describe points close to, but away from, the critical point. We can further inspect the classical FDR at criticality using exact numerics: in Fig.~\ref{FDR GP plot}, we show that, with the exception of short times differences, this relation holds at criticality. We further inspect the behavior at the weakly dissipative critical point $\Gamma \to 0$ in Fig.~\ref{FDR SP plot} and find that the classical FDR holds remarkably well at all times. The agreement between $T_{\text{eff}}$ in the time and frequency domains at the phase boundary further cements the applicability of the fluctuation-dissipation relation near phase transitions.\par 
 \begin{figure}[tp]
     \centering
     \includegraphics[width=\linewidth]{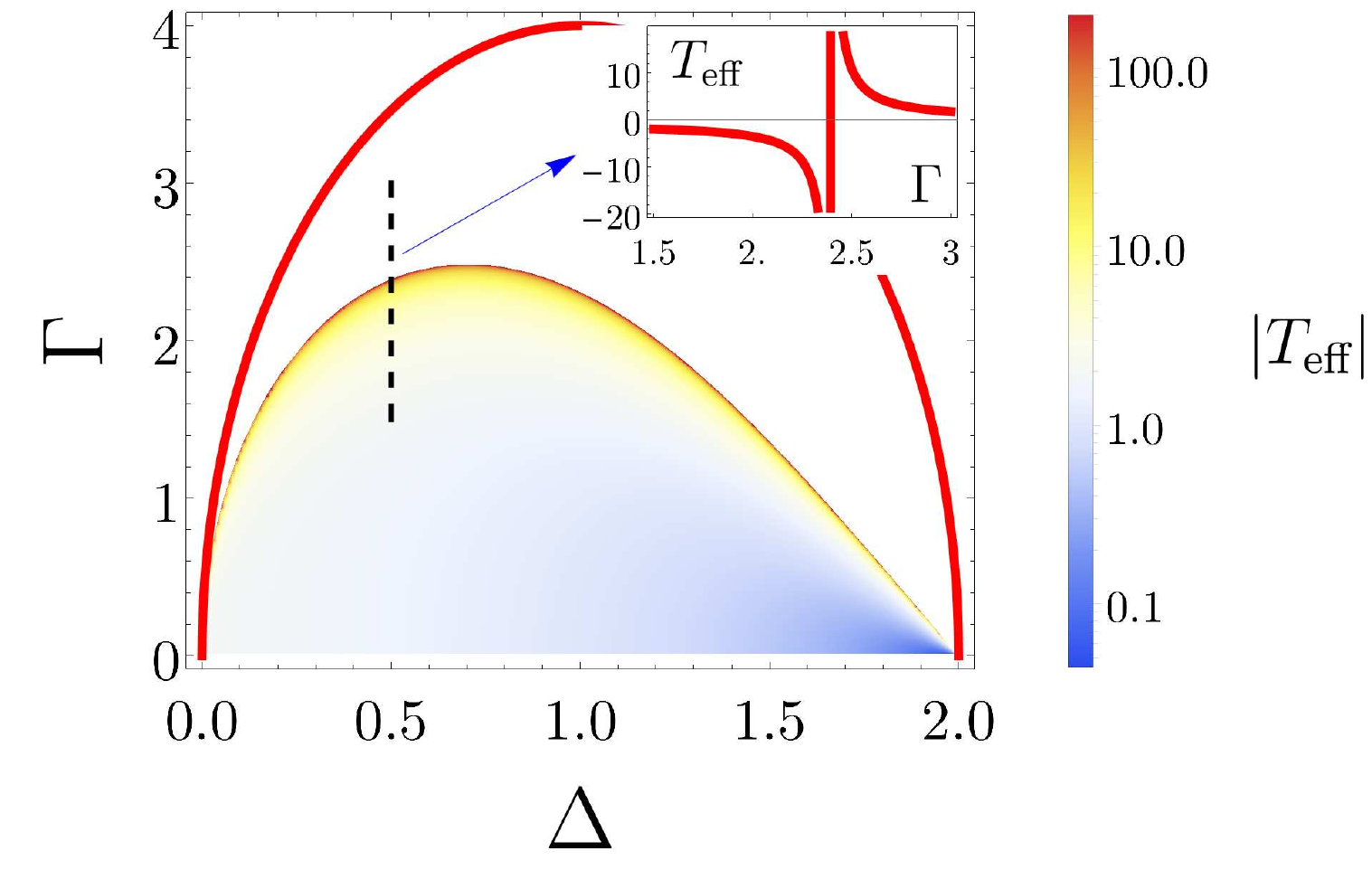}
     \caption{Density plot of $|T_{\text{eff}}|$ in the ordered phase as a function of $\Delta$ and $\Gamma$; we have set $J=1$. The thick curve indicates the phase boundary and the highlighted region indicates the region with negative effective temperature. (Inset) The effective temperature in the ordered phase ($J = 1, \Delta = 0.5$) as a function of $\Gamma$, taken along the dashed line in the main figure. As $\Gamma$ decreases, the effective temperature diverges and then flips sign.}
     \label{resp density plot}
\end{figure}
In the ordered phase, we can numerically evaluate the effective temperature by combining the expressions given in \cref{op inv response,op inv keldysh} together with the definition of the effective temperature in Eq. \eqref{FDR}. Interestingly, as $\Gamma$ is lowered, the effective temperature diverges deep in the ordered phase and then flips sign; see the inset of Fig. \ref{resp density plot}. This behavior occurs due to the change in sign of the low-frequency behavior of $\chi(\omega)$ as was pointed out in Fig.~\ref{ordered resp fig}(a). The curve corresponding to infinite temperature ends at the weakly dissipative critical point $\Gamma\to 0$ and $\Delta = 2J$. We can thus employ our field-theoretical toolbox to analytically investigate the origin of this behavior. At a technical level, we want to characterize the fluctuations around the ordered field, $m$, within the ordered phase. To this end, we consider the action describing the fluctuations around the ordered field as $S=\int_{\omega} m_q P^R_{\text{ord}}(\omega) \delta m_c +\cdots$ where $\delta m_c(t)= m_c(t)-m$ and $P^R_{\text{ord}}(\omega)$ is given exactly by Eq. \eqref{op inv response}. 
\begin{figure}[tp]
    \centering
    \includegraphics[width=\linewidth]{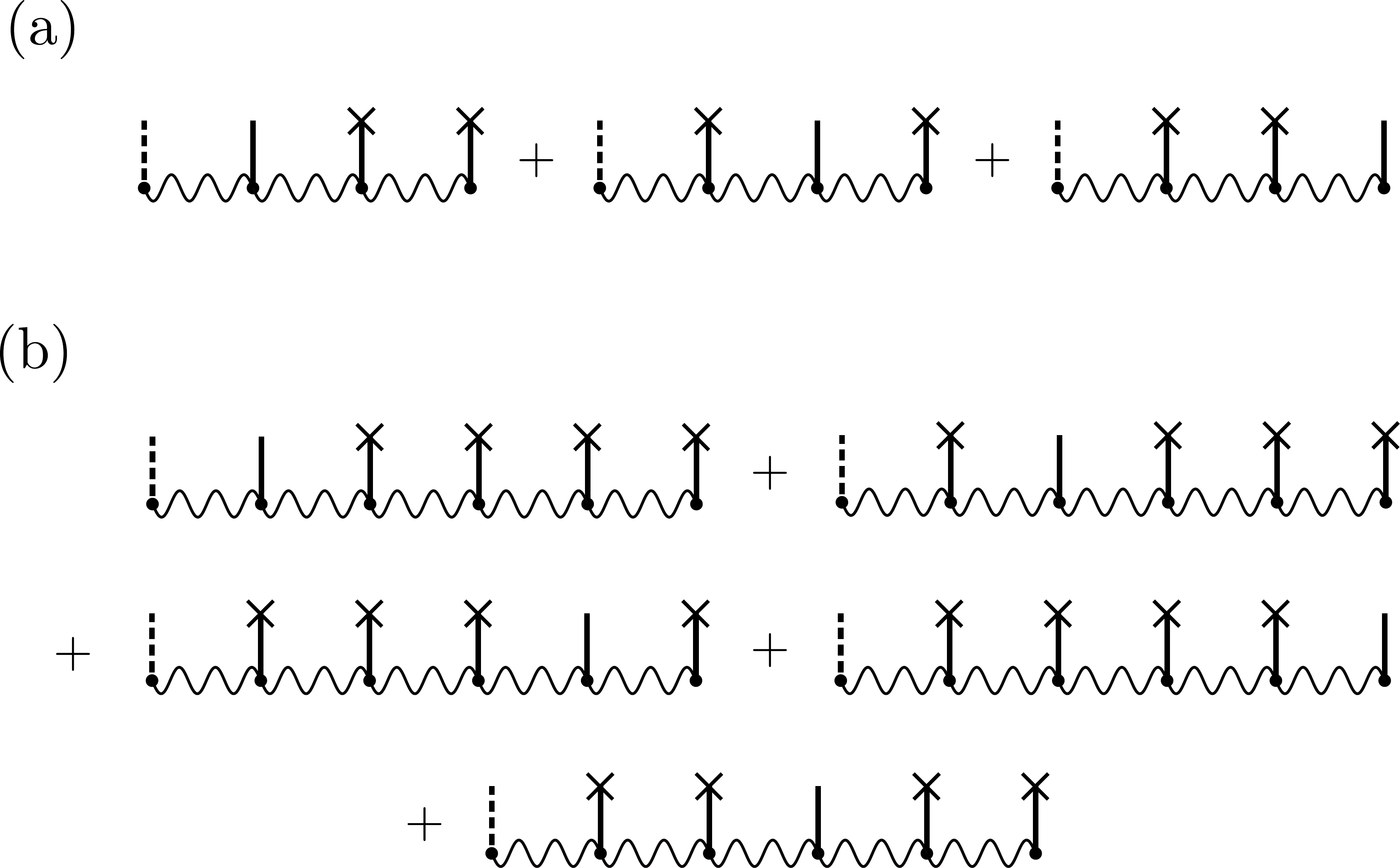}
    \caption{Diagrams contributing to (a) $P_1^R$ and (b) $P^R_2$ in calculating $\gamma_{\text{ord}}$. A cross (x) at the end of a leg corresponds to evaluating the corresponding classical field at its saddle-point value.}
    \label{damping diagrams}
\end{figure}
To probe the effective temperature $T_{\text{eff}}$, we must expand $P^R_{\text{ord}}(\omega)$ at low frequencies as $P_{\text{ord}}^R(\omega) \sim -r + i\gamma_{\rm ord} \omega+\cdots$. Now, $\gamma_{\rm ord} >0$ indicates dissipation, while $\gamma_{\rm ord}<0$ implies gain as this coefficient characterizes friction in the low-frequency dynamics. While the full expression for $P^R$ in the ordered phase is not analytically tractable, we can utilize the diagrammatics developed in the previous section: the diagrams that contribute to $P_{\text{ord}}^R$ in the ordered-phase can be found in Fig.~\ref{damping diagrams}. The explicit forms of the interaction coefficients are reported in Appendix~\ref{interaction coefficients appendix}. It turns out that to capture the negative temperature, we must also include the sixth-order terms in the diagrammatic expansion. We find
\begin{equation}\label{Pr ordered}
    P^R_{\text{ord}}(\omega) = P^R(\omega) + m^2 P_1^R(\omega) + m^4 P_2^R(\omega)+\cdots\,, 
\end{equation}
where $m$ is given by Eq. \eqref{magnetization}, $P^R(\omega)$ by Eq.~\eqref{inverse response frequency}, $P^R_1(\omega)$ is given by 
\begin{align}
    i P_1^R(\omega) = &\overline{u}_{qccc}(-\omega, \omega, 0, 0) + \overline{u}_{qccc}(-\omega, 0, \omega, 0)\nonumber\\
    &+\overline{u}_{qccc}(-\omega, 0, 0, \omega)\,,
\end{align}
and $P_2^R(\omega)$ is given by
\begin{align}
    &i P_2^R(\omega) =  \overline{u}_{qccccc}(-\omega, \omega, 0, 0, 0, 0)\\
    &+ \overline{u}_{qccccc}(-\omega, 0, \omega, 0, 0, 0) + \overline{u}_{qccccc}(-\omega, 0, 0, \omega, 0, 0)\nonumber \\
    &+ \overline{u}_{qccccc}(-\omega, 0, 0, 0, \omega, 0)+ \overline{u}_{qccccc}(-\omega, 0, 0, 0, 0, \omega) \nonumber \,.
\end{align}
Expanding Eq. \eqref{Pr ordered} to first order in $\omega$, we find the friction coefficient
\begin{equation}\label{gamma ordered}
\begin{split}
    \gamma_{\text{ord}} = &\frac{127 J^2 \Delta}{\Gamma (\Gamma^2 + 16\Delta^2)^4}\bigg[ 26\Gamma^6 - 4096\Delta^4 (\Delta - 2J)^2+ \\
    16\Gamma^4\Delta & (53\Delta - 84J) 
    +256\Gamma^2\Delta^2(68J^2 - 80J\Delta + 25\Delta^2)\bigg]\,,
\end{split}
\end{equation}
which indeed captures the negative effective temperature in the ordered phase near the phase boundary at $\Delta=2J$ and $\Gamma \to 0$; see Fig.~\ref{resp density plot}. Indeed, we find that the infinite-temperature curve near the weakly dissipative critical point is given by the line $\Gamma = 2\sqrt{2}(2J-\Delta)$, in harmony with Fig.~\ref{resp density plot}. We finally remark that, for $\Delta <2J$, the effective temperature $T_{\text{eff}} \to 0^-$ in the limit $\Gamma \to 0$.
\par

Before closing this subsection, a remark about the effective temperature is in order. The latter temperature characterizes fluctuations and dissipation of the system at low frequencies. However, it does not imply that the steady state is a thermal state, $\exp(-H/T_\text{eff})$. This can be seen by comparing the equilibrium phase diagram versus the equilibrium phase diagram  \cite{das_infinite-range_2006} in Fig. \ref{phase space}. Specifically, the infinite-range Ising model in a transverse field undergoes a phase transition at a critical value of the transverse field that is $\Delta_c(T) <2 J$ at any finite temperature, a behavior that should be contrasted with our driven-dissipative model whose phase transition extends all the way to $\Delta=2J$.\par

\section{Critical Behavior}
Just like their equilibrium counterparts, non-equilibrium steady states may undergo phase transitions and exhibit critical phenomena. A characteristic feature of criticality is a diverging correlation length, the dynamical analog of which is manifested as a diverging time scale and the associated critical slowdown \cite{tauber_critical_2014}. While there is no intrinsic length scale in an infinite-ranged model, we will identify the dynamical critical behavior of the model considered here and investigate the finite-size scaling with the system size $N$ \cite{botet_size_1982, botet_large-size_1983}. Interestingly, we shall see that two distinct dynamical critical behaviors emerge depending on the strength of dissipation. 

\subsection{Criticality at Finite \texorpdfstring{$\Gamma$}{G}}\label{finite gamma FSS}
Before investigating the finite-size scaling, we first determine the scaling dimensions of the fields at the quadratic level of the action. A low-frequency expansion of Eq. \eqref{action expansion} yields the quadratic action
\begin{equation}
    \mathcal{S} \sim  \int_t  m_q(- \gamma \partial_t- r )m_c + \frac{1}{2} D m_q^2\,,
\end{equation}
with $r$ the distance from the critical point, $\gamma$ a damping parameter, and $D$ the strength of the noise:
\begin{subequations}
\label{langevin parameters}
\begin{align}
    r = -P^R(\omega = 0) = \frac{2J [\Gamma^2 - 16\Delta(2J-\Delta)]}{\Gamma^2 + 16\Delta^2},\\
    \gamma = -i \partial_\omega P^R(\omega)|_{\omega = 0} = \frac{256 J^2 \Gamma \Delta}{(\Gamma^2 + 16\Delta^2)^2},\\
    D = P^K(\omega = 0) = \frac{32 i J^2 \Gamma}{\Gamma^2 + 16\Delta^2}.
\end{align}
\end{subequations}
To find the scaling dimensions of the fields, we demand that the action be scale invariant at the critical point ($r = 0$). One can see that the action is invariant upon rescaling \cite{lundgren_nature_2020}
\begin{equation}\label{eq. before}
    t \to \lambda t, \quad
    m_c \to \sqrt{\lambda}m_c\,,\quad m_q \to \frac{1}{\sqrt{\lambda}}m_q\,,
\end{equation}
\begin{figure}[tp]
    \centering
    \includegraphics[scale=.25]{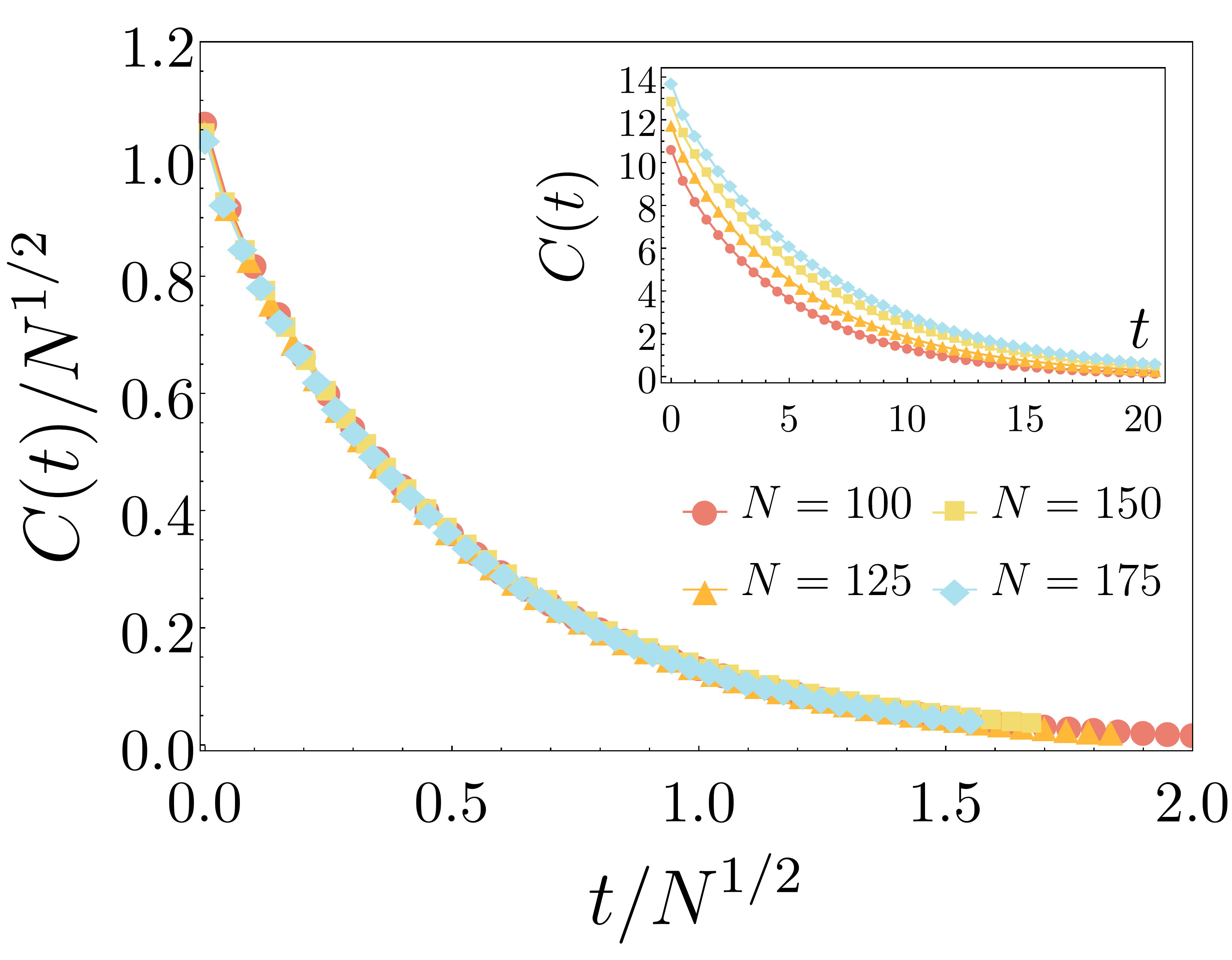}
    \caption{Exact numerics of the finite-size scaling behavior of the correlation function at a generic critical point ($J=1, \Delta = 1, \Gamma=4$). The critical dynamics is overdamped and is governed by a characteristic time scale that scales as $t \sim N^{1/2}$, typical of critical driven-dissipative systems.}
    \label{finite size GP plot}
\end{figure}%
which determines the scaling dimensions of the fields as $[m_c] = \frac{1}{2}$, $[m_q] = -\frac{1}{2}$. These scaling dimensions in turn determine the scaling behavior of the correlation and response functions, and are consistent with \cref{correlation function,response function} in the limit $\Gamma \to \Gamma_c$; see also \cite{paz_critical_2019}. 

To determine the finite-size scaling behavior of the model, we must include finite-size corrections to the quadratic action in Eq.~\eqref{action expansion}. To lowest order in $\mathcal{O}(N^{-1})$, the finite-size corrections are given by the 4-legged diagrams derived in Sec. \ref{Diagrammatics}. Furthermore, it follows from the above scaling dimensions that the most relevant correction (in a renormalization-group sense) is the classical vertex which has a low-frequency limit of \cite{torre_keldysh_2013} 
\begin{equation}\label{classical vertex}
    {\cal S}_{\rm int}= \frac{-u}{2N} \int_{t}  m_c^3 m_q+\cdots\,,
\end{equation}
with
\begin{equation}\label{u coeff}
    u = 2i\overline{u}_{cccq}(\bm{\omega} = 0) = \frac{2048 J^4 \Delta}{(\Gamma^2 + 16\Delta^2)^2}\,.
\end{equation}
We now demand that the full low-frequency expansion of the action, with the inclusion of the classical vertex and at a finite distance from the critical point ($r \neq 0$), remain scale invariant. This is achieved upon rescaling 
\begin{align}
    t \to \lambda t,& \quad m_c \to \sqrt{\lambda}m_c\,,\quad m_q \to \frac{1}{\sqrt{\lambda}}m_q, \nonumber\\
    &r \to \frac{1}{\lambda} r\,, \quad N \to \lambda^2 N\,,
\end{align}
where the first line, also given by \cref{eq. before}, is included for completeness.
Equipped with these scaling dimensions, the correlation function takes on the scaling form
\begin{equation}\label{GP correlation scaling}
    C(t) = \langle m_c(t) m_c(0) \rangle = \lambda^{-1} \hat C (\lambda |t|,  \lambda^{-1} r, \lambda^{-2} N^{-1})\,,
\end{equation}
with $\hat C$ a scaling function and $\lambda$ an arbitrary scaling parameter which can be chosen freely. 
Setting $\lambda = r$ at equal times, $t = 0$, and in the thermodynamic limit $N \to \infty$, we obtain the ``photon-flux'' exponent as
\begin{equation}\label{GP photon flux}
    C(0) = \frac{1}{r}\hat C (0, 1, 0)\,,
\end{equation}
\begin{figure}[tp]
    \centering
    \includegraphics[scale=.22]{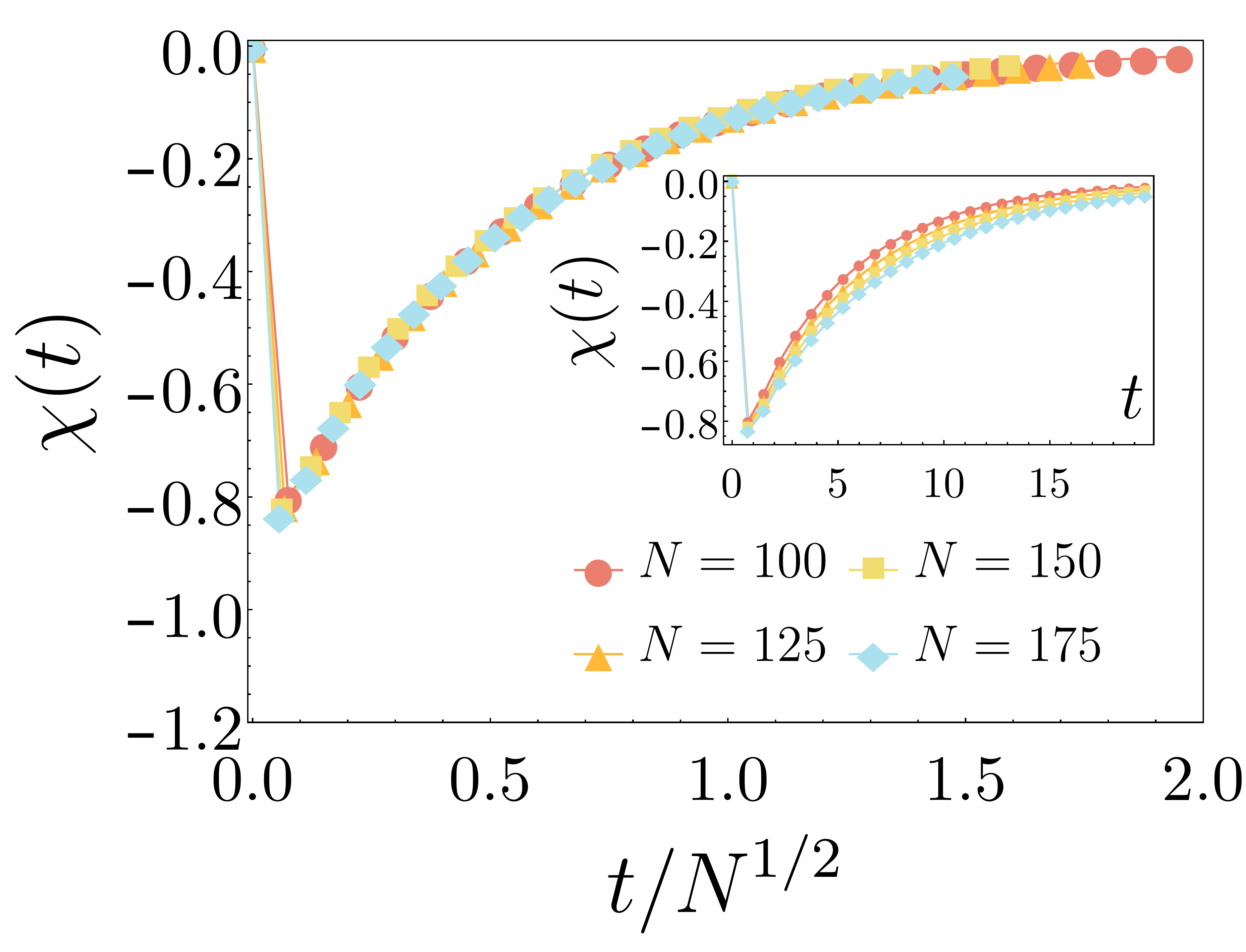}
    \caption{Finite-size scaling behavior of the response function at a generic critical point ($J = 1, \Delta = 1, \Gamma = 4$) from exact numerics. The amplitude of the response function does not scale with $N$, while  the characteristic time scale of the dynamics scales as $t\sim N^{1/2}$, identifying the dynamical exponent $\zeta=1/2$.}
    \label{finite size GP resp}
\end{figure}%
which establishes the exponent $\nu = 1$ \cite{torre_keldysh_2013}. Next we determine the finite-size scaling at criticality ($r=0$). Here, we set $\lambda = N^{-1/2}$ in \cref{GP correlation scaling}, which leads to the scaling form
\begin{equation}\label{GP finite size scaling}
    C(t) = \sqrt{N}\hat C (t/\sqrt{N}, 0, 1)\,.
\end{equation}
This equation identifies both static and dynamic finite-size critical exponents: the amplitude of correlations (i.e., fluctuations) scale as $C \sim N^\alpha$ with the exponent $\alpha = 1/2$, while a critical time scale emerges as $t \sim N^{\zeta}$ with the dynamical exponent $\zeta = 1/2$.
A similar analysis yields the scaling form of the response function:
\begin{equation}\label{GP response finite size scaling}
    \chi(t) = \hat \chi (t/\sqrt{N}, 0, 1)\,.
\end{equation}
We thus see that the amplitude of the response function does not scale with $N$. We confirm the (static as well as dynamic) scaling behavior of both the correlation and response functions in Figs.~\ref{finite size GP plot} and  \ref{finite size GP resp}, respectively. 
Additionally, we see that the critical dynamics observed here is purely relaxational. In the next section, we show that a distinct dynamical critical behavior emerges at low dissipation.\par

\subsection{Criticality at \texorpdfstring{$\Gamma \to 0$}{GtoZero}}\label{weakly dissipative critical properties}
The effective classical behavior at a generic critical point is due to the competition of  drive and dissipation. It is then interesting to consider the limit $\Gamma \to 0$ where dissipation is small compared to the energy scales in the system. Interestingly, the phase transition persists in this limit and occurs at $\Delta = 2J$ as $\Gamma \to 0$; see Fig.~\ref{phase space}. One must be careful when considering this point as setting $\Gamma$ to zero would make the problem unphysical since dissipation is required to find a unique non-equilibrium steady state. Rather, we shall consider the asymptotic behavior in the limit $\Gamma \to 0$ at the level of the low-frequency expansion of Eq.~\eqref{action expansion}. The resulting action then becomes 
\begin{equation}
    \mathcal{S} \sim  \int_t  m_q( -a \partial_t^2 -\gamma \partial_t -r)m_c + \frac{1}{2}D m_q^2\,,
\end{equation}
where the parameters $\gamma$, $r$, and $D$ are provided in Eq.~\eqref{langevin parameters} upon taking the appropriate limit; the new parameter $a$ is given by
\begin{equation}\label{alpha coeff}
    a = \lim_{\Gamma \to 0} \partial_{\omega}^2 P^K(\omega)|_{\omega = 0 } = \frac{J^2}{\Delta^3}\,.
\end{equation}
Indeed the inertial term in the action (proportional to $a$) is required in the limit of vanishing dissipation.
This is because the damping parameter $\gamma \sim \Gamma$ and the noise $D \sim \Gamma$ both vanish with $\Gamma$. To determine the new scaling dimensions of the fields, we once again seek a scaling transformation that keeps the action scale invariant, but this time we also should include the scaling of $\Gamma$ itself. We find that the quadratic action at the critical point is invariant under
\begin{equation}\label{SP finite size transformation}
\begin{split}
    t \to \lambda t\,, \quad &\Gamma \to \frac{1}{\lambda} \Gamma\,, \quad 
    m_c \to \lambda m_c\,, \quad m_q \to m_q\,,
\end{split} 
\end{equation}
establishing the new scaling dimensions $[m_c] = 1$, $[m_q] = 0$. The new scaling dimensions alter the original scaling dimensions of the correlation and response functions, again in harmony with their behavior in the limit $\Gamma \to 0$; see also Ref.~\cite{paz_critical_2019}.

To obtain the finite-size scaling behavior, we once again include the classical vertex, which remains the most relevant interaction term. The full action (including the mass term) remain invariant if we impose the rescaling 
\begin{equation}
    r \to \frac{1}{\lambda^2}r, \quad N \to \lambda^4 N,
\end{equation}
in addition to those in \cref{SP finite size transformation}.
From this, we find the scaling form for the correlation function as
\begin{equation}\label{SP correlation scaling}
    C(t) = \frac{1}{\lambda^2} \hat C_0 (\lambda |t|, \lambda^{-1} \Gamma, \lambda^{-2}r, \lambda^{-4} N^{-1})\,,
\end{equation}
where the subscript $0$ denotes the scaling function near the weakly dissipative critical point. Also, notice the dependence of the nontrivial scaling of $\Gamma$ in contrast with a generic critical point; cf. \cref{GP correlation scaling}. 

First, we consider the point $\Delta=2J$ at finite yet small $\Gamma$. Setting  $\lambda = \Gamma$ and $t = 0$ in the thermodynamic limit, we find
\begin{equation}\label{SP photon flux}
    C(0) = \frac{1}{\Gamma^2} \hat C_0 (0, 1,{\rm const}, 0) \propto \frac{1}{r}\,,
\end{equation}
where the scaling behavior in the last step follows from the fact that $r \sim \Gamma^2$, rendering the same photon-flux exponent $\nu = 1$.

\begin{figure}[tp]
    \centering
    \includegraphics[scale=.24]{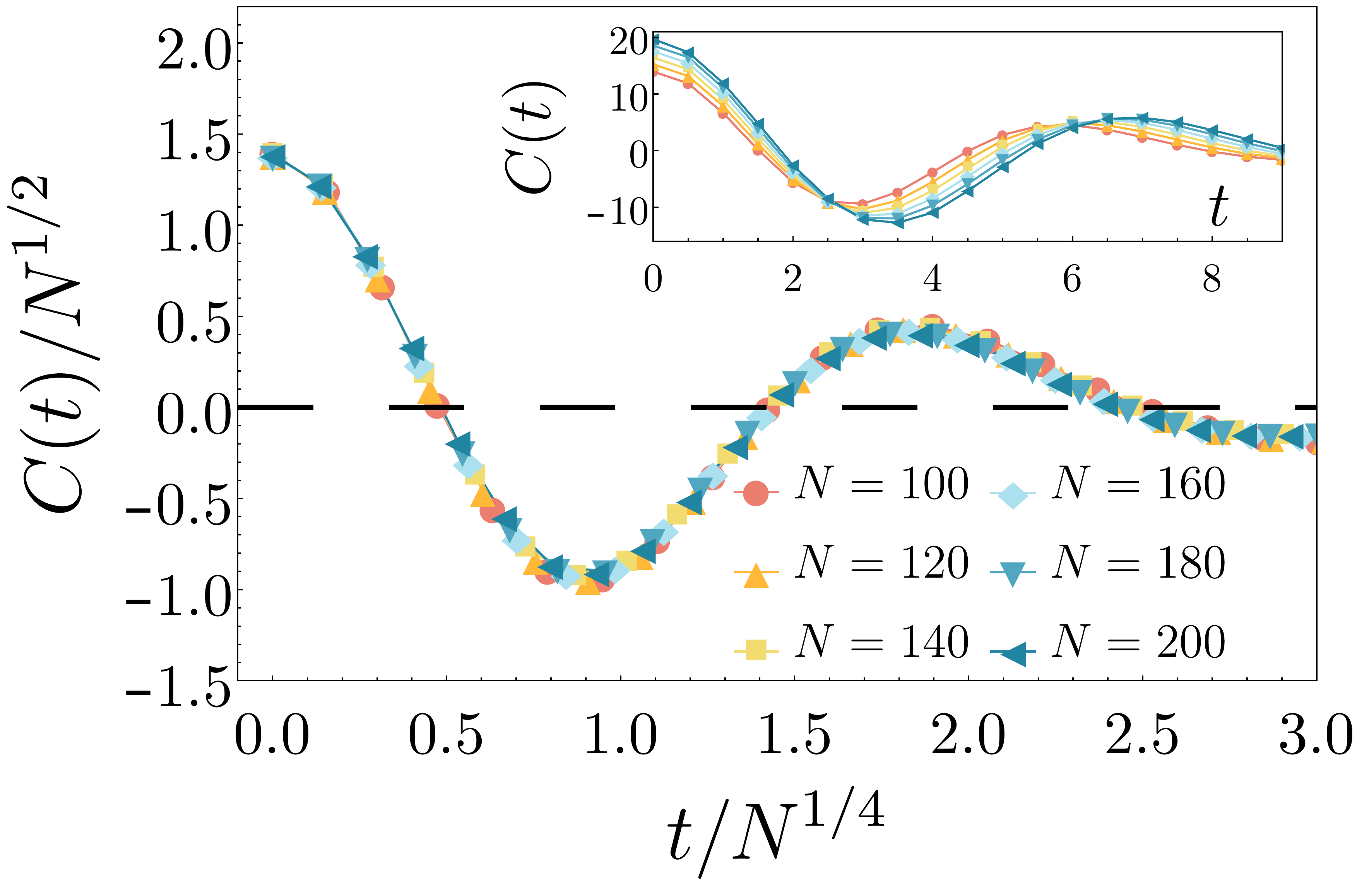}
    \caption{Finite-size scaling of the correlation function near the weakly dissipative critical point ($J = 1, \Delta = 2, \Gamma = 0.1$). 
    The dynamics is underdamped in contrast with the purely relaxational behavior at a generic driven-dissipative phase transition, and exhibits the critical scaling $t \sim N^{1/4}$ to be contrasted with $t \sim N^{1/2}$ of relaxational dynamics; cf. Fig.~\ref{finite size GP plot}.}
    \label{finite size SP plot}
\end{figure}
\begin{figure}[tp]
    \centering
    \includegraphics[scale=.19]{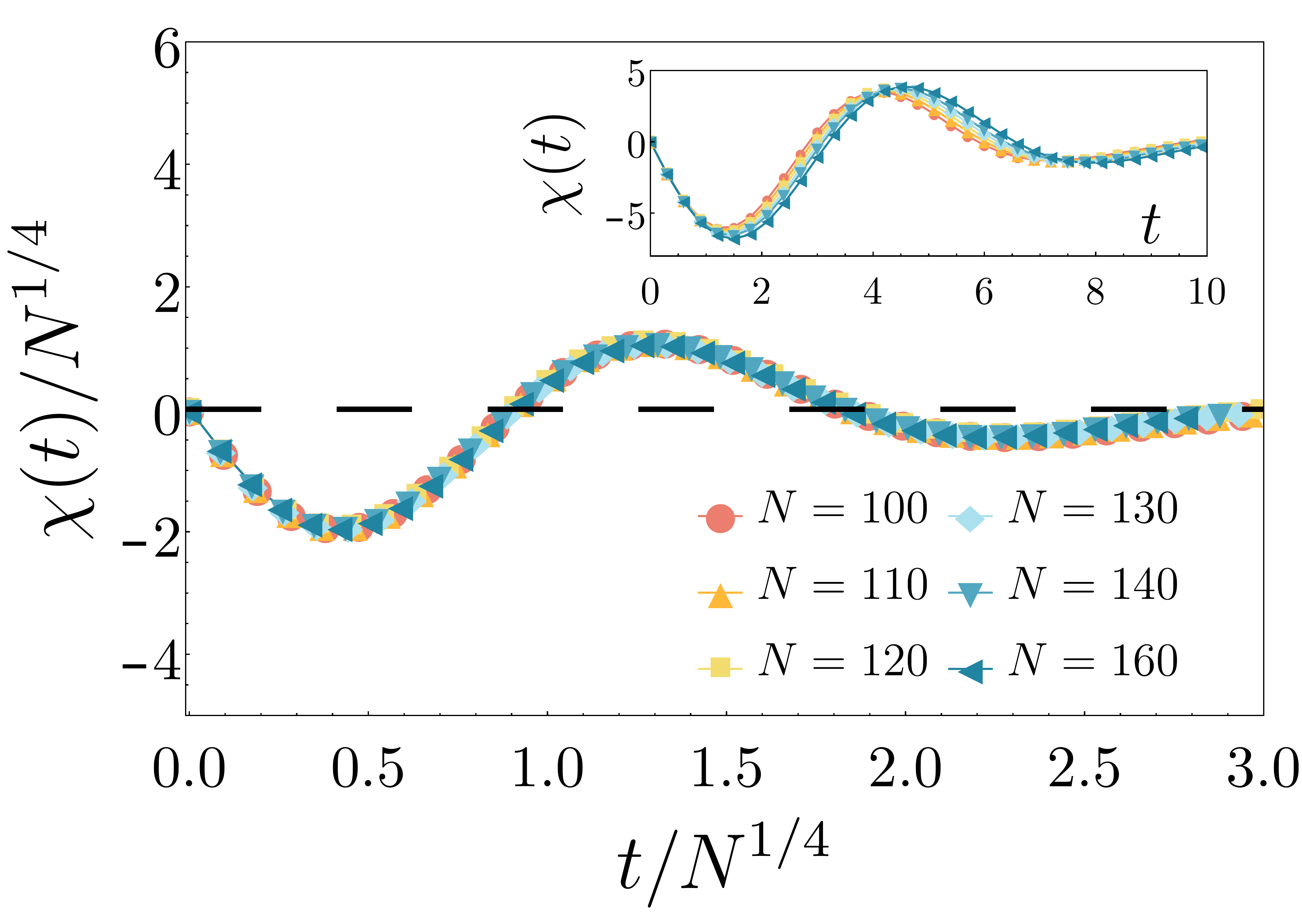}
    \caption{Finite-size scaling of the response function near the weakly dissipative critical point from exact numerics ($J = 1, \Delta = 2, \Gamma = 0.1$). The dynamics is distinguished from a generic critical point in that the dynamical critical exponent is different, $\zeta=1/4$ and that it is underdamped; cf. Fig.~\ref{finite size GP resp}.}
    \label{finite size SP resp}
\end{figure}

Next, we shall focus on finite-size scaling. To this end, we consider a weakly-dissipative critical point at finite yet small $\Gamma$; we shall choose $\Delta\lessapprox 2J$ to ensure criticality. Now, we set $\lambda^4 = N^{-1}$ together with $r= 0$ to find 
\begin{equation}\label{SP finite size scaling}
    C(t) = \sqrt{N} \hat C_0 (|t|N^{-\frac{1}{4}}, \Gamma N^{\frac{1}{4}}, 0, 1)\,.
\end{equation}
From this equation, we find that the weakly-dissipative limit does not affect the static scaling exponent, $\alpha=\frac{1}{2}$, but it \textit{does} change the dynamical exponent to $\zeta = \frac{1}{4}$. We thus conclude that a weakly dissipative point changes the dynamical critical behavior. Repeating the above analysis for the response function, we find the finite-size scaling form
\begin{equation}\label{SP response finite size scaling}
    \chi(t) = N^{\frac{1}{4}}\hat \chi_0 (tN^{-\frac{1}{4}}, \Gamma N^{\frac{1}{4}},0, 1)\,.
\end{equation}
In contrast with a generic critical point (Eq. \eqref{GP response finite size scaling}), the amplitude of the response function in the above equation grows with the system size as $\chi \sim N^{\frac{1}{4}}$. \Cref{finite size SP plot,finite size SP resp} show the finite-size critical behavior of the correlation and response function, respectively, and confirm the prediction of the scaling analysis. 
In conclusion, while the static exponent $\alpha$ and the flux exponent $\nu$ remain the same everywhere on the phase boundary, the dynamical exponent $\zeta$ takes a different value in the weakly-dissipative limit.

\begin{figure}[tp]
    \centering
    \includegraphics[scale=.5]{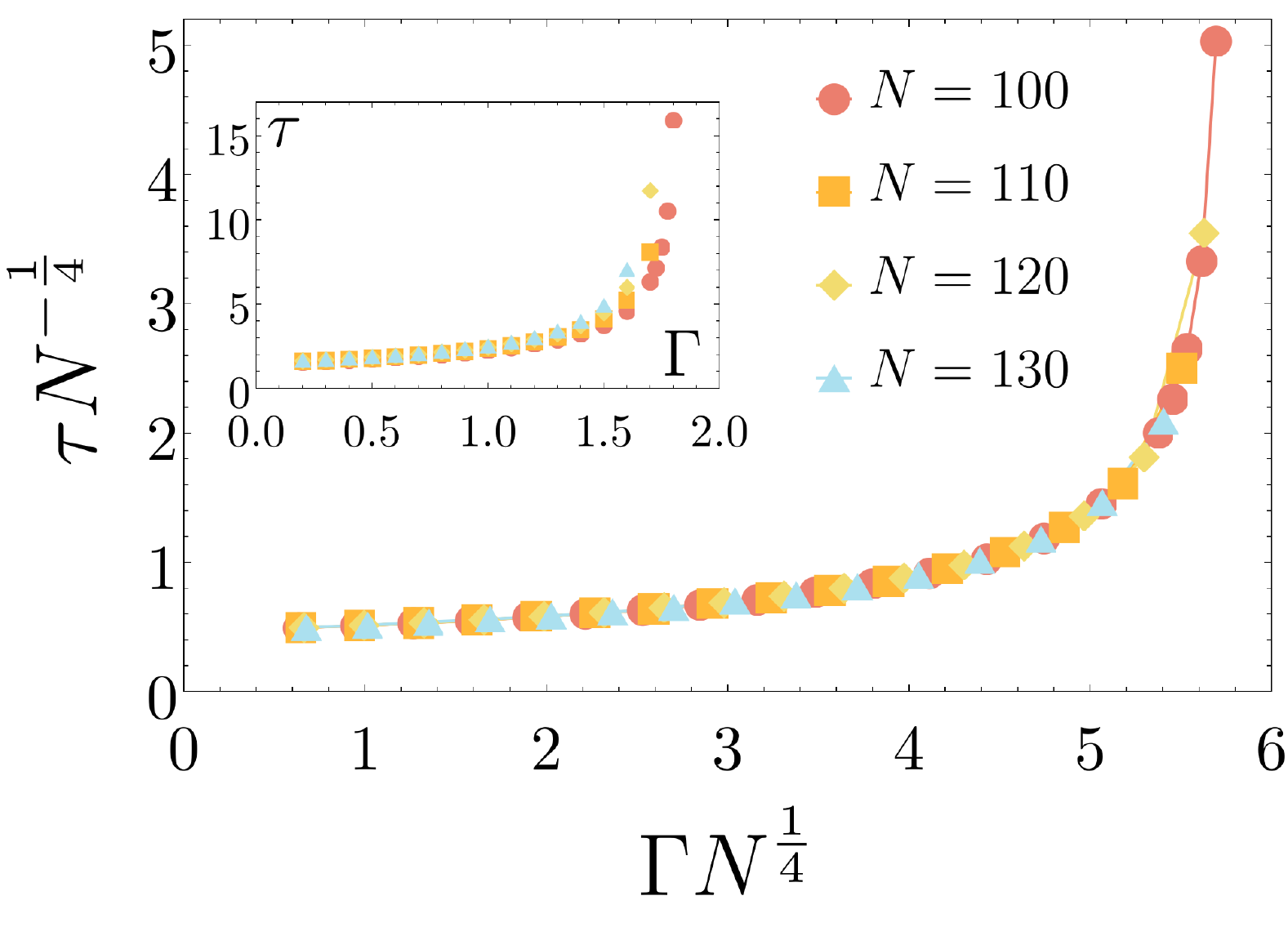}
    \caption{The first zero of the correlation function, $\tau$, as a function of the dissipation rate $\Gamma$ and at different system sizes. Both quantities are scaled with system size to make the scaling behavior manifest. 
    The time scale $\tau$ diverges at sufficiently large $\Gamma$ approximately when $\Gamma N^{1/4} \approx 6$. The inset shows the unscaled plots for comparison.}
    \label{crossover}
\end{figure}

What further distinguishes the weakly dissipative critical point is the fact that the dynamics is \textit{underdamped} (see \cref{finite size SP plot,finite size SP resp}) in contrast with the typical relaxational/overdamped dynamics seen at a generic critical point, and generally in driven-dissipative systems. 
As $\Gamma$ is further increased along the phase boundary, one should expect a crossover to overdamped critical dynamics. This is somewhat analogous to the quantum critical region and the crossover to thermal critical behavior \cite{sachdev_quantum_2011}.
In the context of the infinite-range model that we have considered in this work, the crossover behavior becomes manifest as a function of system size. Indeed, we can determine the crossover behavior from \cref{SP finite size scaling}: for $\Gamma t \lesssim 1$ and $\Gamma \lesssim N^{-\frac{1}{4}}$, the critical dynamics is underdamped, while for large times and/or large $\Gamma$ the system experiences a dynamical crossover where we recover the usual relaxational behavior (while remaining on the phase boundary). To quantitatively investigate the crossover, we define the first zero of the correlation function, denoted by $\tau$, as a measure of the oscillatory behavior of the underdamped dynamics. 
In Fig. \ref{crossover}, we plot $\tau$ as a function of $\Gamma$ and for different system sizes. Indeed, we find that for sufficiently large values of $\Gamma$, this time scale diverges where the dynamics becomes overdamped. Furthermore, this figure shows that this time scales as $\tau \sim N^{\frac{1}{4}}\hat{\tau}(\Gamma N^{\frac{1}{4}})$ with $\hat{\tau}$ a universal scaling function, hence it confirms the scaling of the crossover value, $\Gamma_{\rm cr}\sim N^{-1/4}$.\par 

One can gain some intuition for the underdamped critical behavior near the weakly-dissipative critical point from several different angles. First, the point $\Delta = 2J$ is exactly where $\Gamma_c$ switches from real to imaginary, as a result of which Eq.~\eqref{correlation function} shows underdamped dynamics even away from the phase boundary (when $\Delta>2J$). Second, one can imagine that the underlying coherent dynamics generated by the first term in Eq. \eqref{DDIM} could have a stronger effect in the limit $\Gamma \to 0$. Additionally, the the infinite-range Ising model is integrable in the absence of dissipation; while dissipation generically spoils integrability, the dynamics is approximately integrable in the limit $\Gamma \to0$, which could lead to nontrivial dynamics \cite{lange_pumping_2017, lange_time-dependent_2018, lenarcic_perturbative_2018, shirai_thermalization_2020}. Nevertheless, in  Sec. \ref{integrability}, we show that the underdamped dynamics survives to the first nontrivial order of integrability-breaking perturbations. 
\par

\subsection{Comparison with Equilibrium}
From the scaling dimensions and critical exponents, we can place each phase transition in its respective universality class. Remarkably, both finite-$\Gamma$ and $\Gamma \to 0$ phase transitions are in equilibrium universality classes, albeit with a classical and quantum flavor, respectively.
For a generic critical point at finite $\Gamma$, the scaling dimensions are $[m_c] = \frac{1}{2}$, $[m_q] = -\frac{1}{2}$ with the critical exponents $\alpha = 1/2$, $\zeta = 1/2$. These quantities place this phase transition in the same universality class as the classical infinite-ranged Ising model at finite temperature with Glauber-type dynamics (i.e. non-conserving dynamics) \cite{oh_monte_2005}, which itself belongs to the ``model A'' class of Hohenberg \& Halperin \cite{hohenberg_theory_1977}.
Despite the microscopic quantum dynamics, the combination of drive and dissipation render the critical behavior  effectively classical and equilibrium-like. This appears to be the generic behavior in driven-dissipative phase transitions \cite{torre_keldysh_2013, maghrebi_nonequilibrium_2016, sieberer_keldysh_2016, maghrebi_nonequilibrium_2016, owen_quantum_2018, chan_limit-cycle_2015, wilson_collective_2016, le_boite_steady-state_2013, foss-feig_emergent_2017, vicentini_critical_2018, overbeck_multicritical_2017, mitra_nonequilibrium_2006, wouters_absence_2006}. However, there are exceptions such as classical yet truly non-equilibrium critical behavior \cite{young_non-equilibrium_2019}, as well as the emergence of quantum criticality in the limit of weak dissipation and drive \cite{Marino2016,rota_quantum_2019}.

\begin{figure}[tp]
    \centering
    \includegraphics[width=\linewidth]{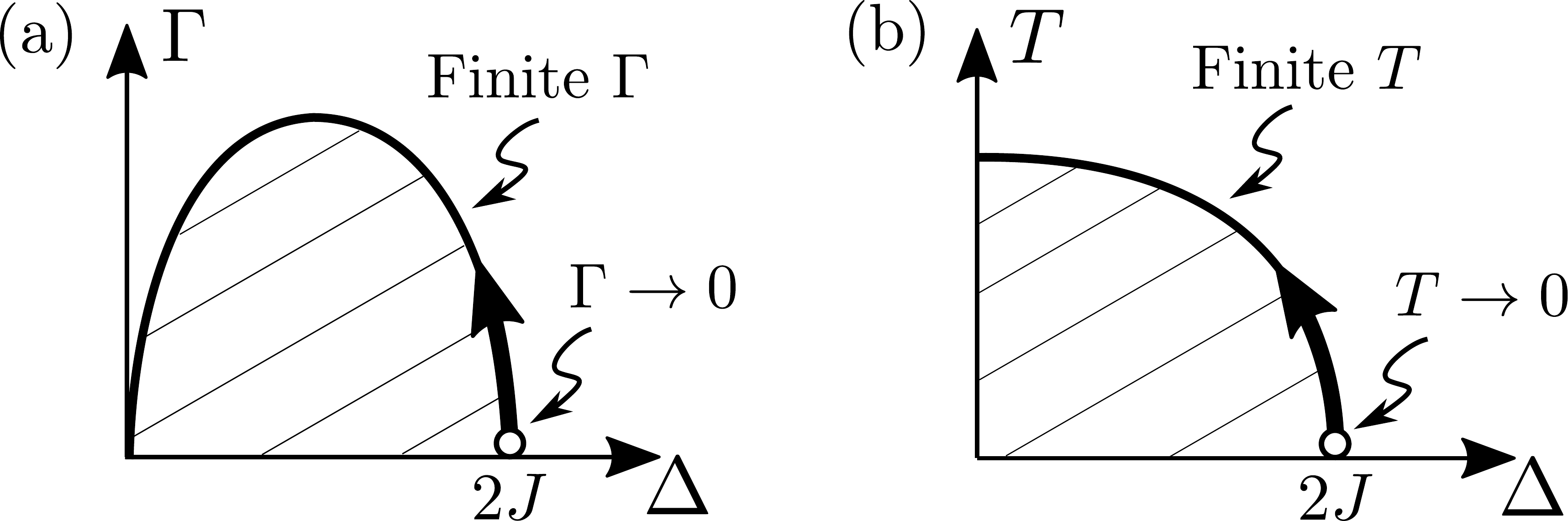}
    \caption{Phase diagrams of the infinite-range (a) driven-dissipative Ising model (DDIM), and (b)  equilibrium Ising model in a transverse field. The shaded regions denote the ordered phase. The weakly dissipative critical point of the DDIM, $\Gamma \to 0$ in (a), exhibits underdamped dynamics in contrast with the relaxational dynamics at a generic critical point. Analogously, the equilibrium model in (b) exhibits distinct (quantum and thermal) dynamics at zero and finite temperature. Both $\Gamma\to 0$ and $T\to 0$ define unstable fixed points but with respect to dissipation and thermal fluctuations, respectively. The weakly dissipative dynamics in (a) exhibits identical critical scaling to a finite-temperature critical point in (b)}
    \label{phase space}
\end{figure}%

In the weakly-dissipative limit, we have found the scaling dimensions $[m_c] = 1$, $[m_q] = 0$, which are distinct from both classical ($[m_c] = \frac{1}{2}$, $[m_q] = -\frac{1}{2}$) and quantum ($[m_c] = \frac{1}{2}$, $[m_q] = \frac{1}{2}$) cases \cite{titum_non-equilibrium_2019, torre_keldysh_2013}. These scaling dimensions lead to the new set of critical exponents $\alpha = 1/2$, $\zeta = 1/4$, as opposed to the quantum critical exponents $\alpha = 1/3$, $\zeta = 1/3$ \cite{titum_non-equilibrium_2019}. The former exponents place this phase transition in the same universality class as the finite-temperature transverse-field infinite-range Ising model, i.e. the Hamiltonian in Eq. \eqref{Hamiltonian}. Therefore, while the phase transition is equilibrium-like, it resembles the quantum Ising model at finite temperature rather than the classical stochastic Ising model. For comparison, see Fig. \ref{TFIM scaling} in Appendix \ref{TFIM appendix}. Various exponents and the comparison against classical and quantum equilibrium settings can be found in Table \ref{table1}. 

The comparison between the driven-dissipative and equilibrium behaviors can be taken one step further due to the existence of a dynamical crossover in both cases. As shown previously, the weakly-dissipative point is an unstable fixed point with respect to dissipation, where upon renormalization the critical dynamics undergoes a crossover from underdamped to overdamped dynamics; see Fig. \ref{phase space}(a). 
This crossover can be understood due to $\Gamma$ scaling inversely as time upon rescaling at the weakly-dissipative critical point, which then sets a crossover time scaling as $\sim N^{1/4}$. 
The equilibrium analog of a dynamical crossover occurs at finite temperature; see Fig. \ref{phase space}(b). Upon renormalization, the (perfectly oscillatory) coherent quantum critical dynamics undergoes a crossover to underdamped dynamics. Similarly to our driven-dissipative system, the temperature $T$ scales inversely as that of time; one can see this from the equilibrium fluctuation-dissipation relation $C = i\coth (\omega/2 T) \chi$ where $\omega$ and $T$ scale in the same way \cite{tauber_critical_2014, torre_keldysh_2013}. A similar line of reasoning indicates a crossover time $\sim N^{1/3}$. In short, the dynamical crossover of the driven-dissipative Ising model is distinguished from its equilibrium analog not only by the critical exponents but also by the nature of the crossover (underdamped-to-overdamped vs coherent-to-underdamped crossover, respectively).

\begin{table}[t!]
\centering
\renewcommand{\arraystretch}{1.5}
\setlength\tabcolsep{1.5mm}
\begin{tabular}{c|c c|c|c c}
     & \multicolumn{2}{c|}{Driven-Diss.} & Class. & \multicolumn{2}{c}{Quantum}\\
     & \multicolumn{1}{c}{$\Gamma>0$} & $\Gamma \to 0$ &  $T>0$ & \multicolumn{1}{c}{$T>0$}& $T \to 0$ \\
     \hline
     $t \sim N^\zeta$ & {\cellcolor{anti-flashwhite}}$\frac{1}{2}$ & {\cellcolor{beaublue}}$\frac{1}{4}$ & {\cellcolor{anti-flashwhite}}$\frac{1}{2}$ & {\cellcolor{beaublue}}$\frac{1}{4}$ & $\frac{1}{3}$ \\
     $C \sim N^\alpha$ &  {\cellcolor{anti-flashwhite}}$\frac{1}{2}$ & {\cellcolor{beaublue}}$\frac{1}{2}$ & {\cellcolor{anti-flashwhite}}$\frac{1}{2}$ & {\cellcolor{beaublue}}$\frac{1}{2}$ & $\frac{1}{3}$\\
\end{tabular}
\caption{Driven-dissipative vs. equilibrium classical/ quantum Ising models. A generic (finite-$\Gamma$) critical point exhibits the same critical behavior as the classical stochastic Ising model, while the weakly dissipative ($\Gamma \to 0$) critical point can be identified with the quantum Ising model at finite temperature.}
\label{table1}
\end{table}

\subsection{Langevin Description}
An alternative way of understanding the critical behavior of a driven-dissipative system is through the lens of the Langevin equation, a stochastic differential equation used to describe noisy systems \cite{gardiner_quantum_2004}. Near a critical point, where we have shown the classical vertex is the most relevant finite-size correction, we can map the low-frequency limit of the Keldysh action to a Langevin equation \cite{kamenev_field_2011, torre_keldysh_2013, sieberer_keldysh_2016}. Putting together the quadratic action from \cref{action expansion} with the interaction in \cref{classical vertex}, the action reads
\begin{equation}\label{low freq action}
    \mathcal{S} \sim  \int_t  \big[ -( \gamma \partial_t+ r)m_c(t)- \frac{u}{2N} m_c^3 (t) + \frac{1}{2} D m_q(t)  \big] m_q (t)\,,
\end{equation}
with the action parameters given by \cref{langevin parameters,u coeff,alpha coeff}.
The first step in mapping to the Langevin equation is a Hubbard-Stratonovich transformation of the quantum field $m_q$ to introduce a noise field $f(t)$ as
\begin{align}
    \mathcal{S} =  &\int_t  \big[ -( \gamma \partial_t+ r)m_c(t) - \frac{u}{2 N} m_c^3 (t) + \sqrt{2}f(t)\big] m_q (t)\nonumber \\
    &- \int_t \frac{1}{D}f(t)^2\,. \label{eq. action 79}
\end{align}
Now, integrating over $m_q$ yields a delta function which is nothing but the Langevin equation ($m = m_c/\sqrt{2}$):
\begin{equation}\label{overdamped langevin}
    \gamma \partial_t m(t) = -r m(t) -\frac{1}{N}u m^3 (t) + f(t)\,.
\end{equation}
The term $f(t)$ characterizes a white noise with a Gaussian distribution, mean $\langle f(t) \rangle = 0$, and variance $\langle f(t) f(t') \rangle = -i\frac{1}{2} D \delta(t-t') = 2\gamma T_{\text{eff}} \delta(t-t')$. It is now clear that Eq.~\eqref{low freq action} near criticality is equivalent to an overdamped Langevin equation, with an effective temperature $T_{\text{eff}}$ and in an effective potential given by
\begin{equation}
    \mathcal{H} = \frac{1}{2}r m^2 + \frac{1}{4 N}u m^4\,.
\end{equation}
Indeed, \cref{overdamped langevin} reproduces the overdamped critical dynamics discussed in Sec.\ref{finite gamma FSS}. The stochastic Langevin equation can be turned to a Fokker-Planck equation that describes the evolution of the probability distribution \cite{kamenev_field_2011, tauber_critical_2014}; with the effective equilibrium dynamics, the steady-state probability distribution of $m$ takes the form 
\begin{equation}\label{GP distribution}
    P_{\text{eq}}(m) \sim e^{-\mathcal{H}/T_{\text{eff}}}\,.
\end{equation}

The nature of the dynamics changes in the limit $\Gamma \to 0$. In this case, dissipation is vanishingly small, $\gamma \sim \Gamma \to 0$, therefore we should also include the term proportional to $\omega^2$ in the low-frequency expansion of $P^R(\omega)$. Following a similar procedure in this limit, we arrive at the Langevin equation
\begin{equation}\label{underdamped langevin}
    a \partial_t^2 m(t) + \gamma \partial_t m(t) = -r m(t) -\frac{u}{N} m^3 (t) + f(t)\,,
\end{equation}
with the parameters taken from \cref{langevin parameters} in the same limit. Incidentally, we have identified underdamped dynamics and persistent oscillations in Sec.~\ref{weakly dissipative critical properties}. Now, we can see that these oscillations are due to the inertial term that can be of the same order as dissipation (since $\gamma \to 0$). Again, one can identify the corresponding Fokker-Planck equation, also known as the Kramers-Chandrasekhar equation, whose steady-state solution is just the Maxwell-Boltzmann distribution \cite{hannes_risken_focker-planck_nodate}:
\begin{equation}\label{SP distribution}
    P(m, \dot{m}) \sim e^{-(\mathcal{H} + \frac{1}{2}a\dot{m}^2)/T_{\text{eff}}}\,.
\end{equation}
This distribution only differs from Eq. \eqref{GP distribution} in the multiplicative Gaussian distribution of $\dot{m}$. The probability distribution of $m$ in \cref{SP distribution} is identical to that of \cref{GP distribution} upon integrating out $\dot{m}$. In other words, the static properties are identical irrespective of dissipation. In contrast, the critical dynamics is markedly different as we have seen in the previous subsections. \par

Before closing this section, We emphasize that the Langevin equations derived here are only valid near the phase boundary and outside the heated region, since they are based on the dynamics of the slow mode. 

\section{Beyond the Infinite-Range Model}\label{integrability}
The infinite-range Ising model is rather special as the dynamics of the order parameter is exactly determined by mean field, although fluctuations at or close to criticality require a separate treatment as discussed in previous sections. 
In this section, we investigate the effects of non-mean-field perturbations, and specifically short-range interactions, on the dynamics. Most importantly, we show that the underdamped dynamical critical behavior in the limit $\Gamma \to 0$ persists even in the presence of the  short-range interactions.\par 

To investigate the role of integrability at the weakly-dissipative critical point, we add a nearest-neighbour interaction to the Hamiltonian in \cref{Hamiltonian}: 
\begin{equation}
H_{\rm NN} = H - \lambda \sum_i \sigma^x_i \sigma^x_{i+1}.
\end{equation}
We shall consider the perturbative limit $\lambda \ll J, \Delta $ and assume periodic boundary conditions. The short-term interaction alters the mean-field structure of the infinite-range Ising model, breaks its integrability \cite{lerose_chaotic_2018}, and could modify the phase boundary. 
Such perturbations can be viewed as spin-wave fluctuations, and have been investigated in other non-equilibrium settings such as quantum quenches \cite{lerose_chaotic_2018, lerose_impact_2019, zhu_dicke_2019}. 
While our model is distinct due to its driven-dissipative dynamics, we can still resort to a similar picture in terms of spin waves
\[\tilde{\sigma}^\alpha_k = \sum_{j=1}^N e^{-ikj} \sigma^\alpha_j\,,\]
where $k = 2\pi n/N$ with $n \in \{0,1, \cdots, N-1\}$. We shall identify the collective spin as the $k = 0$ mode; without short-range interactions, there is no coupling between this and other modes with $k\ne0$, however, the short-range interaction couples them and thus spoils the mean-field nature of the model. Naively, one might expect that spin waves act as an effective bath for the ``large spin'' corresponding to the $k = 0$ mode, which would lead to an effective dissipation (even in the limit $\Gamma\to0$). However, we will show using the diagrammatic techniques that this is not the case, and therefore the underdamped critical dynamics at the weakly dissipative critical point is robust against short-range interactions.

\subsection{Short-Range Perturbation via Field Theory}
The quantum-to-classical mapping process is not altered much by the inclusion of short-range interactions. The steps leading to the Hubbard-Stratonovich transformation in Eq.~\eqref{hs transformation} are identical, except now we must also perform a multi-dimensional Hubbard-Stratonovich transformation on the short-range interaction terms in the vectorized Liouvillian. The short-range Ising perturbation is diagonalized in the same basis as  Eq.~\eqref{basis}. The Hubbard-Stratonovich transformation reads as
\begin{equation}\label{nn hs transformation}
\begin{split}
   &\pm \frac{i}{2} \lambda \delta t(\bm{\sigma}^{(u/l)})^T  {\mathbf D}^{-1} \bm{\sigma}^{(u/l)}\\ &\to \mp \frac{i}{2  \lambda \delta t} (\bm{m}^{(u/l)})^T \bD \,\bm{m}^{(u/l)} \pm i \,(\bm{m}^{(u/l)})^T \bm{\sigma}^{(u/l)}\,,
\end{split}
\end{equation}
where $\bm{\sigma} = (\sigma_1,...,\sigma_N )$ represents the spins, while $\bm{m}^{(u/l)} = (m_1^{(u/l)},...,m_N^{(u/l)})$ denotes the scalar fields on the upper/lower leg of the ladder, respectively. The kernel ${\mathbf D}^{-1}$, representing the nearest-neighbour interaction, is given by
\begin{equation}
\bD^{-1} =
\begin{tikzpicture}[baseline=(current bounding box.center)]
\matrix[matrix of math nodes,nodes in empty cells,right delimiter=),left delimiter=(](m)
{
0 & 1 & & 1\\
 1 &  &  & \\
 &  &  &  1 \\
 1 &  & 1 &  0\\
};
\draw[loosely dotted, thick] (m-1-1)-- (m-4-4);
\draw[loosely dotted, thick] (m-1-2)-- (m-3-4);
\draw[loosely dotted, thick] (m-2-1)-- (m-4-3);
\end{tikzpicture},
\end{equation}
with 1 next to the diagonal (note the periodic boundary conditions) and 0 everywhere else.
This kernel is invertible for odd $N$ or even $N$ not divisible by 4; for simplicity, we take $N$ to be odd. After tracing out the spins, redefining the local fields ${\bm m}^{(u/l)} / (2 \lambda \delta t) \to {\bm m}^{(u/l)}$, rotating to the Keldysh basis ${\bm m}_{c/q} = ({\bm m}^{(u)} \pm {\bm m}^{(l)})/\sqrt{2}$, and taking the continuum limit, we find the exact Keldysh action including the short-range interaction:
\begin{equation}\label{Action}
\begin{split}
    \mathcal{S} = -2JN& \int_t m_c m_q - 2\lambda\int_t \bm{m}^T \widetilde{\bP}\, \bm{m}\\
    & - i \sum_i \ln \tr \left(\mathcal{T}e^{\int_t \mathbb{T} + \mathbb{T}'_i}\right)\,.
\end{split}
\end{equation}
Here, $\bm{m}$ denotes a column vector with $\bm{m}_c$ stacked on top of $\bm{m}_q$.
The kernel for the local fields is given by 
\begin{equation}
    \widetilde{\bP} =
   \begin{pmatrix}
    \bm{0} & \bD \\
    \bD & \bm{0}
    \end{pmatrix}.
\end{equation}
Furthermore, the short-range interaction leads to an additional matrix added to the matrix $\mathbb{T}$ in the exponential:
\begin{equation}
\mathbb{T}'_i = i 2\sqrt{2} \lambda \,\text{diag}(m_{i,q}, m_{i,c}, -m_{i,c}, -m_{i,q})\,.
\end{equation}
Ideally, we must integrate out the local fields to obtain an effective action in terms of only the original collective field $m$. In order to switch to a picture in terms of spin waves, we introduce the Fourier transform of the local fields as
\begin{equation}
    m_j = \frac{1}{N}\sum_{k=0}^{N-1} e^{i k j} m_k\,.
\end{equation}
The action too can be recast in the Fourier space. In this basis, the matrix $\mathbf D={\rm diag}\{D_k\}$ takes a diagonal form with the matrix elements (recalling that $N$ is odd)
\begin{equation}\label{momentum eigenvalues}
  D_k = \sum_{j=0}^{N-1} e^{-i k j} D_j =\frac{1}{2}\sec k\,,
\end{equation}
where $j = m-n$ with $m$ and $n$ the row and column labels of the matrix $\bD$, respectively; here, we have used the translational invariance due to periodic boundary condition.

Finally, we remark that $m_{k=0}$ too represents the collective field $m$ that is originally introduced through the Hubbard-Stratonovich transformation of the infinite-rang Ising interaction. Indeed, it can be shown by introducing source fields that $m$ and $m_{k=0}$ are redundant (see Sec.~\ref{field spin relationship}). 
Therefore, to simplify the subsequent treatment, we introduce the new fields
$\overline{m} = \sqrt{N}m + \frac{\lambda}{J\sqrt{N}}m_0$, $\widetilde{m} = \sqrt{N}m - \frac{1}{\sqrt{N}}m_0$, where $\overline{m}$ serves as the new order parameter, while $\widetilde{m}$ is entirely decoupled from all other fields and appears quadratically, and can be simply integrated out. Absorbing a factor of $\sqrt{\lambda/JN}$ into $m_k$, we find the total action 
\begin{equation}\label{new action}
\begin{split}
    \mathcal{S} = &\frac{-2J^2}{J + \lambda}\int_t \overline{m}_c(t)\overline{m}_q (t) - i \sum_j \ln \tr \left(\mathcal{T}e^{\int_t \mathbb{T} + \mathbb{T}'_j}\right)
    \\
    &- 2J\sum_{k \neq 0} 
    \begin{pmatrix}m_{-k,c},\,
      m_{-k,q}\end{pmatrix}
      \begin{pmatrix}0 & D_k\\ D_k & 0\end{pmatrix}
      \begin{pmatrix}m_{k,c}\\m_{k,q}\end{pmatrix}\,,
\end{split}
\end{equation}
where the matrices in the log-trace are given by 
\begin{equation}
    \mathbb{T} = \mathbb{T}_0 + i2\sqrt{2} \frac{J}{\sqrt{N}}\text{diag}(\overline{m}_{q}, \overline{m}_{c}, -\overline{m}_{c}, -\overline{m}_{q})\,,
\end{equation}
with $\mathbb{T}_0$ defined in Eq. \eqref{T matrix no m}, and
\begin{equation}
    \mathbb{T}'_j = i \sqrt{\frac{8J \lambda}{N}} \sum_{k\neq 0}^{N-1} e^{i k j}\,\text{diag}(m_{q,k}, m_{c,k}, -m_{c,k}, -m_{q,k})\,. 
\end{equation}
Notice that $m_{k=0}$ does not appear in the action, and the collective field is completely characterized through $\overline{m}_{c/q}(t)$.

\subsection{Quadratic Action}
We now follow a similar procedure as before and expand Eq.~\eqref{new action} to quadratic order in both $\overline{m}$ and $m_k$:
\begin{align}\label{quadratic action expansion}
     &\mathcal{S} = \frac{1}{2}\int_{t,t'} \begin{pmatrix}\overline{m}_{c},\,
      \overline{m}_{q}\end{pmatrix}_t
      \begin{pmatrix}0 & P^A\\ P^R & P^K\end{pmatrix}_{t-t'} \begin{pmatrix}\overline{m}_{c}\\\overline{m}_{q}\end{pmatrix}_{t'}
\\
     &+\frac{1}{2}\sum_{k\neq 0}\int_{t,t'} \begin{pmatrix}m_{-k,c},\,
      m_{-k,q}\end{pmatrix}_t 
      \begin{pmatrix}0 & P_k^A\\ P_k^R & P_k^K\end{pmatrix}_{t-t'} \begin{pmatrix}m_{k,c}\\m_{k,q}\end{pmatrix}_{t'}. \nonumber 
\end{align}
The quadratic action takes the Keldysh structure with the elements (recalling that $P^R(t) = P^A(-t)$)
\begin{equation}\label{New collective elements}
\begin{split}
P^{R}(t)&= \frac{-2J^2}{J+\lambda}\delta(t) + \Theta(t)8J^2 e^{-\frac{\Gamma}{2} |t|}\sin{(2\Delta t)}, \\
P^K(t) &= i8J^2 e^{-\frac{\Gamma}{2} |t|}\cos{(2\Delta t)},
\end{split}
\end{equation}
and
\begin{equation}\label{New momentum elements}
\begin{split}
P^{R}_k (t)&= -4JD_k\delta(t) + \Theta(t)8J\lambda e^{-\frac{\Gamma}{2} |t|}\sin{(2\Delta t)}, \\
P^K_k(t) &= i8J\lambda e^{-\frac{\Gamma}{2} |t|}\cos{(2\Delta t)}.
\end{split}
\end{equation}
One can immediately see that the collective field $\overline{m}$ is decoupled from spin waves $m_k$ at the level of Eq. \eqref{quadratic action expansion}. 
This is because any (bi)linear coupling between $\overline{m}$ and $m_k$ is forbidden by momentum conservation. To investigate the effect of spin waves, we need to go to higher-order terms that characterize the interaction between these fields. 
As we shall see, the nonlinear coupling will dramatically change the effect of spin waves on the collective mode: while linear coupling of the two fields will mimic a thermal bath (of spin waves) at finite temperature \cite{zwanzig_nonequilibrium_2001}, the nonlinear coupling will have no such effect. For another setting where nonlinear coupling changes the nature of dissipation, see Ref.~\cite{Maghrebi-flight-2016}.

Next, we take advantage of the perturbative nature of spin waves and calculate their contribution to the self energy whose low-frequency behavior determines how spin waves impact the dynamics of the order parameter $\overline{m}$. To this end, we first list the free Green's functions describing spin waves in the time domain:
\begin{equation}\label{spinwave response}
    G^R_{k}(t) = -\frac{1}{4 J D_k}\delta(t) - \frac{2\lambda \Delta}{J D_k^2 \Delta_k}\Theta(t) e^{-\Gamma t/2}\sin \left(\frac{\Delta_k t}{2} \right)\,,
\end{equation}
and
\begin{align}\label{spinwave correlation}
    &G^K_{k}(t) = \frac{-i \lambda e^{-\Gamma |t|/2}}{4J D_k^2 \Delta_k(\Gamma^2 + \Delta_k^2)}\times\\
    & \bigg[\Delta_k (2 \Gamma^2 + \Delta_k^2 + 16 \Delta^2)\cos \frac{\Delta_k t}{2} \nonumber - \Gamma(\Delta_k^2 - 16\Delta^2) \sin \frac{\Delta_k |t|}{2}\bigg]\,,\nonumber
\end{align}
where $\Delta_k = 4\sqrt{\Delta(\Delta - \lambda/D_k)}$. It is also useful to cast the Green's functions in frequency space: 
\begin{equation}
    G^R_{k}(\omega) = \frac{1}{P^R_k(\omega)} = \frac{-1}{4 J D_k} \frac{(\omega + \omega^+)(\omega + \omega^-)}{(\omega - \omega_a)(\omega - \omega_b)}\,,
\end{equation}
and
\begin{align}
    &G^K_{k}(\omega) = - P^K_{k}(\omega)|G^R_{k,0}(\omega)|^2\\
    &=  \frac{ -i \lambda \Gamma}{4 J D^2_k}\frac{(\omega + \omega^+)(\omega + \omega^{+*}) + (\omega + \omega^-)(\omega + \omega^{-*})}{(\omega - \omega_a)(\omega - \omega_b)(\omega - \omega_a^*)(\omega - \omega_b^*)}\,, \nonumber
\end{align}
with $\omega^{+/-} = i\Gamma/2 \pm 2\Delta$ and $\omega_{a/b} = (-i\Gamma \pm \Delta_k)/2$.
 
Finally, we identify the low-frequency effective temperature of spin waves:
\begin{equation}\label{T eff spin waves}
    T_{\text{eff},k}= \lim_{\omega \to 0} \frac{\omega}{2} \frac{G^K_k(\omega)}{G^R_k(\omega) - G^A_k(\omega)} = \frac{\Gamma^2 + 16\Delta^2}{32\Delta}\,.
\end{equation} 
Interestingly, this effective temperature is $k$-independent and is in fact equal to the effective temperature of the collective mode; cf. \cref{effective temperature}. This provides an interesting picture where the collective mode is in effective equilibrium with a thermal bath of spin waves. 

\begin{figure}[tp]
    \centering
    \includegraphics[scale=.3]{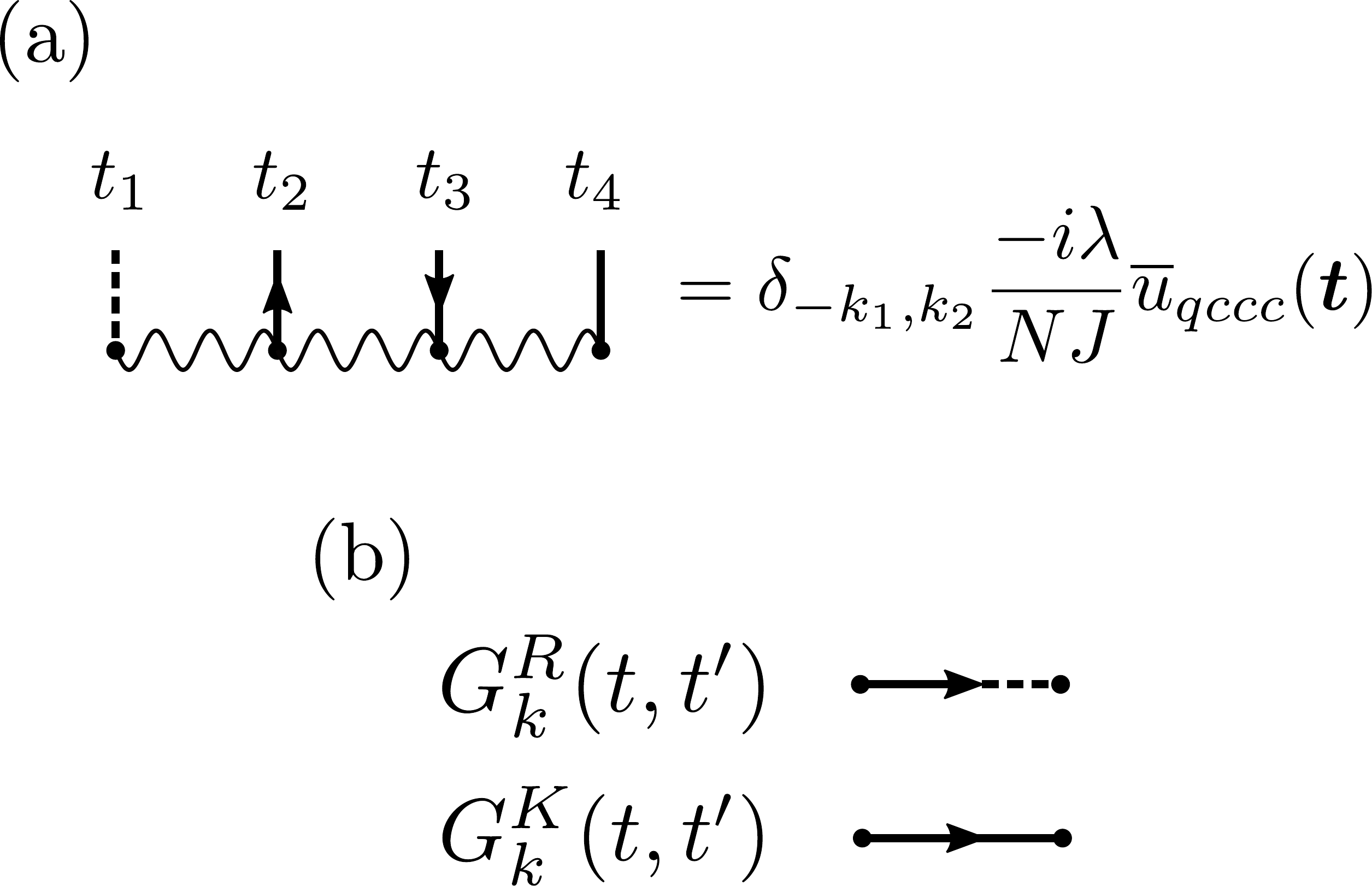}
    \caption{(a) A representative diagram involving spin waves, $m_{k,c/q}$. An additional prefactor of $\sqrt{\lambda/J}$ arises for each appearance of spin wave compared to that of the collective field; cf. Fig.~\ref{feynman rules}. The Kronecker delta enforces momentum conservation. The field's dependence on time and momenta has been suppressed for compactness. (b) Contracted arrowed legs represent spin-wave Green's functions.}
    \label{mk classical vertex}
\end{figure}

\subsection{Self-Energy}
In this section, we compute the correction to the self energy due to spin waves and investigate their effect on the phase diagram and the dynamics, particularly at the weakly dissipative point. Our starting point is the Keldysh form of the familiar Dyson equation \cite{kamenev_field_2011}, 
\begin{equation}
     {\mathbfcal G}^{-1} = \bG^{-1} - \bm{\Sigma}\,,
\end{equation}
where ${\mathbfcal G}$ is the exact Green's function for the collective field, and $\bG^{-1}$ is given by the kernel in the first term in Eq. \eqref{quadratic action expansion}. The self-energy $\bm{\Sigma}$ has the typical Keldysh structure and takes the form
\begin{equation}
\bm{\Sigma} = \begin{pmatrix}
0 & \Sigma^A \\
\Sigma^R & \Sigma^K
\end{pmatrix}\,.
\end{equation}
The low-frequency expansion of the retarded and Keldysh elements of the self energy will renormalize the parameters describing the dynamics of the collective field as
$\Sigma^R(\omega) \sim -\delta r + i\delta \gamma \omega$ and $\Sigma^K(\omega = 0) = \delta D$.
At any generic critical point, the spin waves will simply provide a correction $\delta \gamma$ and $\delta D$ to the otherwise finite values of dissipation and fluctuations, respectively. However, the weakly-dissipative critical point where $\Gamma \to 0$ is particularly susceptible to the coupling to the spin waves. This is because spin waves provide an effective thermal bath for the collective mode [see \cref{T eff spin waves}], which could very well generate dissipation (even when $\Gamma \to 0$).

To calculate the self energy, we utilize the diagrammatic representation developed in Sec.~\ref{Diagrammatics};
we also include lines with an arrow to denote spin waves with a nonzero momentum in addition to those without an arrow which refer to the collective field. 
The connected diagrams inside the logarithm in Eq. \eqref{log diagrams} are modified accordingly: we include an additional prefactor of $\sqrt{\lambda /J}$ for each appearance of the $m_k$ fields, and keep track of momentum indices. The diagrams resultant from expanding the logarithm in Eq. \eqref{interaction vertex} should be summed over all momenta, with an overall Kronecker delta enforcing momentum conservation. An example of the classical vertex for the spin waves can be found in Fig. \ref{mk classical vertex}.\par
\begin{figure}[tp]
    \centering
    \includegraphics[scale=.24]{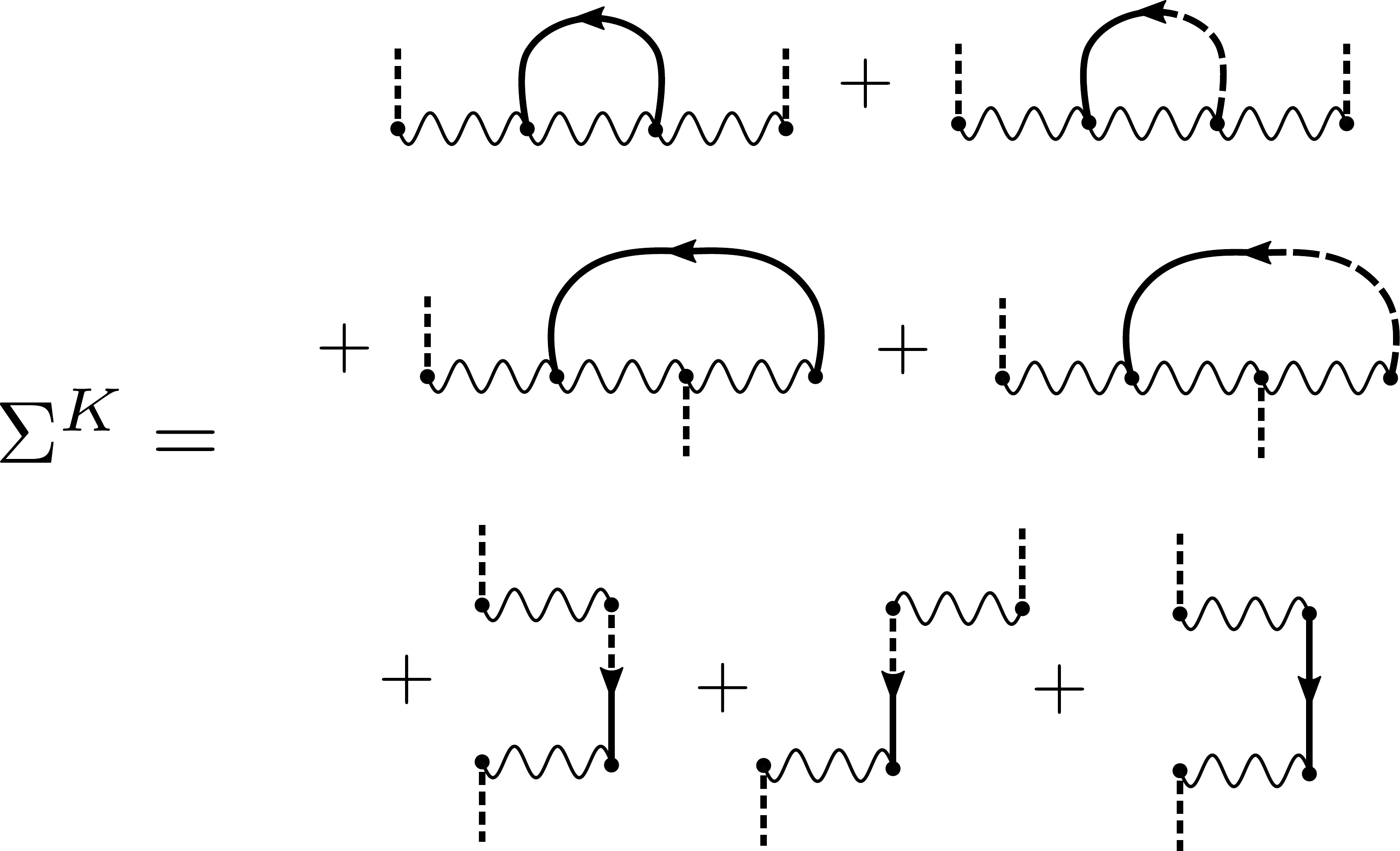}
    \caption{Diagrams contributing to the Keldysh component of the self energy, $\Sigma^K$, to the order $\sim \lambda^2$.}
    \label{keldysh self energy diagrams}
\end{figure}
\begin{figure}[tp]
    \centering
    \includegraphics[scale=.24]{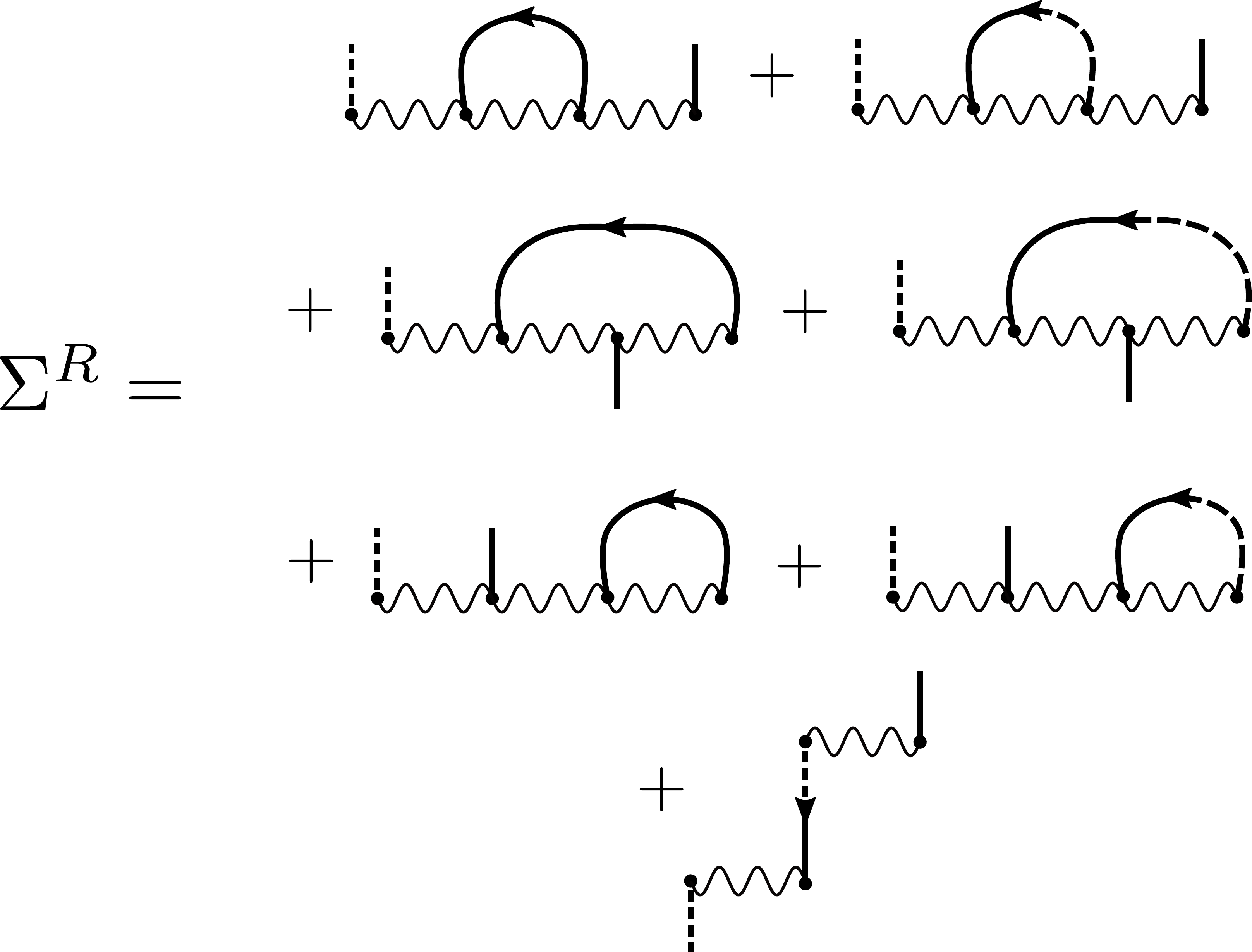}
    \caption{Diagrams contributing to the retarded component of the self energy, $\Sigma^R$, to the order $\sim \lambda^2$.}
    \label{response self energy diagrams}
\end{figure}
The lowest nontrivial correction to the self energy arises at the order $\mathcal{O}(\lambda^2)$ due to a combination of momentum conservation and the fact that $\bD^{-1}$ is traceless. The one-loop diagrams contributing to $\Sigma^K$ to the order $\mathcal{O}(\lambda^2)$ are depicted in Fig.~\ref{keldysh self energy diagrams}, while those contributing to $\Sigma^R$ are given in Fig.~\ref{response self energy diagrams}. All other diagrams are either higher order in $\lambda$ or are suppressed as $\mathcal{O}(1/N)$. Note that only diagrams with two external quantum legs contribute to $\Sigma^K$, while those with one external quantum and another classical leg contribute to $\Sigma^R$, in harmony with the Keldysh structure of the action at the quadratic level. As an example calculation, the self-energy contribution to $\Sigma^K$ due to the $\overline{u}_{qccq}$ one-loop diagram is given by 
\begin{equation}\label{freq example}
\begin{split}
    \Sigma^K_{(qccq)}(\omega) = &\frac{-\lambda}{J N}\sum_{k \neq 0} \int_{\omega'}\big[\overline{u}_{qccq}(-\omega, \omega', -\omega', \omega)\\
    &+ \overline{u}_{qccq}(\omega, \omega', -\omega', -\omega)\big]G^K_{k,0}(\omega')\,.
\end{split}
\end{equation}
The overall minus follows from a factor of $-i$ from the perturbative expansion of the path integral multiplied by another factor of $-i$ from the connected four-legged diagrams in Eq. \eqref{interaction vertex}. 
The above expression must be symmetrized with respect to the external frequency due to the same symmetry of the Keldysh component $P^K$. The interaction coefficient in frequency space is then given by
\begin{equation}
    \overline{u}_{qccq}(\bm{\omega}) = \frac{i 256\Delta^2 J^4}{\omega_1 + \omega_2 - i \Gamma}f(\omega_1, \omega_4)\,,
\end{equation}
where $f(x,y) = 1/[(x - \omega^+)(x - \omega^-)(y - \omega^{+*})(y - \omega^{-*})]$.
Setting $\omega_1 = \omega_4 = 0$ and expanding to lowest non-zero order in $\lambda$, we find the correction
\begin{equation}
    \Sigma^K_{(qccq)}(0) = \frac{-i49152 J^2 \Delta^2 \Gamma \lambda^2}{(\Gamma^2 + 16\Delta^2)^2 (9 \Gamma^2 + 16\Delta^2)}\,,
\end{equation}
where we have used the fact that $\sum_{k\neq0} 1/D_k^2 = 2N -1/2$ and neglected terms of $\mathcal{O}(1/N)$. 
Repeating this calculation for all of the diagrams in Fig. \ref{keldysh self energy diagrams}, we find that the Keldysh component of the self-energy at low frequencies is given by
\begin{equation}\label{K self energy}
    \Sigma^K(0) = \delta D = \frac{i16384 J^2 \Delta^2 \Gamma}{(\Gamma^2 + 16\Delta^2)^3}\lambda^2\,.
\end{equation}
Similarly, the retarded component of the self energy is determined by considering the diagrams in Fig.~\ref{response self energy diagrams}; we find
\begin{equation}\label{R self energy}
\begin{split}
    \Sigma^R(\omega) &\sim + \delta r  + \delta\gamma i \omega \\
    &= \frac{1536 \lambda^2J^2 \Delta }{(\Gamma^2 + 16\Delta^2)^2} + \frac{ 8192 \lambda^2J^2 \Delta \Gamma}{(\Gamma^2 + 16\Delta^2)^3}i \omega  \,.
\end{split}
\end{equation}
The above equations produce the first nontrivial correction to the self energy due to the coupling to spin waves. At a generic critical point, these corrections remain finite and simply act as shifts to the noise and dissipation, as expected. Interestingly, we find from \cref{K self energy,R self energy} that $\delta \gamma$ and $\delta D$ vanish in the limit $\Gamma \to 0$.
In other words, while spin waves renormalize the low-frequency parameters, they do not qualitatively change the nature of the dynamics even in the limit $\Gamma\to0$. We thus conclude that the underdamped critical dynamics at the weakly-dissipative point is robust against generic perturbations exemplified by short-range interactions in Eq.~\eqref{Hamiltonian}, at least to the lowest nontrivial order ($\sim \lambda^2$).

Finally, we can inspect the effect of spin waves on the phase boundary of the model. These effects can be seen by setting the renormalized mass $r_{\rm ren} \equiv r + \delta r$ to zero, where $r$ is the bare mass defined in Eq.~\eqref{langevin parameters}:
\begin{equation}\label{perturbed boundary}
    [\Gamma^2 + 16\Delta(\Delta-2\overline{J})](\overline{J}-\lambda)^2  + \frac{768\overline{J}^3 \Delta }{(\Gamma^2+16\Delta^2)}\lambda^2 = 0\,,
\end{equation}
where we have defined $\overline{J} = J + \lambda$ and dropped terms of $O(\lambda^3)$ or higher. We have redefined $J$ to include the contribution of the short-range interaction to the collective mode, and to solely separate out the effect of spin waves.
\begin{figure}[tp]
    \centering
    \includegraphics[scale=.45]{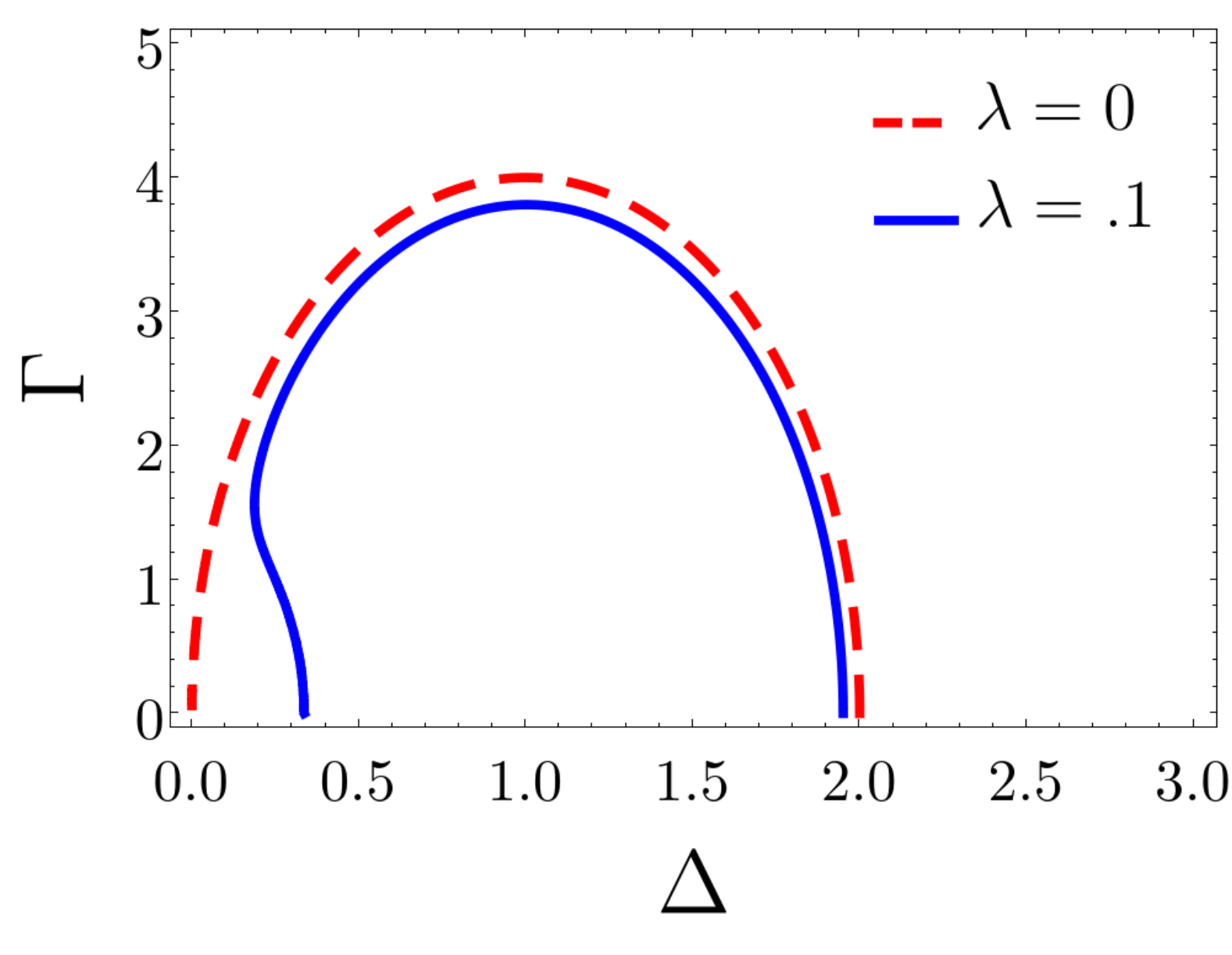}
    \caption{Phase boundary in the presence and absence of short-range interactions; here, $J=1$ and an exaggerated value of $\lambda=.2$ is chosen for better visualization. The feature around $\Delta \sim \lambda$ could be an artifact of our perturbation scheme, which requires $\lambda \ll \Delta$.}
    \label{boundary shift}
\end{figure}%
In Fig.~\ref{boundary shift}, one can see that ordered region shrinks due to the coupling to spin waves. This is expected as spin waves introduce more fluctuations and thus disfavor ordering. Finally, we remark that Eq.~\eqref{perturbed boundary} should not be trusted near $\Delta \to 0$ since it was implicitly assumed that $\Delta \gg \lambda$ in our perturbative calculation (to expand $\Delta_k$ in powers of $\lambda$); however, it is possible that short-range interactions dramatically alter the phase boundary near the origin, in a fashion that could be related to the predicted first-order phase transition in the DDIM in $d$ dimensions \cite{maghrebi_nonequilibrium_2016, overbeck_multicritical_2017}.

\section{Conclusion}\label{conclusion}
In this work, we have performed a thorough field-theoretical and numerical analysis of the driven-dissipative Ising model with infinite-range interactions using a non-equilibrium variant of the quantum-to-classical mapping. This mapping has allowed a tractable field-theoretical framework even in the presence of local spontaneous emission. While other techniques for dealing with local dissipation exist \cite{dalla_torre_dicke_2016, shchadilova_fermionic_2018}, our formalism is a powerful alternative which naturally reflects the underlying Ising symmetry of the driven-dissipative spin model. With this technology, we have shown that the DDIM exhibits qualitatively different critical dynamics in the weakly-dissipative limit.
While one might naively expect that weak dissipation could give rise to quantum criticality, we have seen that the critical exponents characterizing the phase transition remain classical; see however Ref.~\cite{rota_quantum_2019} for quantum behavior in the weakly dissipative limit of a short-range model. Using the diagrammatic language, we have also shown that the underdamped critical behavior persists in the presence of short-range perturbations.

Our results can be observed in experimental platforms where the open Dicke model is realized \cite{baumann_dicke_2010, baden_realization_2014}. Alternatively, this model can be realized directly using ion-trap platforms with sufficiently long-range interactions between atoms \cite{kim_entanglement_2009}. While weak dissipation may be hard to access experimentally, we have shown that the phenomena discussed in this paper emerge at a moderate value of dissipation rate for experimentally accessible system sizes.

In a sequel paper, we will extend this work to the full open Dicke model with local spontaneous emission, and perform a comprehensive analysis of the critical properties and effective temperature using similar techniques to what we have developed here. In another future paper, we will examine the entanglement properties of the driven-dissipative Ising model to fully characterize its many-body characteristics. Calculating entanglement in non-equilibrium many-body systems is highly non-trivial, but it is possible analytically in the context of the model presented here. A possible future direction is using the quantum-to-classical mapping and the diagrammatics developed in this work to study the short-ranged DDIM, a model that has been previously treated using approximate schemes or small-size numerics \cite{lee_unconventional_2013, maghrebi_nonequilibrium_2016,overbeck_multicritical_2017,Jin2018,Kilda2019}.

\appendix
\section{Large Detuning Limit of the Open Dicke Model}\label{ODM to DDIM}
Here, we show that the model in Eq.~\eqref{DDIM} follows from the open Dicke model in the limit of the large cavity detuning. Beginning with Eq. \eqref{ODM}, the full quantum master equation takes the form
\begin{equation}
\begin{split}
    \frac{d \rho}{d t} = &-i[H_{\text{Dicke}}, \rho] + \kappa\left(a \rho a^\dagger - \frac{1}{2}\{a^\dagger a, \rho\}\right)\\
    &+ \Gamma \sum_i \left(\sigma^-_i \rho \sigma^+_i - \frac{1}{2}\{\sigma^+_i \sigma^-_i, \rho\} \right).
\end{split}
\end{equation}
Following the same steps as outlined in Sec. \ref{mapping} (combined with a coherent-state representation for the cavity field), we obtain an action that consists of cavity, atomic and interaction terms:
\begin{equation}
    \mathcal{S}_D = \mathcal{S}_{\text{cav}} + \mathcal{S}_{\text{int}} + \mathcal{S}_{\text{spin}}\,.
\end{equation}
The cavity term in the action is given by
\begin{equation}
\mathcal{S}_{\text{cav}} = \int_\omega \begin{pmatrix} a_c \\ a_q \end{pmatrix}^\dagger \begin{pmatrix}0 & \omega - \omega_0 - i \frac{\kappa}{2}\\
    \omega - \omega_0 + i\frac{\kappa}{2} & i \kappa \end{pmatrix} \begin{pmatrix}a_c \\ a_q \end{pmatrix}\,.
\end{equation}
Defining $a = (x - i p)/2$, we can integrate out the imaginary component of the cavity field, $p$, exactly as $S_{\text{int}}$ does not depend on $p$. Tracing out the spins (see Sec. \ref{mapping}), we then find an \textit{exact} expression for the action 
\begin{equation}
     \mathcal{S}_D = \int_\omega \bx^{T}(-\omega)\bD(\omega)\bx(\omega) -i N \ln\tr{\left(\mathcal{T}e^{\int_t \mathbb{T}_D(x_{c/q}(t)}\right)}\,,
\end{equation}
where $\bx(\omega) = (x_c(\omega), x_q(\omega))^T$ and the kernel $\bD(\omega)$ is given by
\begin{equation}
\begin{split}
\renewcommand*{\arraystretch}{1.8}
    &\bD(\omega) \equiv \begin{pmatrix}0 & D^A(\omega) \\ D^R(\omega) & D^K(\omega) \end{pmatrix}\\
    &= \begin{pmatrix}0 & \frac{1}{4}\left(-\frac{(\kappa + 2 i \omega)^2}{4\omega_0}-\omega_0 \right)\\
    \frac{1}{4}\left(-\frac{(\kappa - 2 i \omega)^2}{4\omega_0}-\omega_0 \right) & \frac{i \kappa (\kappa^2 + 4(\omega^2 +\omega_0^2))}{16\omega_0^2}\end{pmatrix} \,.
\end{split}
\end{equation}
The matrix $\mathbb{T}_D$ is rather similar to that in Eq. \eqref{Keldysh Action}:

\begin{widetext}
\begin{equation}
\renewcommand*{\arraystretch}{1.5}
     \mathbb{T}_D(x_c(t), x_q(t)) = 
    \begin{pmatrix}
     -\frac{\Gamma}{4}-i\frac{2\sqrt{2}g}{\sqrt{N}} x_q(t) & i\Delta & -i\Delta & \frac{\Gamma}{4} \\
    i\Delta - \frac{\Gamma}{2} & -\frac{3\Gamma}{4} - i \frac{2\sqrt{2}g}{\sqrt{N}} x_c(t)  & -\frac{\Gamma}{4} & -i\Delta -\frac{\Gamma}{2} \\
    -i\Delta - \frac{\Gamma}{2} &-\frac{\Gamma}{4} &- \frac{3\Gamma}{4} + i\frac{2\sqrt{2}g}{\sqrt{N}}x_c(t) & i\Delta - \frac{\Gamma}{2} \\
    \frac{\Gamma}{4} & -i\Delta & i\Delta & -\frac{\Gamma}{4} + i\frac{2\sqrt{2}g}{\sqrt{N}}x_q(t)
    \end{pmatrix}\,.
\end{equation}
\end{widetext}
We then make the transformation $m_c \equiv D^R_0 x_c/\sqrt{N}g$ and $m_q \equiv D^R_0 x_q/\sqrt{N}g$ with $\bD_0\equiv \bD(\omega = 0)$, and further define $J \equiv -g^2/D^R_0 =  16g^2 \omega_0/(\kappa^2 + 4\omega_0^2)$ and $\Gamma_x \equiv J\kappa/\omega_0$.
The action is then cast as
\begin{equation}
    \mathcal{S}_D = \int_\omega \bmm^{T}(-\omega)\bP(\omega)\bmm(\omega) - iN \ln \tr \left(\mathcal{T}e^{\int_t \mathbb{T}(m_{c/q}(t))}\right)\,,
\end{equation}
where $\bmm(\omega) = (m_c(\omega), m_q(\omega))^T$, the kernel $\bP$ is given by
\begin{equation}\label{Eq. P Dicke}
\renewcommand*{\arraystretch}{1.5}
    \bP(\omega) = N \begin{pmatrix}
    0 & -J(1 + \frac{4i\kappa \omega - 4\omega^2}{\kappa^2 + \omega_0^2}) \\
    -J(1 - \frac{4i\kappa \omega + 4\omega^2}{\kappa^2 + \omega_0^2}) & i\Gamma_x(1+ \frac{4\omega^2}{\kappa^2 + 4 \omega_0^2})
    \end{pmatrix}\,,
\end{equation}
and the matrix $\mathbb{T}(m_c(t), m_q(t))$ is identical to that in Eq.~\eqref{Keldysh Action}.

Now we consider the limit of large $\omega_0$ and $\kappa$, in which case we can ignore those terms in Eq.~\eqref{Eq. P Dicke} that are suppressed by a factor of $1/(\kappa^2+\omega_0^2)$. This eliminates the frequency-dependent terms and yields the kernel  
\begin{equation}
\renewcommand*{\arraystretch}{1.5}
    \bP(\omega) \approx N \begin{pmatrix}
    0 & -J \\
    -J & i\Gamma_x
    \end{pmatrix}\,.
\end{equation}
Using the quantum-to-classical mapping, one can show that the diagonal term ($\sim i \Gamma_x$) can be identified with dephasing in the form of the Lindblad operator $L_x = \sqrt{\Gamma_x/N}S_x$. Indeed, this agrees with the large-detuning limit discussed in Ref.~\cite{damanet_atom-only_2019}.
Our model is different, however, due to the atomic spontaneous emission, which allows for a nontrivial non-equilibrium steady state. To obtain the DDIM, we can consider the detuning $\omega_0$ to be the largest frequency frequency scale even compared to $\kappa$. In this limit, we can neglect the dephasing term, since $\Gamma_x =  J \kappa/\omega_0 \ll J$, and recover the driven-dissipative Ising model introduced in Eq.~\eqref{Action}.

\section{Interaction coefficients}\label{interaction coefficients appendix}
There are many relevant interaction coefficients necessary to compute the diagrams in Sec. \ref{Diagrammatics}. They are defined by Eq. \eqref{interaction coefficients} in the time domain and in Eq. \eqref{u ft} in the frequency domain. Here, we list the relevant interaction coefficients for the four-legged one-loop diagrams in \cref{keldysh self energy diagrams,response self energy diagrams} in the frequency domain:
\begin{equation}
     \overline{u}_{qccc}(\bm{\omega}) = \frac{-128\Delta J^4( \Gamma/2 - i\omega_4)}{\omega_1 + \omega_2 -i \Gamma}f(\omega_1, \omega_4)\,,
\end{equation}
\begin{equation}
    \overline{u}_{qcqq}(\bm{\omega}) = \frac{i 128\Delta J^4 ( \Gamma/2 - i\omega_4)}{(\omega_1 + \omega_2 - i \epsilon)(\omega_1 + \omega_2 - i \Gamma)}f(\omega_1, \omega_4) \,,
\end{equation}
\begin{equation}
    \overline{u}_{qcqc}(\bm{\omega}) =  \frac{ 256\Delta^2 J^4 \Gamma}{(\omega_1 + \omega_2 - i \epsilon)(\omega_1 + \omega_2 - i \Gamma)} f(\omega_1, \omega_4)\,,
\end{equation}
\begin{equation}
     \overline{u}_{qccq}(\bm{\omega}) = \frac{ i 256\Delta^2\lambda J^3}{\omega_1 + \omega_2 - i \Gamma}f(\omega_1, \omega_4)\,,
\end{equation}
\begin{equation}
\begin{split}
    \overline{u}&_{qc}(\omega_1, \omega_2)\overline{u}_{qc}(\omega_3, \omega_4)\\
    &= - 256 \Delta^2 J^4 \times 2\pi \delta(\omega_1 + \omega_2)f(\omega_1, \omega_4)\,.
\end{split}
\end{equation}
\begin{align}
     \overline{u}&_{qq}(\omega_1, \omega_2)\overline{u}_{qc}(\omega_3, \omega_4)\\
    &= -i128\Delta J^4 (i\omega_1 + \Gamma/2) \times 2\pi \delta(\omega_1 + \omega_2)f(\omega_1, \omega_4)\,. \nonumber
\end{align}
We also list here the interaction coefficient for the six-legged classical vertex used to calculate the damping parameter in the ordered phase in Sec. \ref{thermalization},
\begin{equation}
\begin{split}
    \overline{u}&_{qcccccc}(\bm{\omega}) \\
    &= \frac{-256 J^6 \Delta (\Gamma + 2i (\omega_1 + \omega_2 + \omega_3)(\Gamma - 2i\omega_6)}{(\omega_1 + \omega_2 + \omega_3 - \omega^+)(\omega_1 + \omega_2 + \omega_3 -\omega^-)}\\
    &\,\,\,\,\times\frac{f(\omega_1, \omega_6)}{(\omega_1+\omega_2 -i \Gamma) (\omega_5 + \omega_6 + i \Gamma)}\,,
\end{split}
\end{equation}
where $\bm{\omega} = (\omega_1, ..., \omega_n)$ (for an $n$-legged diagram), $f(x,y) = 1/[(x - \omega^+)(x - \omega^-)(y - \omega^{+*})(y - \omega^{-*})]$, and $\omega^{+/-} = i\Gamma/2 \pm 2\Delta$. For self-energy calculations, it is useful to recall that $\sum_{k\neq 0} 1/D_k = -1/2$ and $\sum_{k \neq 0} 1/D_k^2 = 2N - 1/2$, where $D_k$ is defined in Eq. \eqref{momentum eigenvalues}.

\section{Numerical Methods}\label{numerics}
Quantum many-body systems are difficult to simulate numerically due to exponential growth of the Hilbert space with the system size. This growth is even worse when dissipation is considered, making an exact simulation beyond a few spins almost impossible. However, Eq. \eqref{DDIM} exhibits a permutation symmetry which can be taken advantage of to reduce the size of the relevant Hilbert space to $\mathcal{O}(N^3)$. This symmetry breaks the Liouvillian matrix $\mathbb{L}$ into a block-diagonal structure, where each block corresponds to a different symmetry sector. The non-equilibrium steady state resides in the fully symmetric sector, therefore we introduce a permutation symmetric basis as \cite{gong_steady-state_2016,kirton_suppressing_2017}
\begin{widetext}
\begin{align}\label{symmetric basis}
\begin{split}
     \rho_{N_x, N_y, N_z} = \frac{1}{\mathcal{N}}\sum_\mathscr{P}&{\rm P}_\mathscr{P}(\sigma^{x}_1\otimes...\otimes\sigma^{x}_{N_x} \otimes\sigma^{y}_{N_x+1}\otimes...\otimes\sigma^{y}_{N_x+N_y}\\
     &\otimes\sigma^{z}_{N_x+N_y+1}\otimes...\otimes\sigma^{z}_{N_x+N_y+N_z} \otimes I_{N_x+N_y+N_z+1}\otimes...\otimes I_{N})\,,
\end{split}
 \end{align}
\end{widetext}
with the normalization factor $\mathcal{N} = \sqrt{N!N_x!N_y!N_z!N_I!}$. The sum is over all permutations $\mathscr{P}$ of the indices, where the operator $P_\mathscr{P}$ permutes the indices according to $\mathscr{P}$. These basis elements are normalized as $\tr(\rho_\mu \rho_{\nu})/2^N = \delta_{\mu, \nu}$ where $\mu = (N_x, N_y, N_z)$, and the Liouvillian matrix elements are given by
\begin{equation}\label{liouvillian matrix}
    \mathbb{L}_{\mu, \nu} = \frac{1}{2^N}\tr\left(\rho_\mu \mathcal{L}[\rho_\nu]\right)\,.
\end{equation}
In this basis, the dimensionality grows polynomially with the system size as $N(N+1)(N+2)/6 \sim \mathcal{O}(N^3)$ in contrast with the exponential growth in a generic many-body system. This scaling can also be contrasted with the $\mathcal{O}(N^4)$ growth of the usual Dicke (angular-momentum) basis. Because the Liouvillian is permutation symmetric, action of $\mathcal{L}$ on this basis will keep us in the fully symmetric subspace. To efficiently construct the matrix $\mathcal{L}_{\mu \nu}$ from the permutation symmetric basis given by Eq. \eqref{symmetric basis}, we should identify how the basis itself is transformed by $\mathcal{L}$ defined in Eq. \eqref{DDIM}.
The action of the Liouvillian on a state can be determined analytically by inspecting how the total-spin operators act on one of our basis elements:
\begin{align}
\begin{split}
    S_x\rho_{N_x,N_y,N_z} =& \sqrt{N_x(N_I + 1)}\,\rho_{N_x-1, N_y, N_z}\\
    &+ i\sqrt{N_y (N_z + 1)}\,\rho_{N_x, N_y-1, N_z+1}\\
    &- i\sqrt{(N_y+1)N_z}\,\rho_{N_x, N_y+1, N_z -1}\\
    &+ \sqrt{(N_x+1)N_I}\,\rho_{N_x+1, N_y, N_z}\,, 
\end{split}
\end{align}
\begin{align}
\begin{split}
    S_y\rho_{N_x,N_y,N_z} =& \sqrt{N_y(N_I + 1)}\,\rho_{N_x, N_y-1, N_z}\\
    &+ i\sqrt{N_z (N_x + 1)}\,\rho_{N_x+1, N_y, N_z-1}\\
    &- i\sqrt{(N_z+1)N_x}\,\rho_{N_x-1, N_y, N_z+1}\\
    &+ \sqrt{(N_y+1)N_I}\,\rho_{N_x, N_y+1, N_z}\,, 
\end{split}
\end{align}
\begin{align}
\begin{split}
    S_z\rho_{N_x,N_y,N_z} =& \sqrt{N_z(N_I + 1)}\,\rho_{N_x, N_y, N_z-1} \\
    &+ i\sqrt{N_x (N_y + 1)}\,\rho_{N_x-1, N_y+1, N_z}\\
    &- i\sqrt{(N_x+1)N_y}\,\rho_{N_x+1, N_y-1, N_z} \\
    &+ \sqrt{(N_z+1)N_I}\,\rho_{N_x, N_y, N_z+1}\,, 
\end{split}
\end{align}
where $N_I = N - N_x - N_y - N_z$, and the action from the right can be found by taking the adjoint of the RHS. The only other non-trivial term is the dissipative term $\sum_i \sigma^-_i \rho \sigma^+_i$, whose action on the basis elements is given by
\begin{align}
\begin{split}
    \sum_i\sigma^{-}_{i}&\rho_{N_x, N_y, N_z}\sigma^{+}_{i}\\
    &= \frac{1}{2}\big[(N_I - N_z)\rho_{N_x, N_y, N_z}\\
    &+ \sqrt{N_z(N_I + 1)}\rho_{N_x, N_y, N_z-1} \\
    &- \sqrt{N_I(N_z+1)}\rho_{N_x, N_y, N_z+1}\big]\,.
\end{split}
\end{align}
Using the above relations, we find the action of the Liouvillian on our basis as

\begin{widetext}
\begin{align}
\begin{split}
    \mathcal{L}[\rho_{N_x,N_y,N_z}] =& \frac{4J}{N}\Big(\sqrt{(N_x+1)(N_y+1)N_z N_I}\,\rho_{N_x+1,N_y+1,N_z-1} + \sqrt{N_x(N_y+1)N_z(N_I+1)}\rho_{N_x-1,N_y+1,N_z-1}\\
    &- \sqrt{N_x N_y (N_z+1)(N_I+1)}\,\rho_{N_x-1,N_y-1,N_z+1} -\sqrt{(N_x+1)N_y(N_z+1)N_I}\,\rho_{N_x+1,N_y-1,N_z+1}\Big) \\
    &+ 2\Delta \Big(\sqrt{N_x(N_y+1)}\, \rho_{N_x-1, N_y+1,N_z} - \sqrt{N_y(N_x+1)}\,\rho_{N_x+1, N_y-1, N_z}\Big)\\
    &+\frac{\Gamma}{2}\Big((N_I-N_z-N)\rho_{N_x,N_y,N_z} - 2\sqrt{(N_z+1)N_I}\rho_{N_x,N_y,N_z+1}\Big)\,.
\end{split}
\end{align} 
\end{widetext}
From here, it is possible to construct the Liouvillian matrix as defined in Eq. \eqref{liouvillian matrix}.\par 
Equipped with Eq. \eqref{symmetric basis}, we can efficiently construct the Liouvillian matrix, and the non-equilibrium steady state can be then obtained through the shifted-inverse-power method \cite{nation_steady-state_2015}. However, for larger system sizes ($N \gtrapprox 90$)  finding the steady state by direct LU decomposition becomes inefficient. At that point, it is more efficient to use linear solvers such as BICGSTAB to approximate the steady state.\par
To characterize the dynamics, we investigate the correlation function $C(t) = \langle \{ S_x(t), S_x(0) \} \rangle/N$ and response function $\chi(t) = -i\langle [S_x(t), S_x(0)]\rangle/N$. The two-time expectation values can be calculated as \cite{gardiner_quantum_2004}
\begin{align}
    \langle \{ S_x(t), S_x(0)\} \rangle &= \tr\left(S_x e^{t\mathcal{L}}[S_x \rho_{ss}] + S_x e^{t \mathcal{L}}[\rho_{ss} S_x]\right)\nonumber \\
    &= 2\text{Re}\tr\left(S_x e^{t\mathcal{L}}[S_x \rho_{ss}]\right)\,, \\
     \frac{1}{i}\langle [ S_x(t) S_x(0) ]\rangle  &= \frac{1}{i}\tr\left(S_x e^{t\mathcal{L}}[S_x \rho_{ss}]  - S_x e^{t \mathcal{L}}[\rho_{ss} S_x]\right)\nonumber \\
    &= 2\text{Im}\tr\left(S_x e^{t\mathcal{L}}[S_x \rho_{ss}]\right)\,,
\end{align}
with $\rho_{ss}$ being the steady state density matrix. We can instead represent the above equations in a vectorized form using our permutation symmetric basis:
\begin{equation}\label{numerical correlation}
   C(t) =  \frac{2}{N}\text{Re}\tr\left(S_x e^{t\mathcal{L}}[S_x \rho_{ss}]\right) = \frac{2}{N}\text{Re}\frac{\sbra{S_x}e^{t\mathbb{L}}\sket{S_x \rho_{ss}} }{ \sbraket{I}{\rho_{ss}} }\,,
\end{equation}
\begin{equation}\label{numerical response}
    \chi(t) =  \frac{2}{N}\text{Im}\tr\left(S_x e^{t\mathcal{L}}[S_x \rho_{ss}]\right) = \frac{2}{N}\text{Im}\frac{\sbra{S_x}e^{t\mathbb{L}}\sket{S_x \rho_{ss}} }{ \sbraket{I}{\rho_{ss}} }\,,
\end{equation}
where we have defined the vectorized state
\begin{equation}
    \sket{\rho(t)} = \sum_\mu c_{\mu}(t) \sket{\rho_\mu}\,.
\end{equation}
The denominator in Eq. \eqref{numerical correlation} is due to the normalization of the steady state (this is equivalent to dividing the state by $c_{0,0,0}$). In the case of static correlations, one can see that the auto-correlation function takes the simple form
\begin{equation}
    C(0) = \frac{2}{Nc_{0,0,0}}\left(\sqrt{2N(N-1)}c_{2,0,0} + Nc_{0,0,0} \right)\,.
\end{equation}
Using these techniques, we are able to numerically investigate dynamical correlations with system sizes up to $N = 200$.

\section{Equilibrium Quantum Ising Model}\label{TFIM appendix}
\begin{figure}[tp]
    \centering
    \includegraphics[scale=.23]{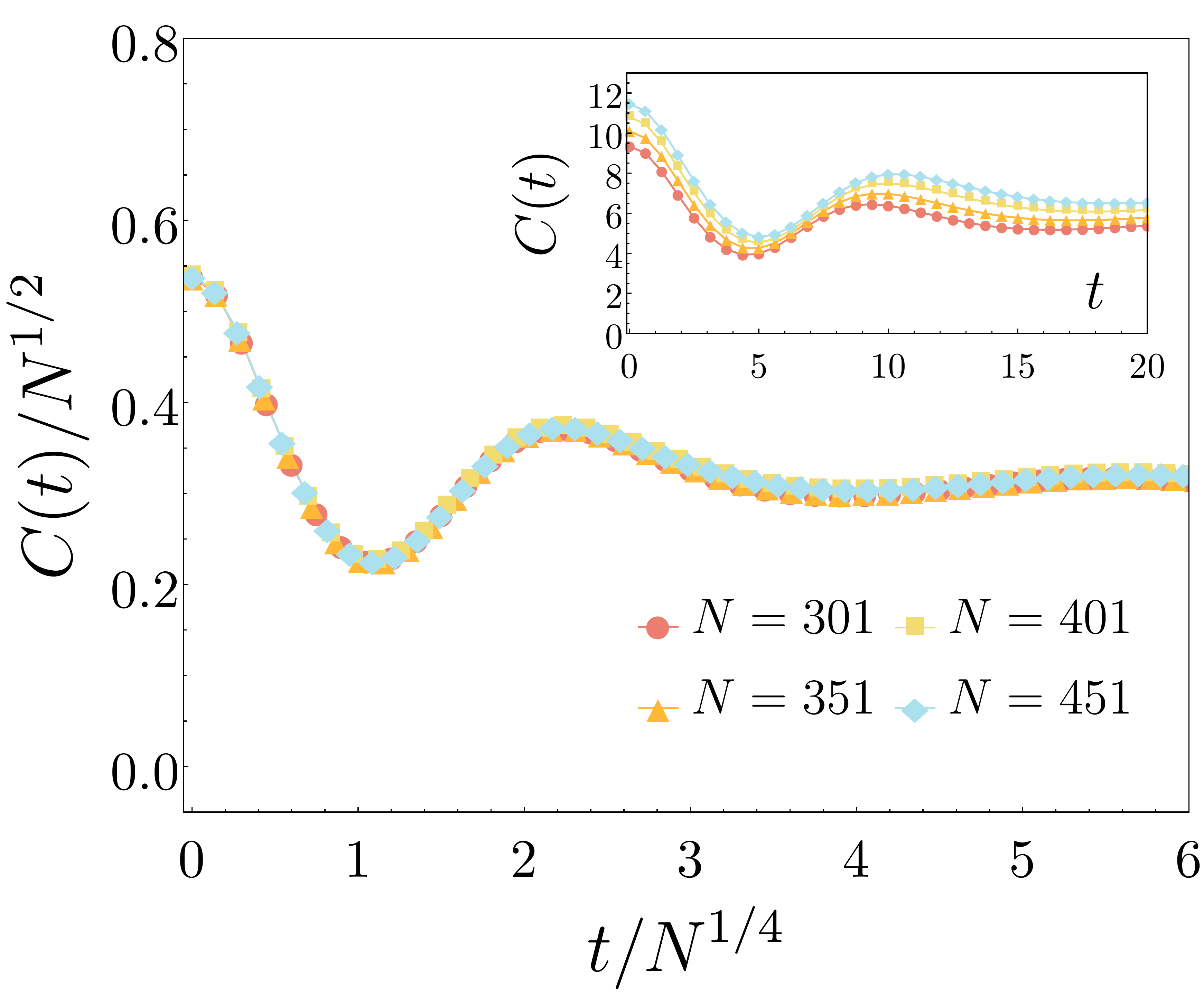}
    \caption{Finite-size scaling behavior of the infinite-range Ising model at a thermal critical point ($J = 1, \Delta = 1, T=1.82048$). At this critical point, fluctuations scale as $N^{1/2}$, while the critical dynamics is underdamped and is governed by a characteristic time scale $t \sim N^{1/4}$. These exponents are identical to those of the driven-dissipative Ising model in the weakly dissipative regime (see Fig. 3 of the main text).}
    \label{TFIM scaling}
\end{figure}

In this section, we report the dynamics of the equilibrium infinite-range Ising model at finite temperature (in the absence of dissipation).
Specifically, we demonstrate via exact numerical simulation that the thermal critical point of this model belongs to the same (static and dynamic) universality class as the driven-dissipative Ising model in the weakly dissipative regime.
We start with the same Hamiltonian 
\begin{equation}
    H = -\frac{J}{N}S_x^2 + \Delta S_z\,.
\end{equation}
This Hamiltonian features a thermal phase transition to an ordered phase where the Ising $Z_2$ symmetry is broken at the critical temperature  \cite{das_infinite-range_2006} 
\begin{equation}\label{TFIM phase boundary}
    T_c = \frac{2 \Delta}{\ln\left(\frac{1+\Delta/2J}{1-\Delta/2J}\right)}\,.
\end{equation}
The Hamiltonian conserves the total spin (i.e., $[H, \Vec{S}] = 0$) which thus defines a good quantum number. In the angular-momentum basis defined by $\ket{S, m}$, the Hamiltonian becomes block diagonal with each block corresponding to a total spin $S$. However, each sector is highly degenerate with a multiplicity of $D(S)$. The multiplicity is given by  $D(N/2) = 1, D(N/2 - 1) = N-1, D(N/2 - 2) = N(N-3)/2$, and
\begin{equation}
    D(N/2 - p) =  \frac{N(N-1)...(N-p+2)}{p!}(N-2p+1)\,,
\end{equation}
for $3 \leq p \leq N/2$ \cite{das_infinite-range_2006}.
The thermal state is then given by
\begin{equation}
    \rho(\beta) = e^{-\beta H} = \bigoplus_{S = 0}^{N/2}\left( \bigoplus_{i=1}^{D(S)} e^{-\beta H_S} \right),
\end{equation}
which is to be understood as the direct sum over each unique spin sector with the corresponding multiplicity $D(S)$. We then numerically calculate the correlation function
\begin{equation}\label{tfim correlation}
\begin{split}
    C(t) &= \frac{1}{N}\langle \{S_x(t), S_x(0)\} \rangle = \frac{2}{N}\text{Re}\langle S_x(t) S_x(0) \rangle\\
    &= \frac{2}{N}\text{Re}\tr\left(e^{-iHt}S_x e^{iHt} S_x \rho(\beta)\right).
\end{split}
\end{equation}
A plot of the correlation function and its finite-size scaling behavior can be found in Fig. \ref{TFIM scaling}. There, we see that the dynamical exponent, defined via $t \sim N^{\zeta}$, is given by $\zeta = 1/4$ and that the dynamics is underdamped just like at the weakly-dissipative critical point of the driven-dissipative Ising model discussed in Sec. \ref{weakly dissipative critical properties}.

\section{Classical (Stochastic) Ising Model}
For completeness, here we introduce the classical stochastic Ising model \cite{oh_monte_2005}. The infinite-range (classical) Ising Hamiltonian is given by 
\begin{equation}
    \mathcal{H} = -\frac{J}{N}S^2\,,
\end{equation}
where $S = \sum_i^N s_i$ with the Ising spin variable $s_i = \pm 1$. While the Hamiltonian (being a $c$ number and commuting with all observables) does not impose any intrinsic dynamics, a stochastic, Glauber-type dynamics can be imposed via the (classical) master equation
\begin{align}
\begin{split}
    \frac{d}{dt} & P(\{s\};t) \\
    &= -\sum_{i=1}^N W(s_i \to -s_i, t)P(s_1, ..., s_i,..., s_N;t) \\
    &+ \sum_{i=1}^N W(-s_i \to s_i, t)P(s_1,...,-s_i,...,s_N;t)\,.
\end{split}
\end{align}
Here, $P(\{s\};t)$ denotes the probability that the system is in a spin configuration $\{s\}$ at time $t$, and $W(s_i \to -s_i, t)$ represents the transition probability rate of a spin flip at site $i$ and at time $t$. Under equilibrium conditions, the probability and transition rates satisfy detailed balance \cite{cardy_scaling_1996},
\begin{equation}
     \frac{W(s_i \to -s_i)}{W(-s_i \to s_i)} = \frac{P(s_1,...,-s_i,...,s_N)}{ P(s_1, ..., s_i,..., s_N)}\,,
\end{equation}
with the transition rate being of the Glauber type (characterizing a non-conserved order parameter),
\begin{equation}
    W(s_i \to -s_i) = \frac{1}{2 \tau_0}[1 - s_i \tanh{(\beta E)}]\,.
\end{equation}
Here, $\tau_0$ defines the characteristic time scale of Glauber dynamics, and $E = -(2J/N) \sum_i^N s_j$.
From here, one can simulate the relaxation of the system from a near-equilibrium state using Monte Carlo methods combined with the transition rate given above. Monte-Carlo simulations of the this model at criticality are consistent with a critical dynamical scaling where $t\sim N^{1/2}$ \cite{oh_monte_2005}.

\bibliographystyle{apsrev4-2}

\end{document}